%% file: main_arXiv.tex
\documentclass[fleqn,10pt]{wlscirep}
\title{A 3-D Projection Model for X-ray Dark-field Imaging}

\author[*,1,2]{Lina Felsner}
\author[*,1,3]{Shiyang Hu}
\author[1,2,3]{Andreas Maier}
\author[1]{Johannes Bopp}
\author[2,4]{Veronika Ludwig}
\author[4]{Gisela Anton}
\author[1]{Christian Riess}

\affil[1]{Pattern Recognition Lab, Department of Computer Science,}
\affil[2]{International Max Planck Research School for Optics and Imaging,}
\affil[3]{Erlangen Graduate School in Advanced Optical Technologies (SAOT),}
\affil[4]{Erlangen Centre for Astroparticle Physics (ECAP),}
\affil[ ]{Friedrich-Alexander University of Erlangen-Nuremberg, Erlangen, Germany}
\affil[*]{Both authors contributed equally to this work}
\affil[ ]{\textit{lina.felsner@fau.de, shiyang.hu@fau.de}}

\usepackage{xspace}
\usepackage{subfig}
\usepackage{amsfonts}
\usepackage{booktabs}
\usepackage{siunitx}
\usepackage{subfig}
\usepackage{dsfont}
\usepackage{tikz,pgfplots}
\usepackage{transparent}

\graphicspath{{./images/}}

\newcommand{\etal}{\textit{et~al.}}
\renewcommand{\vec}[1]{\mathbf{#1}}

\begin{abstract}
The X-ray dark-field signal can be measured with a grating-based Talbot-Lau
interferometer.  It measures small angle scattering of micrometer-sized
oriented structures.  Interestingly, the signal is a function not only of the
material, but also of the relative orientation of the sample, the X-ray beam
direction, and the direction of the interferometer sensitivity.
This property is very interesting for potential tomographically reconstructing
structures below the imaging resolution. However, tomographic reconstruction
itself is a substantial challenge.

A key step of the reconstruction algorithm is the inversion of a forward
projection model. In this work, we propose a very general 3-D projection model.
We derive the projection model under the assumption that the observed scatter
distribution has a Gaussian shape.

We theoretically show the consistency of our model with existing, more
constrained 2-D models. Furthermore, we experimentally show the compatibility
of our model with simulations and real dark-field measurements.  We believe
that this 3-D projection model is an important step towards more flexible
trajectories and, by extension, dark-field imaging protocols that are much
better applicable in practice.
\end{abstract}

\begin{document}

\maketitle  

\section{Introduction}
The probably most studied acquisition system for X-ray phase-contrast imaging 
is the Talbot-Lau grating interferometer.  
This system allows to measure a X-ray absorption image and two additional
images, namely the differential phase image and the dark-field image.
The X-ray dark-field measures ultra-small angle scattering, which is caused 
by inhomogeneities in materials at micrometer scale~\cite{Pfeiffer08:HXD,Wen09:FXS, Revol14:LFS}. 

Recently, X-ray dark-field imaging has received much attention for its potential applications in medical imaging and non-destructive material testing. 
The investigated applications in medical imaging span a wide range. Examples are
the identification of different lung diseases~\cite{weber2012investigation,yaroshenko2016visualization,hellbach2015vivo,hellbach2018depiction}, lung cancer~\cite{scherer2017x}, the identification of micro-calcifications~\cite{michel2013dark}, or the differentiation of kidney stones~\cite{scherer2015non}. 
Other examples are the detection of bone structures~\cite{Wen09:FXS} and fractures~\cite{hauke2018hairline} as well as brain connectivity~\cite{wieczorek2018brain}.
Also for material testing there are a wide range of application of the dark-field signal~\cite{Revol14:LFS,reza2014investigation,ludwig2018non,yang2014dark,prade2017nondestructive}.

The origin of the observed dark-field can have various reasons, such as small-angle x-ray scattering,  
an intra-pixel differential phase contrast that cannot be resolved, or even beam hardening \cite{koenig2016origin}.
While the effects are not clearly separable, we will focus on the dark-field created through small-angle scattering.
Two properties of the dark-field signal are particularly interesting. First,
ultra-small angle scattering is caused by structural variations at the scale of
few micrometers, which is significantly below the resolution of conventional
X-ray imaging systems~\cite{Yashiro10:OOV,Revol11:SPP}.  Second, a grating-based system
allows to measure the 3-D orientation of elongated micrometer-sized structures
such as fibers~\cite{Bayer13:PAD}. Traditional absorption X-ray
systems have to be able to fully resolve a fiber in order to measure its
orientation. In contrast to that, X-ray dark-field imaging enables to deduce
the fiber orientation of considerably smaller structures.

Jensen~\etal~\cite{Jensen10:DXD} and Revol~\etal~\cite{Revol12:OSX} explored the 
fundamentals of the dark-field orientation-dependency.  
In a tomographic setup, either the object or the imaging system rotates during the
acquisition. 
During the rotation, the relative orientation between object and system
changes, which leads to a variation in the signal.  This signal variation
allows to reconstruct the orientation of the structure.
There have been several reconstruction methods proposed in previous
works~\cite{Bayer14:RSV,Hu15:3DT,Malecki14:XTT,Vogel15:CXT,Wieczorek16:xrt,Schaff17:NID}.
However all of them are based on 2-D projection models of the 3-D structure.
This means that the models rely on the reconstruction of several 2-D slices and
are not compatible with true 3-D trajectories.

In this work, we aim to overcome this limitation by proposing a dark-field projection model over the 3-D space. 
This allows to directly estimate the 3-D structure, and to use sophisticated 3-D trajectories such as a helix.

\subsection{Talbot-Lau Interferometer}
\label{sec:talbot_lau}

\begin{figure}[tb]
	\centering
	\def\svgwidth{0.8\textwidth}  
	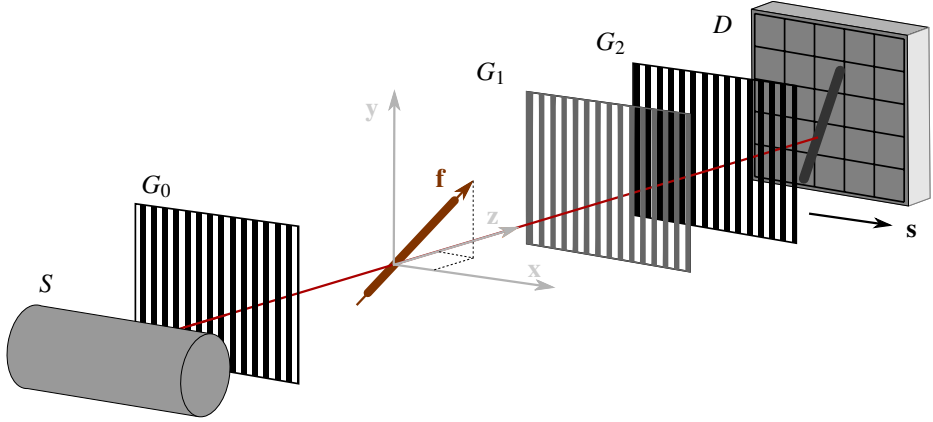
	\caption{Sketch of an X-ray Talbot-Lau interferometer. The setup consists
	of a source~$S$, a detector~$D$, and three
	gratings~$G_0$,~$G_1$, and~$G_2$ in between.  The
	global coordinate system is denoted as $\vec{x}$, $\vec{y}$, $\vec{z}$, an
	example fiber in the beam path is denoted as $\vec{f}$, and the sensitivity
	direction of the setup is denoted as $\vec{s}$.}
	\label{fig:TLI-sketch}
\end{figure}

The Talbot-Lau interferometer is a grating-based phase-contrast setup. 
A sketch of the system is shown in Fig.~\ref{fig:TLI-sketch}.
The system is an extension of conventional X-ray imaging setups, where
three gratings $G_0$, $G_1$, and $G_2$ are placed between
the source and detector. 
X-rays are generated by a conventional X-ray tube $S$. This X-ray tube
can be operated in an X-ray regime that is compatible with medical applications, 
such that a medical X-ray detector $D$~\cite{Pfeiffer06:PRD,Weitkamp06:TGI} can be used.
Grating $G_0$ effectively separates X-rays from the large source into
narrow slit sources that are individually coherent, but mutually incoherent. 
$G_1$ imprints a periodic phase modulation onto the wave front to create an interference pattern at the detector.
Both gratings $G_0$ and $G_1$ have periods that are in the range of few micrometers.
For operation with the much lower resolution of clinical X-ray detectors,
the interference pattern is sampled with the $G_2$ grating in
front of the detector, which also has a period in the range of micrometers. 
The sampling at the detector can be either
performed by slightly detuning the grating $G_2$, which leads to the Moir{\'e} effect~\cite{Takeda82:FTM,Bennett10:GBS,Bevins12:MCX}, or
by performing phase stepping~\cite{Pfeiffer06:PRD,Weitkamp06:TGI}. 
Both approaches
sample points on the interference curve, which can then be fitted by a sine.
In practice, two scans are performed, namely a reference scan without object
in the beam path, and an object scan with the object.  By comparing the
sinusoidal curve of both scans, it is possible to calculate the three quantities
absorption, differential phase, and dark-field. 
As in standard X-ray imaging, absorption is defined as the change in the
average intensity.  The differential phase is the angular shift of the sine.
The dark-field signal is given by the ratio of the amplitude of the sine over
the average intensity.  

For this work, it is important to note that all three signals are
created by sampling the sinosoidal function in one direction. We call this
direction the \textit{sensitivity direction} $\vec{s}$.
The sensitivity direction is perpendicular to the grating bars.

\subsection{Related Work}
\label{sec:related_work}

X-ray Tomography is performed by rotating either the X-ray setup or the object
during the acquisition. This rotation changes the orientation of the object
relative to the sensitivity direction. A key difference between traditional
X-ray absorption and dark-field is the impact of this relative orientation:
X-ray absorption is independent of the relative orientation, while X-ray
dark-field depends on it.

This makes a major difference for the choice of reconstruction algorithm. The
popular filtered backprojection (FBP) algorithm implicitly assumes that the
signal strength is independent of the viewing direction --- which does in
general not hold for X-ray dark-field imaging. 

The tomographic reconstruction, in general, requires the inversion of a projection model.
For the angle-dependent dark-field signal, several 2-D projection models were proposed, 
which are discussed briefly in the following.

\begin{figure}
	\centering 
	\subfloat[Revol~\cite{Revol12:OSX} projection model \label{fig:revol}]{
		\def\svgwidth{0.25\linewidth}
		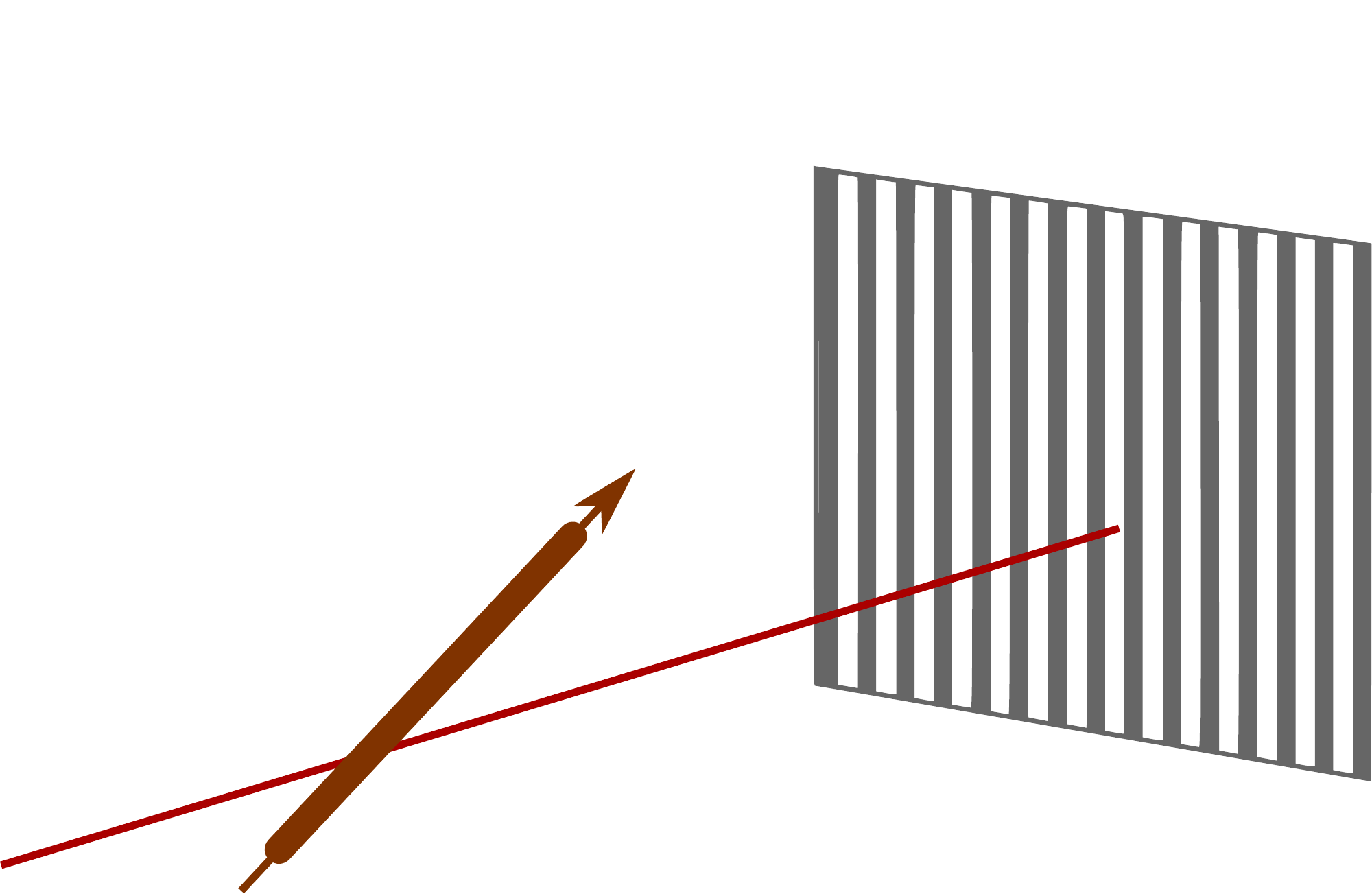
	}
	\qquad
	\subfloat[Bayer~\cite{Bayer13:PAD} projection model \label{fig:bayer}]{
		\def\svgwidth{0.25\linewidth}
		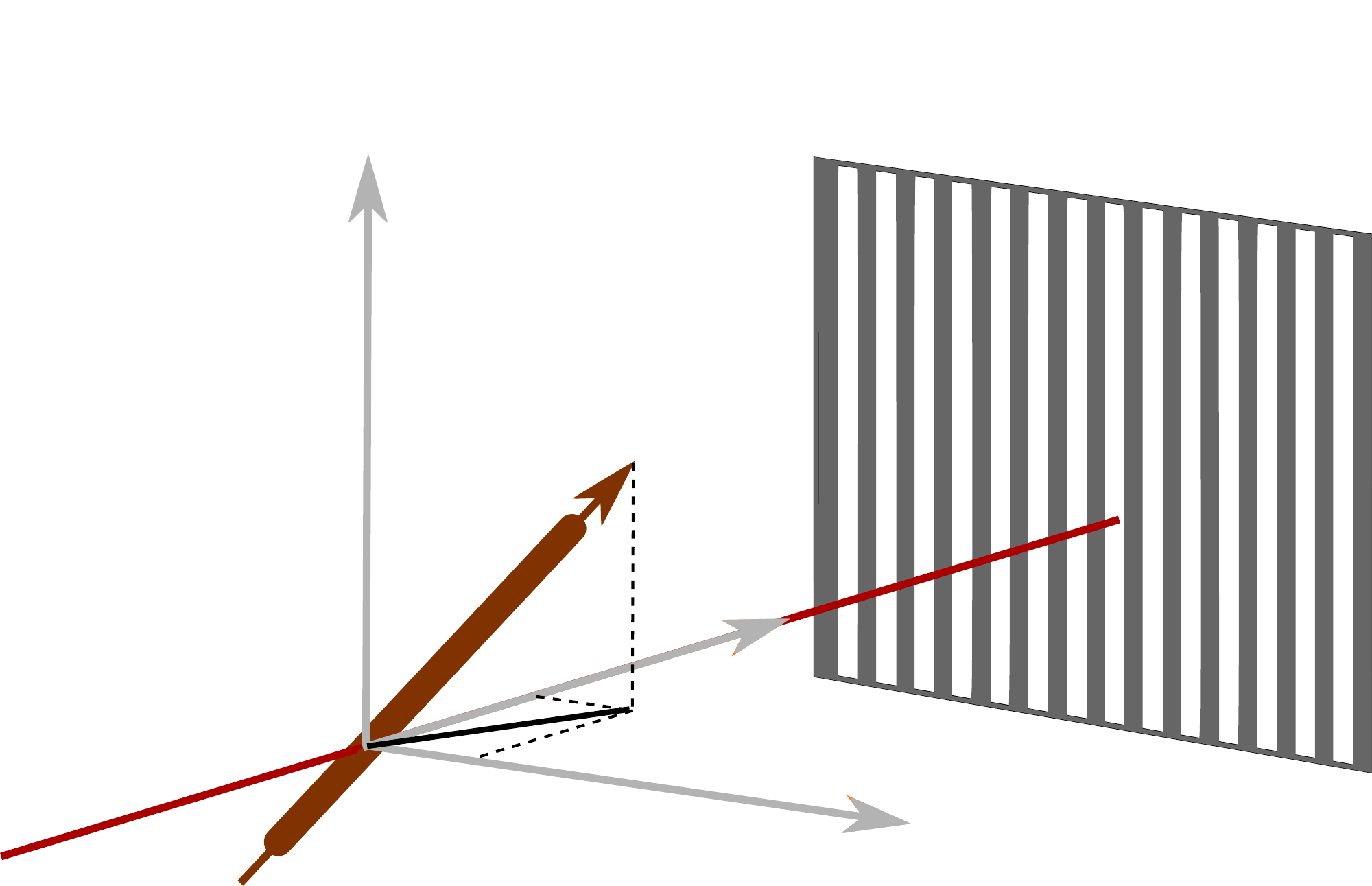
	}
	\qquad
	\subfloat[Schaff~\cite{Schaff17:NID} projection model \label{fig:schaff}]{
		\def\svgwidth{0.25\linewidth}
		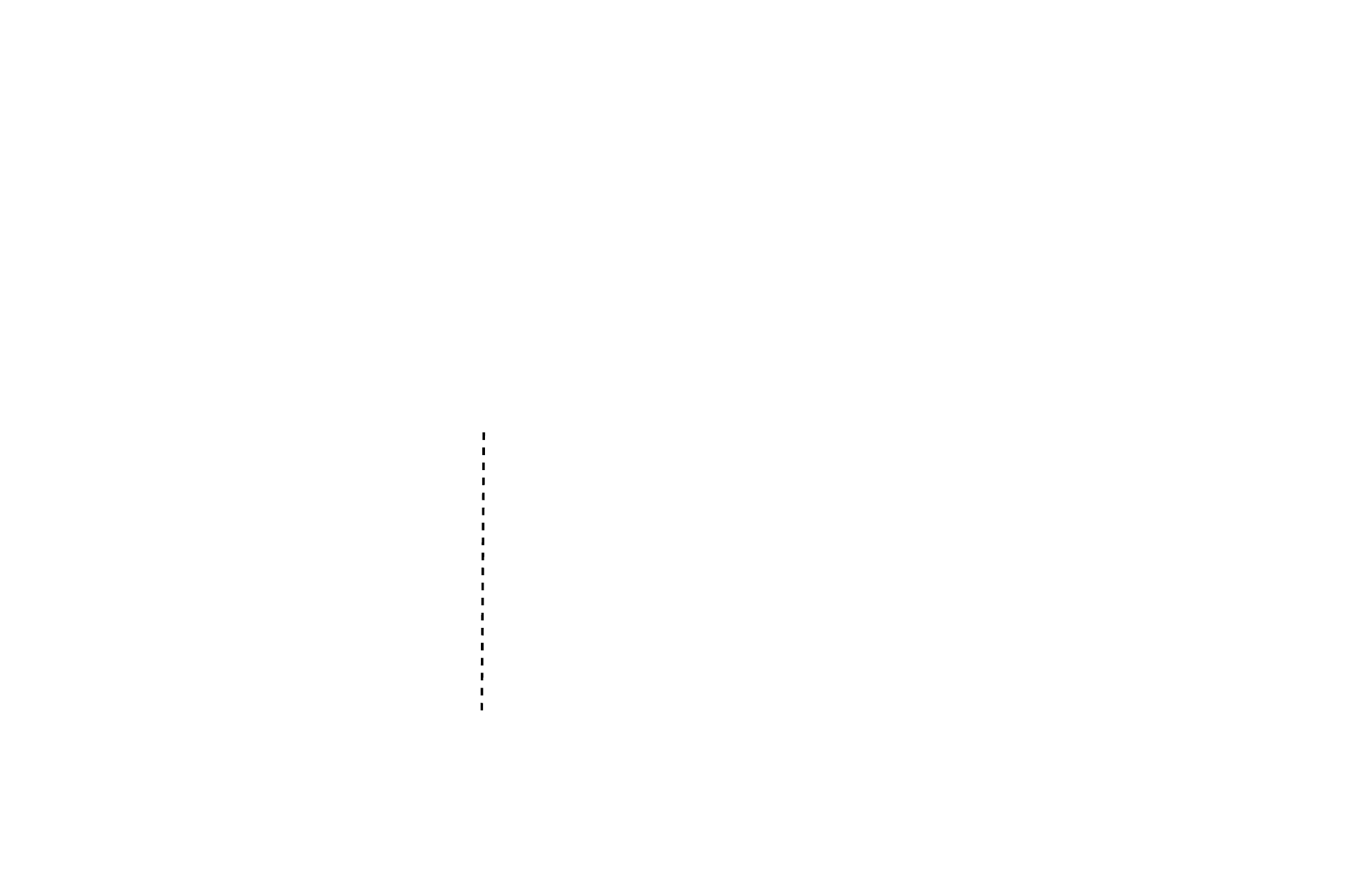
	}
	\caption{Sketch of three different 2-D projection models from previous works. 
		The rotation angle is given as $\omega$. The fiber vector is denoted as $\vec{f}$, 
		and $\theta$ and $\phi$ are the elevation angle and azimuthal angle, respectively. 
		$\vec{s}$ is the sensitivity direction.}
	\label{fig:projection_models}
\end{figure}

Jensen~\etal~\cite{Jensen10:DXD} first showed the angle dependency of
dark-field projections.
They rotated the object around the optical axis of the system, and found that
the variations in visibility can be described by the first two orders of the
Fourier expansion.  Shortly afterwards, Revol~\etal~\cite{Revol12:OSX} modeled
the dark-field scatter by a 2-D Gaussian function and showed that the logarithm
of the dark-field signal can be formulated as
\begin{equation}
\tilde{V}(\omega) = A + B \cdot \sin^2\left(\omega - \theta \right) \enspace,
\end{equation}
where $\omega$ is the rotation angle of the fiber around the optical axis, 
$\theta$ is the starting angle of the fiber in the $\vec{x}\vec{y}$-plane (see Fig.~\ref{fig:revol}) 
and $A$, $B$ are an isotropic and anisotropic contribution of the scatter, respectively. 
The projection models~\cite{Jensen10:DXD,Revol12:OSX} assume that the object is
rotated around the optical axis, which limits these models to thin sample
layers.  Malecki~\etal~\cite{malecki2013coherent} investigated the signal
formation for the superposition of layers with different fiber orientations.
They conclude that the dark-field signal can be represented as the line
integral along the beam direction over the anisotropic scattering components.

In order to describe the dark-field for thicker objects,
Bayer~\etal~\cite{Bayer13:PAD} proposed another projection model.  They showed
that the projection of a fibrous structure also depends on the azimuthal
angle $\phi$.  This corresponds to the angle of the fiber projection in the $\vec{x}\vec{z}$
plane in Fig.~\ref{fig:bayer}.  They derive the dark-field signal as 
\begin{equation}
\tilde{V}(\phi) = A + B \cdot \sin^2\left(\phi - \omega \right) \enspace . 
\end{equation}

The third projection model was proposed by Schaff~\etal~\cite{Schaff17:NID} 
and is shown in Fig.~\ref{fig:schaff}.  Here, the
grating bars are aligned along the 2-D trajectory, and the dark-field signal is
measured along along the rotation axis. Schaff~\emph{et al.} approximate this
signal as constant with respect to the tomographic rotation, such that the the
scattering strength only depends on the angle between the fiber and the
rotation axis.

This approximation simplifies the reconstruction, since a normal FBP algorithm can be used. 
However, for the two other projection models, the resulting signal per voxel
varies along the trajectory.
2-D object orientations are in this case reconstructed via iterative
reconstruction~\cite{Bayer14:RSV,Malecki14:XTT,Hu15:3DT,Vogel15:CXT,Wieczorek16:xrt}.
Among these works, Bayer~\etal~\cite{Bayer14:RSV} proposed a method to reconstruct 2-D in-plane orientations of fibers. 
Hu~\etal~\cite{Hu15:3DT} proposed to reconstruct the 3-D
orientation by combining two 2-D in-plane scans with different
trajectories. X-ray tensor tomography has been proposed by Malecki~\etal~\cite{Malecki14:XTT},
Vogel~\etal~\cite{Vogel15:CXT}, and Wieczorek~\etal~\cite{Wieczorek16:xrt} by combining multiple 2-D planes.

Since all projection models describe the dark-field only as a function
of one angle, it is only possible to reconstruct a 2-D slice.  The
reconstruction of the full 3-D distribution of oriented materials requires the
combination of scans from several trajectories, which overall
leads to quite complex acquisition protocols.
Malecki~\etal~\cite{Malecki14:XTT} reconstructed a scattering tensor by using
the model from Revol~\etal~\cite{Revol12:OSX} and rotated the sample into a
finite number of scattering directions.  Hu~\etal~\cite{Hu15:3DT} used the
model by Bayer~\etal~\cite{Bayer13:PAD,Bayer14:RSV} and used two 2-D
reconstructions to compute the 3-D fiber direction, while Schaff~\etal~\cite{Schaff17:NID} fit
a 3-D ellipse to individually reconstructed 2-D slices.

Previous works take different approaches to describe the 3-D nature of
X-ray dark-field, ranging from Gaussian distributions~\cite{Jensen10:DXD} over
a kartesian basis~\cite{Vogel15:CXT} to a spherical harmonics basis~\cite{Wieczorek16:xrt}.
However, to our knowledge, there exists to date no direct 3-D reconstruction
algorithm. One of the reasons for this may be the fact that 
a reconstruction method requires the inversion of a projection model, which 
to our knowledge has not been defined yet in 3-D.

The definition of a 3-D model makes it possible to use 3-D dark-field trajectories.
For example, the helix is a popular 3-D trajectory with favorable properties in
traditional absorption tomography. In this case, Tuy’s condition for absorption
image can be applied, and the completeness of such a certain trajectory can be
shown~\cite{maier2015discrete}. In principle, a similar system can be pursued
for dark-field tomography if a well-described 3-D trajectory is available.
As long as only 2-D trajectories can be used, the best known acquisition
schemes that fully measure the scattering orientations are still quite
complex~\cite{Sharma2017DesignOA}.

\subsection{Contributions and Organization of this Work}
In this work, we propose a fully three-dimensional X-ray dark-field projection
model.  Previous works are limited to descriptions of 2-D projections of the
dark-field signal, which limits the reconstruction to 2-D scatter projections,
and constrains the trajectories to 2-D. In contrast, the proposed model enables
the use of an arbitrary scanning geometry, and overcomes the need for
combining several 2-D trajectories. The proposed model allows to use established 3-D scanning trajectories
to acquire the 3-D scatter distribution, like for example a helical geometry.
Furthermore, it enables the development of novel 3-D geometries that aim at
optimizing the recovery of directional information for specific clinical
examinations or visual inspection tasks.

Additionally, the proposed model is very general. It allows to freely choose
the ray direction and the sensitivity direction. That way, it overcomes the
restriction of earlier works to parallel beam geometries. Instead, it allows to
model a cone beam, which is of major importance for many popular hardware
designs, like for example a line scanner.

We only use the assumption that the scatter distribution of the dark-field
signal is a 3-D Gaussian, and we derive the general projection model from that.
Furthermore, we discuss the impact of additional constraints if they are
available, and demonstrate the consistency with existing 2-D models. 
In the experiments, we show that the proposed model accurately fits predicted
dark-field values from a wave simulation as well as from real experiments.

The paper is organized as follows. 
Section~\ref{sec:projection_model} provides a mathematical derivation of the proposed model, which 
describes the dark-field signal formation in a very general way.
Afterwards, in Sec.~\ref{sec:additional_constraints}, we discuss the impact of
additional constraints on the model and show that our model is consistent with
the 2-D projection models discussed in Sec.~\ref{sec:related_work}.
Experiments that link the predicted signal to simulations and actual measurements are presented in Sec.~\ref{sec:experiments}.
We conclude the paper in Sec.~\ref{sec:conclusions}.

\section{Proposed X-ray Dark-field Projection Model}
\label{sec:projection_model}

\begin{figure}
	\centering 
	\subfloat[2-D Dark-field Signal\label{fig:2Ddfmodel}] {
		\def\svgwidth{0.25\textwidth}  
		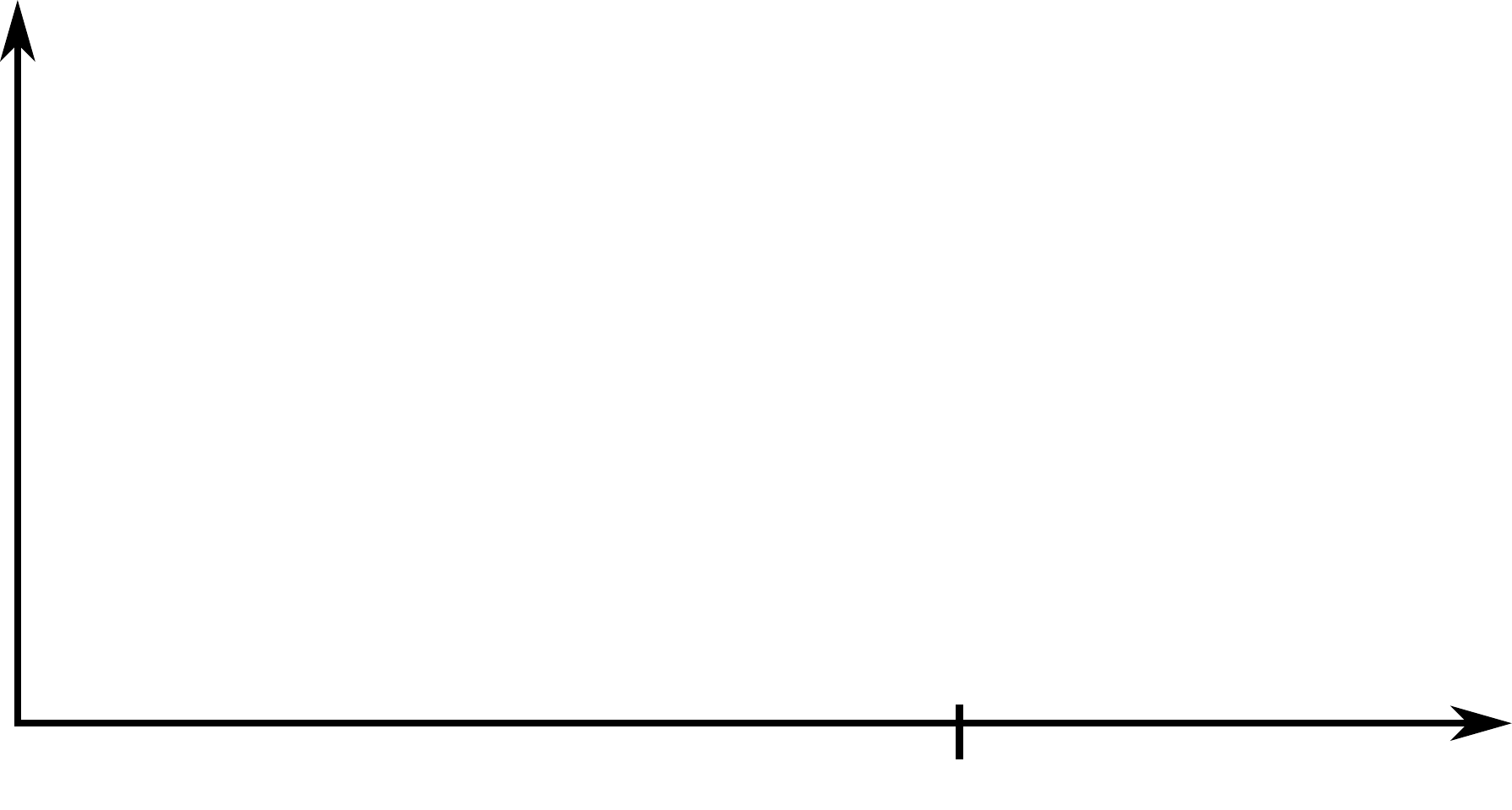
	}
	\qquad
	\subfloat[2-D Scattering Model\label{fig:2Dmodel}] {
		\def\svgwidth{0.25\textwidth}  
		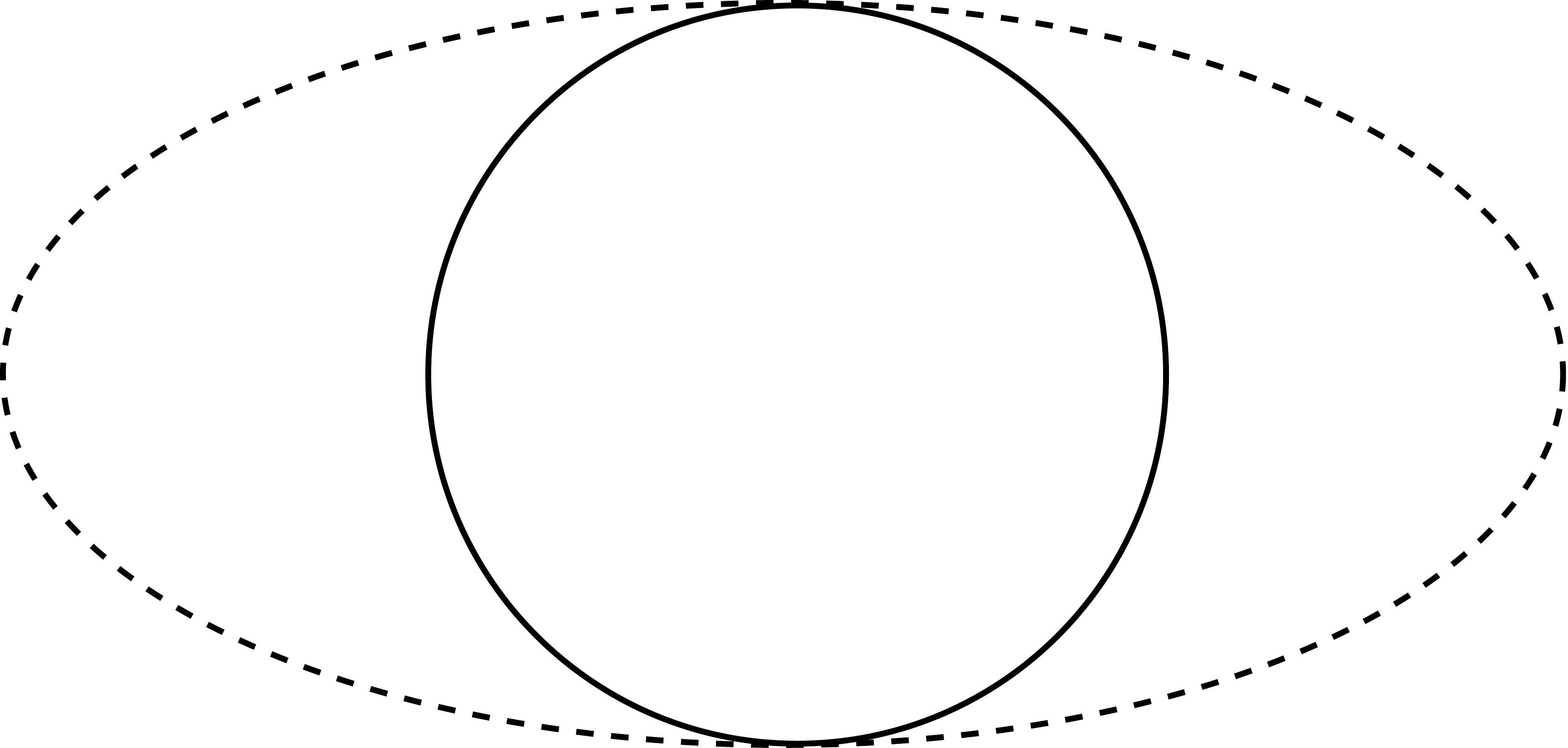
	}
	\caption{Isotropic ($d_\text{iso}$) and Anisotropic ($d_\text{aniso}$) parts for the 2-D dark-field signal and scattering model.}
	\label{fig:models2d}
\end{figure}

The X-ray dark-field signal measures the X-ray small angle scattering of
microstructures in a sample. X-ray dark-field scattering has the special
property that its observed magnitude depends on the relative orientation of
the sample in the setup.
To characterize the signal, we use the notion of \emph{isotropic} and
\emph{anisotropic} scattering components. This notion was originally introduced
for 2-D projection models. A schematic sketch of this model is shown in
Fig.~\ref{fig:models2d}. Here,
the isotropic component scatters 
in all directions equally strongly, independent of the sample or setup
orientation. Conversely, observations of scatter of the anisotropic component
vary with the sample and setup orientation.

Thus, if a sample scatters purely isotropically, its signal is independent of
the orientation.  Such a signal can be 
reconstructed in a similar manner as X-ray absorption. However, if a sample
exhibits partially anisotropic scatter, the signal formation depends on the
orientation and thus becomes considerably more difficult to reconstruct. In
particular, any algorithm for 2-D or 3-D reconstruction has to explicitly take
the direction-dependent signal variation into account.

In order to model the signal formation, we introduce the notion of a
\emph{fiber}. A fiber is a microstructure that exhibits a mixture of isotropic
and anisotropic scattering. The derivation of the model is organized as follows.
First, we expose the relationship between a fiber and its associated scatter
distribution in Sec.~\ref{sec:relationship}.
In Sec.~\ref{sec:fiber_projection},  we show how the fiber is projected by the
X-ray onto the sensitivity direction.
In Sec.~\ref{sec:scattering}, we show how the projected image of the fiber is
converted to a scatter distribution.
In Sec.~\ref{sec:complete_model}, we state the complete model, which is the
actually observed dark-field signature for a sample point.  Afterwards,
Sec.~\ref{sec:line_integrals} shows how the measured signal is expressed as
line integrals.

The dark-field signal formation depends on three quantities, namely the
directions of the X-ray, the dark-field sensitivity direction, and the orientation of
the fiber.  We describe a very general model that considers all three
quantities as arbitrary vectors in 3-D.  This generality has several
advantages. It allows us to model not only a system with parallel beam and a
perpendicular sensitivity direction, but instead arbitrary acquisition
geometries.  Examples for such more general system designs are the use of a
cone-beam scanning geometry, which influences the ray direction, or the use of
a curved X-ray detector, which results in different sensitivity directions. It
also allows to model a 3-D helical scanning trajectory, which requires
flexibility in all these quantities.

\subsection{Relationship between Fiber and Scatter Distribution}
\label{sec:relationship}

We make the simplifying assumption that a fiber has the shape of a cylinder.
More specifically, the fiber cross section is assumed to be a circle, and the
height of the cylinder is assumed to be at least as long as the radius
of that circle. The isotropic scattering component is mainly determined by the
radius of the circle. The anisotropic scattering component is connected to the
size of the cylinder, and will be more rigorously defined in
Sec.~\ref{sec:scattering}.

\begin{figure}[tb]
\centering
	\subfloat[Fiber and its associated Gaussian scatter distribution.\label{fig:scatter}] {
		\def\svgwidth{0.3\textwidth}  
		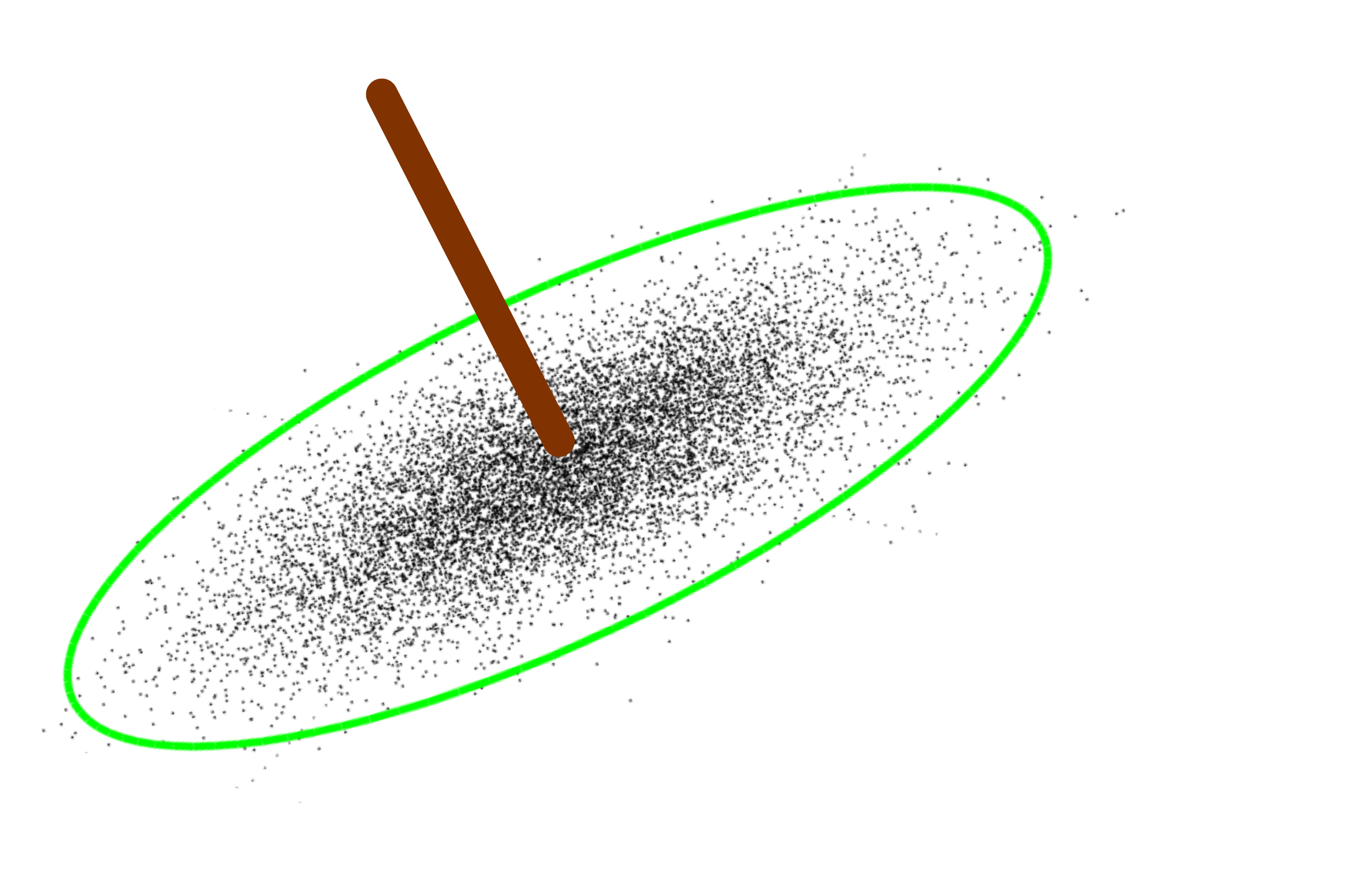
	}
	\qquad
	\subfloat[Basis vectors of the 3-D Gaussian scatter spheroid, \newline where $\sigma_1 = \sigma_2 > \sigma_3$. 
	Adapted from~\cite{Spheroid}.\label{fig:spheroid}] {
		\def\svgwidth{0.4\textwidth}  
			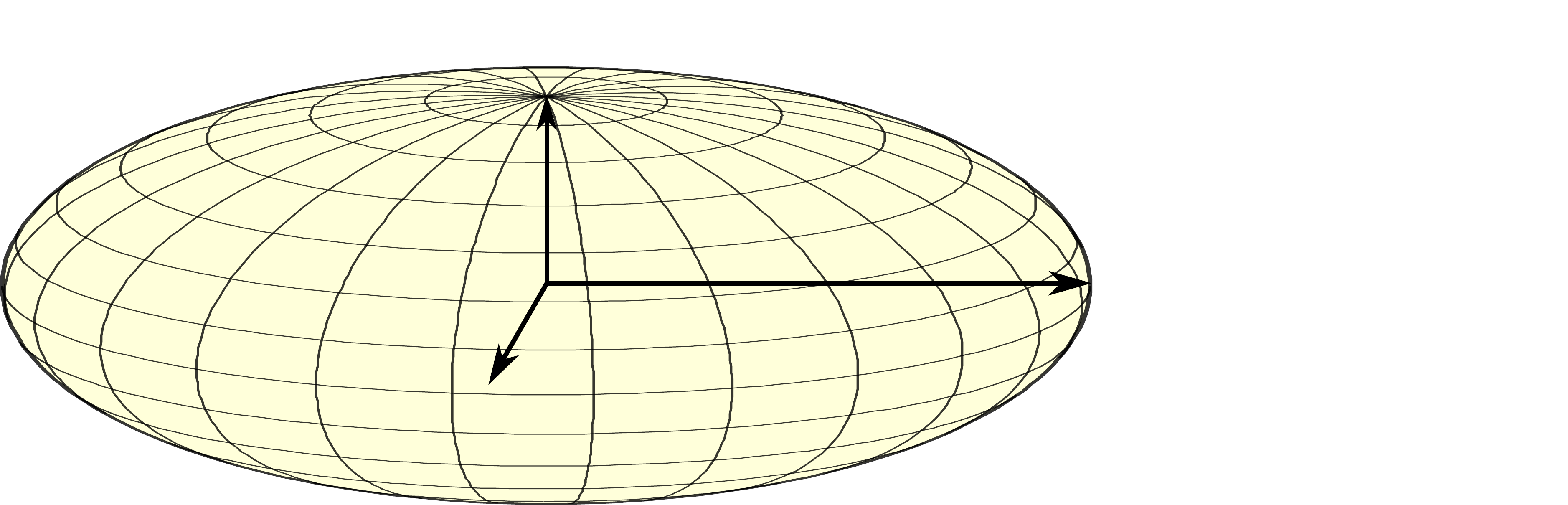	
	}
\caption{Illustrations of the 3-D Gaussian scatter distribution.}
\label{fig:scatter3D}  
\end{figure}

Mathematically, we represent a fiber as a 3-D vector $\vec{f}$ in
$\mathbb{R}^3$, where the vector is parallel to the cylinder axis.
The observed fiber creates dark-field scatter. 
Scatter is not deterministic, and therefore commonly described as a
distribution.

For the following discussion, we are only interested in the relative
orientations of the fiber and its associated scatter. Thus, without loss of
generality, we assume that a fiber and its scatter distribution are rooted in
the origin of the coordinate system. We assume that the shape of a fiber
scatter distribution is a 3-D Gaussian $g(\vec{x})$, which is in line with
earlier models on 2-D scatter distributions~\cite{Jensen10:DXD}. Then,
\begin{equation}
g( \vec{x})= \frac{1}{\sqrt{(2\pi)^3} \left| \Sigma \right|} \cdot 
\exp \left(-\frac{1}{2} (\vec{x}^\top \Sigma^{-1} \vec{x})\right) \enspace ,
\end{equation}
where $\Sigma$ denotes the $3\times 3$ covariance matrix. The shape of
$g(\vec{x})$ is completely described by $\Sigma$.

We make the mild assumption that this covariance matrix $\Sigma$ can be
diagonalized (which is satisfied for any non-trivial 3-D Gaussian scatter
observation).
Then, the eigenvalues of $\Sigma$ describe the scatter strength with respect to
its eigenbasis spanned by the eigenvectors $\vec{b}_1$, $\vec{b}_2$,
$\vec{b}_3$. The eigenvalues correspond to the variances, i.e., the squared
standard deviations along each principal axis of the distribution:
\begin{equation}
	\Sigma = 
	\begin{pmatrix}
	\sigma_1^2 & 0& 0\\
	0 & \sigma_2^2 &0 \\
	 0 & 0& \sigma_3^2\\
	\end{pmatrix} \enspace .
\end{equation}

These variances have a special distribution, which comes from the particular
case of a scattering fiber: the main scattering direction of the fiber is the
2-D subspace that is perpendicular to $\vec{f}$. This is illustrated in
Fig.~\ref{fig:scatter}. All scatter directions within this 2-D subspace are indistinguishable. 
As a consequence, the two largest eigenvalues $\sigma_1^2$ and $\sigma_2^2$ are
identical, i.e., $\sigma_1^2 = \sigma_2^2$. The weakest scattering is observed in
the direction of $\vec{f}$, which is quantified by the smallest eigenvalue
$\sigma_3^2 \le \sigma_1^2$. This is illustrated in Fig.~\ref{fig:scatter}.
The eigenvector $\vec{b}_3$ is associated with the smallest eigenvalue
$\sigma_3^2$ and parallel to $\vec{f}$. More specifically, both vectors are
identical with the exception that their sign might be flipped, i.e., $\vec{b}_3 = \pm \vec{f}$. 
The restrictions on the eigenvalues induces that the shape of the scattering function is a oblate spheroid. 
A 3-D sketch of the eigenvalues is shown in Fig.~\ref{fig:spheroid}.

\subsection{3-D Fiber Projection Model}
\label{sec:fiber_projection}

The dark-field signal formation depends on three geometric vectors, namely the
direction of the X-ray, the dark-field sensitivity direction, and the orientation of the fiber. 
Ultimately, we seek the projection of the fiber along the ray direction onto the sensitivity direction.
This is a mapping from the 3-D fiber vector onto a (1-D) scalar value.
A non-parallel X-ray projection, e.g. from a cone beam, is modelled by a rotation of the fiber.
The sensitivity direction can
have an arbitrary orientation in space. To relate the fiber direction and the sensitivity direction,
we introduce a virtual plane that is perpendicular to the X-ray. Both the fiber and the sensitivity
direction are projected onto that plane. Then, the 2-D projection of the fiber onto the sensitivity
direction in the plane is performed. The resulting equations show that the plane cancels, and that
the projection of the fiber onto the sensitivity direction can be written as a
scalar product. The mathematical details are presented below.

\begin{figure}
	\centering 
		\subfloat[Sketch of the projection of vector $\vec{f}$ and $\vec{s}$ on the plane $\vec{E}$. 
		Please note, that this sketch does not match the exact mathematical description, 
		since the plane $\vec{E}$ does not lie in the origin. \label{fig:plane_sideways}] {
	\def\svgwidth{0.6\textwidth}  
	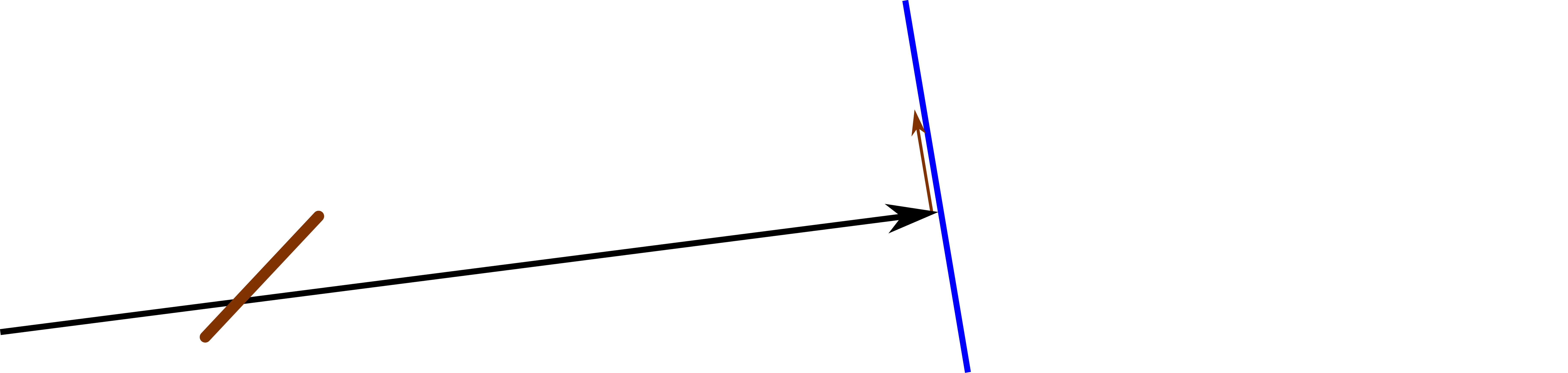
}
	\quad
		\subfloat[Planar view on the plane $\vec{E}$ with the projected vectors $\vec{f}'$ and $\vec{s}'$. 
		\label{fig:plane_planar_view}] {
			\def\svgwidth{0.3\textwidth}  
				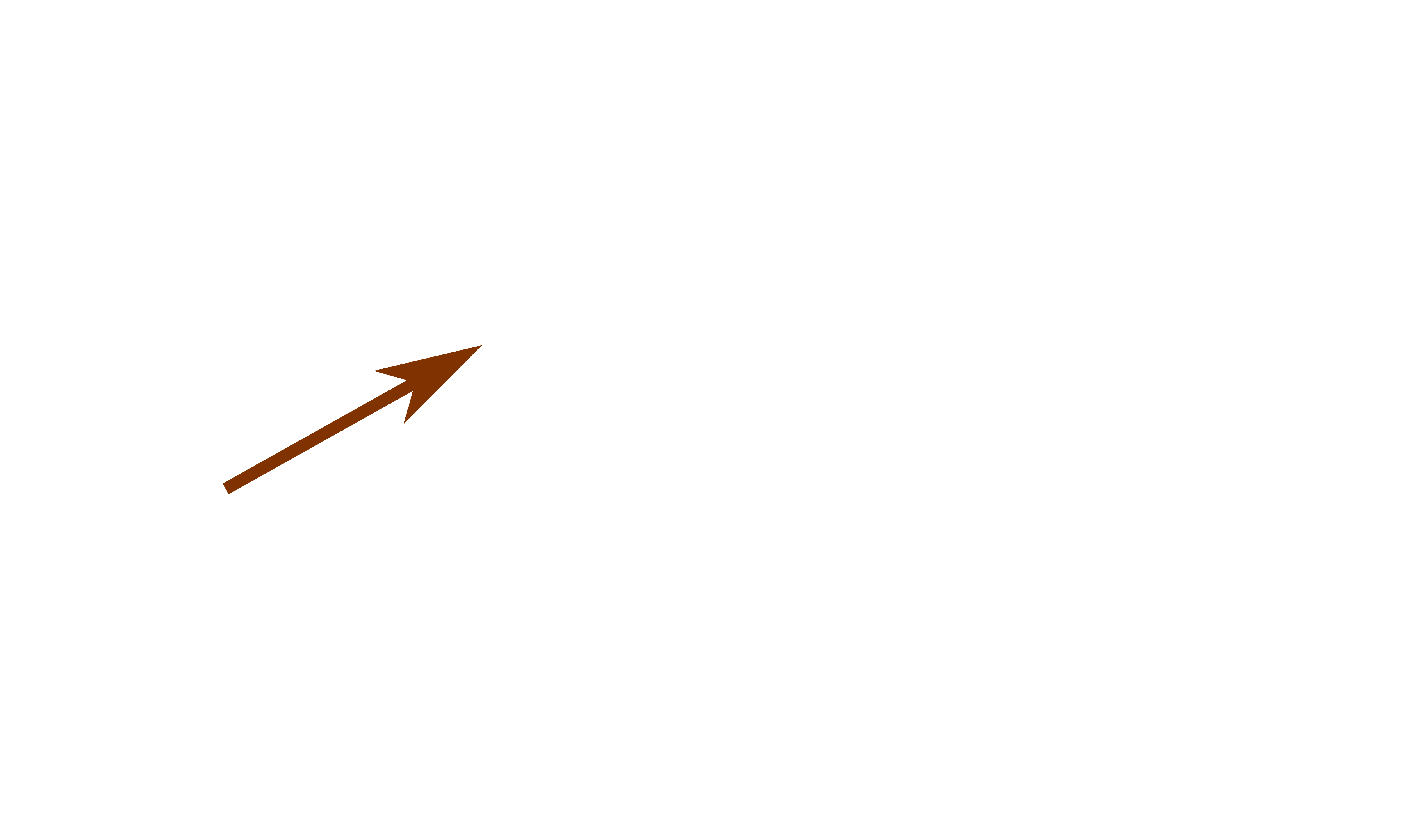
		}

	\caption{Projection of the fiber direction and the sensitivity direction onto the plane.}
	\label{fig:plane}
\end{figure}

Let us consider a single fiber vector $\vec{f}$. Without loss of generality, this
fiber is located in the origin of our world coordinate system. 
The X-ray dark-field projection ray $\vec{r}$ passes through that fiber, and
thereby also the origin of the coordinate system.

In imaging systems, all X-rays that form one projection are typically modelled
as either parallel or diverging from a central ray $\vec{c}$. This changes the
relative orientation between $\vec{r}$ and the fiber vector $\vec{f}$. To
correct for the diverging ray, we denote the angle of divergence as $\alpha$,
and rotate the fiber in the plane spanned by $\vec{c}$ and $\vec{r}$ in the
opposite direction. The corresponding rotation matrix is denoted as
$\vec{R}_{\alpha}$. In the case of parallel projection, $\vec{R}_{\alpha}$ is
the $3\times 3$ identity matrix.

We project the fiber $\vec{f}$ onto a plane~$\vec{E}$ that is perpendicular to the X-ray
direction $\vec{r}$.
For this projection, we use orthogonal projections instead of perspective
projections of the scatter pattern. This is possible, because the projection
of a fiber signature onto the detector is in the range of micrometers, but a
single detector pixel is typically two orders of magnitude larger.

An orthogonal projection of a 3-D vector onto a plane can be performed with an
inner product between the vector and a transformation matrix consisting of the
3-D coordinates of the 2-D basis.
We define the 2-D projection plane as a plane where $\vec{r}$ is the normal
vector. Since $\vec{r}$ passes through the origin, we find it convenient to
choose the plane to also pass through the origin, i.e.,
\begin{equation}
\vec{E} = \left( \vec{r}^{\text{ortho}}_1, \vec{r}^{\text{ortho}}_2 \right)\enspace,
\end{equation}
with $\vec{E} \in \mathbb{R}^{3\times 2}$ where
$\vec{r}^{\text{ortho}}_1$ is a vector perpendicular to $\vec{r}$, i.e., $\vec{r}^\top \vec{r}^{\text{ortho}}_1 = 0$, and
$\vec{r}^{\text{ortho}}_2 = \vec{r} \times \vec{r}^{\text{ortho}}_1$ is the
second vector spanning the plane, also perpendicular to $\vec{r}$. 
This projection is visualized in Fig.~\ref{fig:plane_sideways}. 

The projection of the fiber along the ray and onto the 2-D plane $\vec{E}$ is
then given as product of the rotated fiber $\vec{f}$ with $\vec{E}$, i.e.,
\begin{equation}
\vec{f}' = (\vec{R}_{\alpha}\vec{f})^\top \vec{E}\enspace,
\end{equation}
where $\vec{f}' \in \mathbb{R}^2$ is now a two-dimensional vector in the plane
$\vec{E}$.  

The sensitivity direction $\vec{s}$ denotes the direction along which the X-ray
dark-field signal can be measured. It is a 3-D vector with an arbitrary
orientation. To relate the fiber with the sensitivity direction, we also
project $\vec{s}$ onto plane $\vec{E}$. Analogously to the fiber-plane
projection, we also use here an orthogonal projection. The 2-D projection of $\vec{s}$ on $\vec{E}$ is
\begin{equation}
\vec{s}' = \vec{s}^\top \vec{E}\enspace.
\end{equation}
The projection of both vectors $\vec{f}'$ and $\vec{s}'$ on $\vec{E}$ are shown
in Fig.~\ref{fig:plane_planar_view}.

To determine the alignment of the fiber $\vec{f}$ with sensitivity direction
$\vec{s}$, the inner product is computed, i.e.
\begin{align}
\vec{f}'' & = \vec{f}'^\top \cdot \vec{s}' \\
          & = \left((\vec{R}_{\alpha}\vec{f})^\top \vec{E}\right)^\top \cdot \left( \vec{s}^\top \vec{E} \right) = \left(\vec{E}^\top \left(\vec{R}_{\alpha}\vec{f}\right) \right) \cdot \left( \vec{s}^\top \vec{E} \right) \enspace.
\label{eqn:proj_fiber_to_sensitiviy}
\end{align}
Equation~\ref{eqn:proj_fiber_to_sensitiviy} can be simplified by noting that
the inner product commutes, which leads to
\begin{align}
\vec{f}'' & = \left( \vec{s}^\top \vec{E} \right) \cdot \left(\vec{E}^\top \left(\vec{R}_{\alpha}\vec{f} \right) \right) \\
          & = \vec{s}^\top \cdot \left( \vec{R}_{\alpha}\vec{f}\right) \enspace,
\label{eqn:fiber_to_sensitivity}
\end{align}
since $\vec{E} \vec{E}^\top = \vec{I}$. Equation~\ref{eqn:fiber_to_sensitivity}
shows that the projection of the fiber through the system onto the sensitivity
direction reduces to directly computing the inner product between the fiber and
the sensitivity direction.

Note that in the case of a cone beam, the rotation of the fiber by
$\vec{R}_\alpha$ can also be replaced by a rotation of the sensitivity
direction $\vec{s}$ in the opposite direction. While we believe that the
rotation of the fiber $\vec{f}$ is more intuitive, it may be preferable for
an actual implementation of a reconstruction algorithm to rotate the
sensitivity direction $\vec{s}$, since $\vec{s}$ is a given quantity from the
setup geometry, and $\vec{f}$ is the unknown variable.

\subsection{3-D Projection Model for Scattering}
\label{sec:scattering}

The projection of the fiber onto the sensitivity direction can be translated into the
projection of the scatter. The scatter is the actually observed quantity in the
imaging system. The inverse of this conversion links the observations to the
unknown fiber direction.

As discussed in Sec.~\ref{sec:relationship} 
the scatter distribution for a given fiber $\vec{f}$ is given as
\begin{equation}
	\vec{f} \mapsto \left(\begin{array}{ccc}\sigma_1^2 & & \\ & \sigma_2^2 & \\ & & \sigma_3^2 \end{array}\right) (\vec{b}_1 \quad \vec{b}_2 \quad \vec{b}_3)\enspace,
\end{equation}
where $\sigma_1^2 = \sigma_2^2$, and $\vec{b}_1$, $\vec{b}_2$, $\vec{b}_3$
are an orthogonal basis. 

In Sec.~\ref{sec:fiber_projection} we considered the transformation from the 3-D fiber to a 1-D signal.
We now want to describe this transformation for the scattering distribution.
Since the distribution is described by an orthogonal basis, 
we can transform the basis vectors $\vec{b}_i$ individually to get the transformation.

Since we defined our projection for an arbitrary fiber $\vec{f}$, 
we can use the same mapping for each basis vector $\vec{b}_i$. 
The measured projection of the scattering component $i$ is then given as
\begin{align}
  \sigma_i^{2\,\prime\prime} &= \vec{s}^\top \cdot \left( \vec{R}_{\alpha} \left(\sigma_i^2 \vec{b}_i \right) \right)\\ &=\sigma_i^2 \cdot \left( \vec{s}^\top \cdot \left( \vec{R}_{\alpha}  \vec{b}_i \right) \right) \enspace.
\end{align}

Under the consideration that the scattering distribution is symmetric, the
variance may not depend on the sign of the basis vectors $\vec{b}_i$, and its
oscillation has to be of period $\pi$. In analogy to previous 2-D
models~\cite{Revol12:OSX,Bayer14:RSV}, both requirements are addressed by
squaring the inner product. The projected variance is thus
\begin{equation}
{\hat\sigma}_i^{2\, \prime\prime} = \sigma_i^2 \cdot \left( \vec{s}^\top \cdot \left(\vec{R}_{\alpha}  \vec{b}_i \right) \right)^2 \enspace.
\label{eq:projected_variance}
\end{equation}

\subsection{Complete 3-D Dark-Field Projection Model}
\label{sec:complete_model}

With the individual projections of the fiber and the scattering distribution at
hand, we combine both in this section to directly describe the scatter
distribution for a given fiber. To this end, we use the introduced notions of
\textit{isotropic} and \textit{anisotropic} scattering.
The isotropic part results in an equal amount of scatter in all directions,
while the anisotropic part depends on the relative orientation of the fiber,
ray direction, and sensitivity direction. The goal is to describe the 1-D
dark-field scattering signal in dependency of the fiber, since the fiber is the
quantity that shall eventually be reconstructed.

The observed dark-field signal is modeled as
\begin{equation}
d = d_\text{iso} + d_\text{aniso} \left(\vec{s}^\top (\vec{R}_{\alpha}\vec{f})\right)^2 \enspace .
\label{eqn:3Ddfmodel}
\end{equation}
Here, we again square the scaling factor of the anisotropic part to resemble
the fact that the signal has a period of $\pi$ instead of $2\pi$.
Analogously, the amount of isotropic and anisotropic scattering is also defined
over the variances of the 3-D scattering function. Thus, the isotropic component
is given as
\begin{equation}
d_\text{iso} = \sigma_{1}^2 = \sigma_{2}^2 \enspace ,
\end{equation}
while the anisotropic component is
\begin{equation}
d_\text{aniso} = - \left( \sigma_1^2 - \sigma_3^2 \right) \enspace .
\end{equation}
To define the anisotropic component as the subtraction from the isotropic
scattering may appear counter-intuitive at first glance. However, it allows to
directly represent the fiber $\vec{f}$ in the model. We believe that it is
useful for building a reconstruction algorithm on top of the model to have the 
fiber direction directly accessible, since it is the primary quantity of interest.

The derivation to use the fiber vector in the  of Eq.~\ref{eqn:3Ddfmodel} 
comes from the projected variance in Eq.~\ref{eq:projected_variance}. 
If we consider the smallest scattering component, we observe
\begin{equation}
{\hat\sigma}_3^{2\, \prime\prime} = \sigma_3^2 \cdot \left( \vec{s}^\top \cdot \left(\vec{R}_{\alpha}  \vec{b}_3 \right) \right)^2 \enspace.
\label{eq:sigma_3}
\end{equation}
Since the eigenvector $\vec{b}_3$ and the fiber vector $\vec{f}$ are collinear, 
we can substitute $\vec{b}_3$ in Eq.~\ref{eq:sigma_3} by $\vec{f}$. This leads to
\begin{equation}
{\hat\sigma}_3^{2\, \prime\prime} = \sigma_3^2 \cdot \left( \vec{s}^\top \cdot \left(\vec{R}_{\alpha}  \vec{f} \right) \right)^2 \enspace,
\end{equation}
which is used to get the dark-field model in Eq.~\ref{eqn:3Ddfmodel}.

In 2-D models, the isotopic component is defined as the amount that scatters in all directions
equally, while the anisotropic component is defined as an additional component in the direction
perpendicular to the fiber direction.
In 3-D, a direct adaptation of this approach is somewhat more complicated, 
since the additional scatter of the fiber perpendicular to its main axis forms a 2-D subspace. 
We argue that the concept of isotropic and anisotropic components is not really transferable to the 3-D case.
In 3-D, one can interpret Eq.~\ref{eqn:3Ddfmodel} as the reduction of
observed scatter in the direction of the main axis of the fiber, which is
mathematically correct, yet somewhat counter-intuitive.

The projection of the dark-field does not only depend on the scattering strength described in Eq.~\ref{eqn:3Ddfmodel}, 
but also on the length of the projection rays through the fiber.
Since the fiber is assumed to be smaller than one pixel, the dark-field per voxel $\vec{x}$ can be expressed  as:
	\begin{equation}
		d(\vec{x}) = C(\vec{f},\, \alpha,\, \sigma_1,\, \sigma_3) \cdot d \enspace ,
		\label{eq:fiber_per_voxel}
	\end{equation}
where $C(\vec{f},\, \alpha,\, d_\text{iso},\, d_\text{aniso})$ is a function describing the average length through the fiber cylinder, 
dependent on the fiber direction, the ray direction and linked to isotropic and anisotropic values.

\subsection{Dark-Field Line Integrals}
\label{sec:line_integrals}

In standard X-ray projection imaging, the measured signal intensity is the line integral along the X-ray beam line $L$.
Malecki~\etal~\cite{malecki2013coherent} showed that the superposition of dark-field signals results in a line integral along the beam direction. 
The dark-field signal is similar to the absorption signal given by
\begin{equation}
	D_L = \exp\left[-\int_L d(\vec{x}) \, \mathrm{d}L \right] \enspace .
	\label{eq:3D}
\end{equation}
Here, the line integral is only influenced by the object geometry.

To make the dark-field more interpretable and achieve a non-linear intensity transformation as oftentimes used in X-ray absorption processing, 
several authors use $-\log(D_L)$~\cite{bech2010quantitative}. 

\section{Impact of Additional Constraints on the Model}
\label{sec:additional_constraints}

The proposed projection model is very general. In this section we will show how
specific assumptions allow for simplifications. In particular, we show that the
model is consistent with the more constrained 2-D projection models by
Revol~\cite{Revol12:OSX}, Bayer~\cite{Bayer13:PAD}, and
Schaff~\cite{Schaff17:NID}.  We will now show that we are consistent with these
if we constrain our model to parallel beams and a circular 2-D trajectory. As
sketched in Fig.~\ref{fig:projection_models}, we define for all three 2-D
models the sensitivity direction as $\vec{s} = (1,\, 0,\, 0)^\top$.

In a parallel beam geometry, the rotation matrix $\vec{R}_\alpha$ simplifies to the identity, 
i.e., $\vec{R}_\alpha = \mathds{1}$. 
Consequently, the anisotropic component only depends on the relative orientation of the fiber and sensitivity direction, thus
\begin{equation}
d = d_\text{iso} + d_\text{aniso} \left(\vec{s}^\top \vec{f}\right)^2 \enspace .
\end{equation}

Revol~\etal ~rotates the fiber around the ray direction. 
Thus, the fiber orientation of $\vec{f}$ in the $\vec{xy}$ plane depends on the starting angle $\theta$ and the rotation angle $\omega$.
We will denote this dependency as $\vec{f}(\omega)$. The fiber is then given as
\begin{equation}
	\vec{f}(\omega) = 	
\begin{pmatrix}
\cos (\omega - \theta) &-\sin (\omega - \theta) &0\\
\sin (\omega - \theta) &\cos (\omega - \theta) &0\\
0&0&1
\end{pmatrix}
\begin{pmatrix}
f_x\\f_y\\f_z
\end{pmatrix}	\enspace ,
\end{equation}
with $\vec{f} = (f_x,\, f_y,\, f_z)^\top$.
The dark-field model thus becomes $d = d_\text{iso} + d_\text{aniso} \left((1,\, 0,\, 0)  \vec{f}(\omega) \right)^2$, 
which can be transformed into the original formulation $A + B \cdot \sin^2\left(\omega - \theta \right)$.

The mapping to the model by Bayer~\etal can be performed in a similar manner. 
Here, the fiber is rotated around the $\vec{y}$-axis. Then,
\begin{equation}
\vec{f}(\omega) =
\begin{pmatrix}
\cos (\phi -\omega)  & 0 & \sin (\phi - \omega) \\
0         & 1 &  0          \\
-\sin (\phi - \omega) & 0 & \cos (\phi - \omega)
\end{pmatrix}
	\begin{pmatrix}
f_x\\f_y\\f_z
\end{pmatrix}
\enspace , 
\label{eq:bayer}
\end{equation}
which results in the original formulation $A + B \cdot \sin^2\left(\phi \right)$.

The model by Schaff~\etal constrains the sensitivity direction parallel to the rotation axis. 
Then, the projection of the fiber vector (as stated in Eq.~\ref{eqn:fiber_to_sensitivity}) is given as
\begin{equation}
\vec{f}(\omega)'' = (1,\, 0,\, 0) 
\left(\begin{pmatrix}
1  & 0 & 0 \\
0         & \cos (\omega - \Theta) &  -\sin (\omega - \Theta)         \\
0 & \sin (\omega - \Theta) & \cos (\omega - \Theta)
\end{pmatrix}
\begin{pmatrix}
f_x\\f_y\\f_z
\end{pmatrix}
\right)  = f_x \enspace , 
\end{equation}
which is constant.
It is interesting to note, however, that the model does not consider variations
in intersection lengths through the fiber. In practice, the signal is only then
approximately constant, if the fiber exhibits only a small elevation angle. In
this case, the intersection length is nearly identical for different rotation
angles
$\omega$.

In summary, the proposed model can be transformed into each of the three
existing 2-D models with the addition of suitable constraints.
At the same time, however, the proposed model is general enough to also
represent a full 3-D space with an arbitrarily oriented X-ray, fiber, and
sensitivity.

\section{Experiments and Results}
\label{sec:experiments}

In this section we experimentally evaluate the proposed 3-D dark-field projection model.
We sequentially evaluate different aspects of the model, to mitigate the
combinatorial complexity of evaluating the full parameter space.
The evaluated aspects of the model are
\begin{enumerate}
	\item Dark-field projection model (Equation~\ref{eqn:3Ddfmodel})
	\item Dark-field signal of a single fiber (Equation~\ref{eq:fiber_per_voxel})
	\item Dark-field measurements (Equation~\ref{eq:3D}).
\end{enumerate}
The corresponding experiments are described and discussed in the sections~\ref{sec:res}, \ref{sec:cxi}, and \ref{sec:real}, respectively. 
To evaluate the proposed projection model, we compare the results to simulated
and real dark-field signals in Sec~\ref{sec:cxi} and \ref{sec:real}. 
	
\subsection{Dark-field Projection Model}
\label{sec:res}

\begin{figure}[tb]
	\centering
		\def\svgwidth{0.25\textwidth}  
	\subfloat[\label{Eins}]{			\def\svgwidth{0.25\textwidth}  	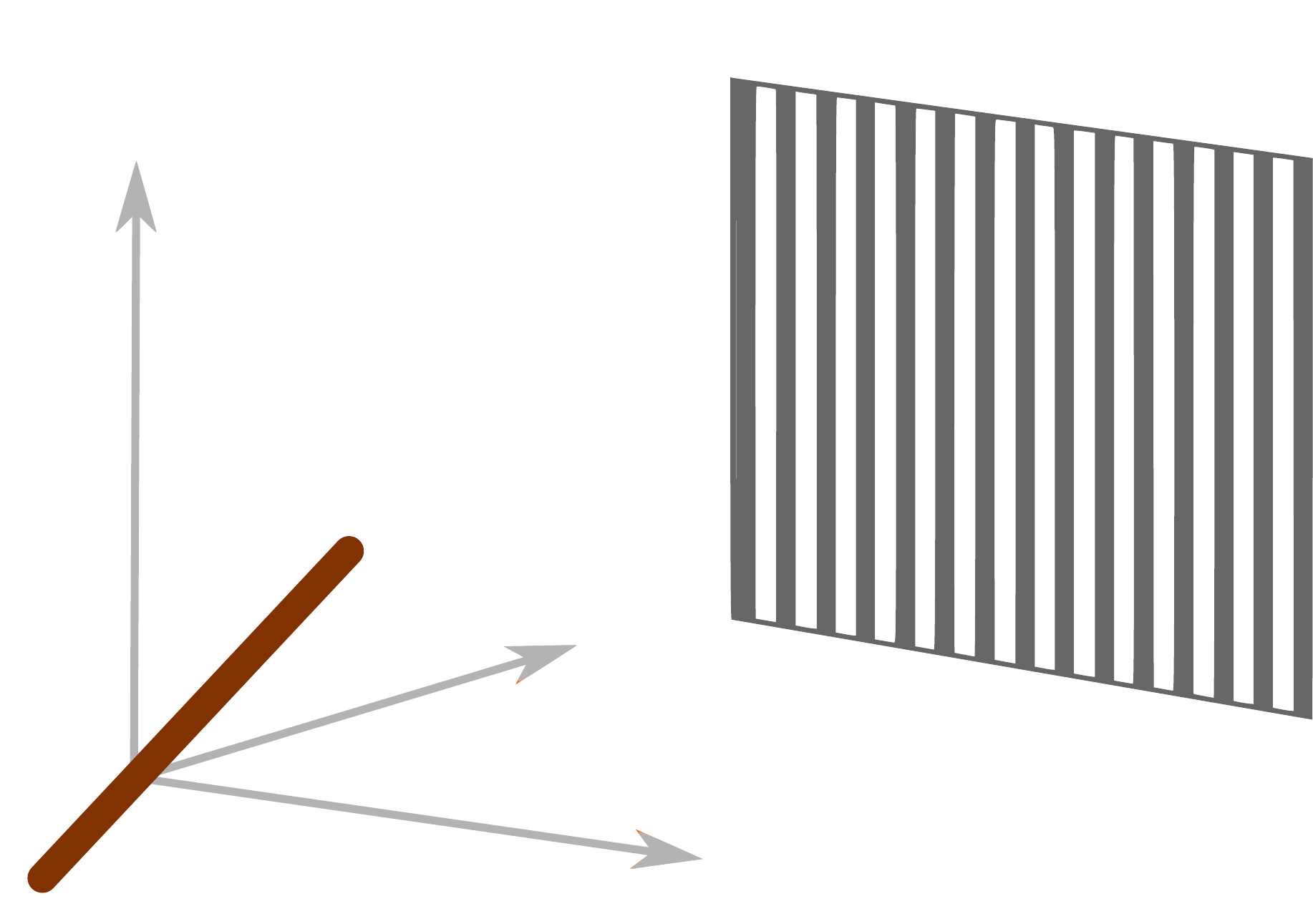	}
	\quad
	\subfloat[\label{Zwei}]{			\def\svgwidth{0.25\textwidth}  	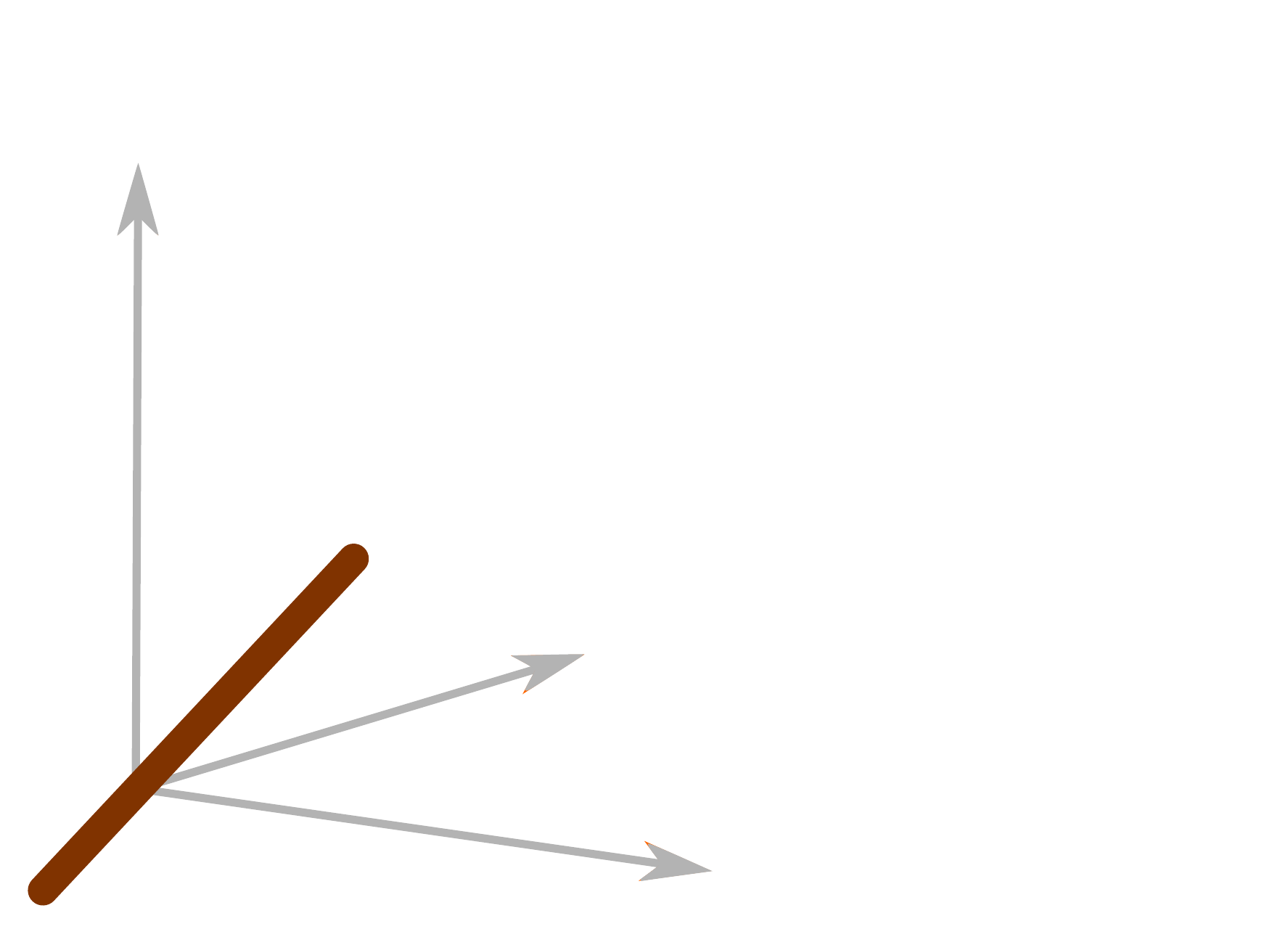	}
	\quad
	\subfloat[\label{Drei}]{			\def\svgwidth{0.22\textwidth}  	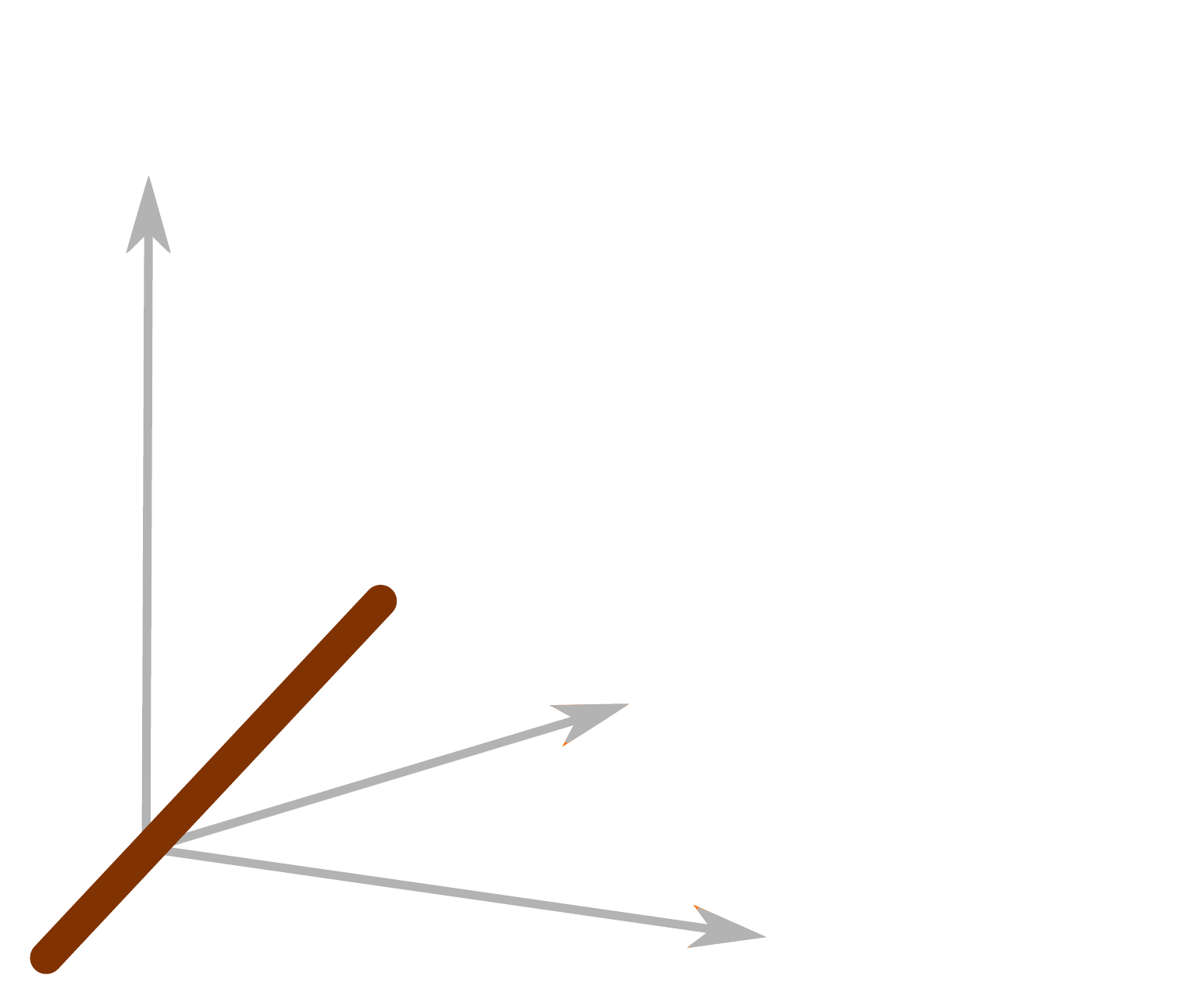	}

	\caption{Three different trajectories to evaluate the dark-field projection model (Sec.~\ref{sec:res}). 
		(a, b) Circle trajectory. (c) Helical trajectory. The sensitivity direction $s$ is shown for each scanning mode.}
	\label{fig:bvm-traj}
\end{figure}

The formulation of the dark-field in Eq.~\ref{eqn:3Ddfmodel} is sufficiently
flexible to describe different trajectories and sensitivity directions.  In
this experiment, we show the dependency of the dark-field on the X-ray
direction and sensitivity direction. 
To this end, we simulate three different trajectories as shown in Fig.~\ref{fig:bvm-traj}.
We evaluate the dark-field for two different fiber vectors, 
both with larger scattering coefficient $\sigma_1 = \sigma_2 = 1$ and smaller scattering coefficient $\sigma_3 = 0.5$.
The fiber directions are $\vec{f} = \{1, 0, 0\}$ and $\vec{f} = \{1, 1, 1\}$,
respectively.

All three trajectories have a source-isocenter distance of \SI{600}{\milli\meter} and source-detector distance of \SI{1200}{\milli\meter}.
We simulate two circular 2-D trajectories over $360^\circ$ with an angular
increment of $1.5^\circ$. Both trajectories have different sensitivity
directions. The sensitivity direction $\vec{s}$ for trajectory (a) in Fig.~\ref{fig:bvm-traj} 
can be represented by the vector $\{1,\, 0\}$ in the detector plane and lies therefore in the rotation plane.
Trajectory (b) in Fig.~\ref{fig:bvm-traj} has the sensitivity direction along the rotation axis, 
i.e. the vector $\{0,\, 1\}$ in the 2-D detector plane.
The third trajectory (c) is a helical 3-D trajectory, also with an angular increment of $1.5^\circ$ and a pitch $h = 0.5$. 
The sensitivity direction is aligned with the helical trajectory. 
The sensitivity direction is in all cases always chosen perpendicularly to the
projection ray in order to not introduce an additional scaling factor from the inner product in Eq.~\ref{eqn:3Ddfmodel}. 

\begin{figure}[tb]
	\centering
	\subfloat[Line plot of dark-field for different trajectories and 
	fiber vectors. The corresponding grating orientations are shown in Fig.~\ref{fig:bvm-traj} and $\sigma_1 = 1$, $\sigma_3 = 0.5$.  \label{fig:bvm-result}] {
		\def\svgwidth{0.15\textwidth}  
\input{line_plot_trajectories.tikz}	
	}
	\qquad
	\subfloat[Projected vectors shown on the spheroid for the fiber vector $\vec{f} = \{1, 1, 1\}$. 
	Magenta ist the trajectory (a).  Light blue shows the signal for trajectory (b). \label{fig:iso_aniso}] {
\includegraphics[height=0.25\textwidth]{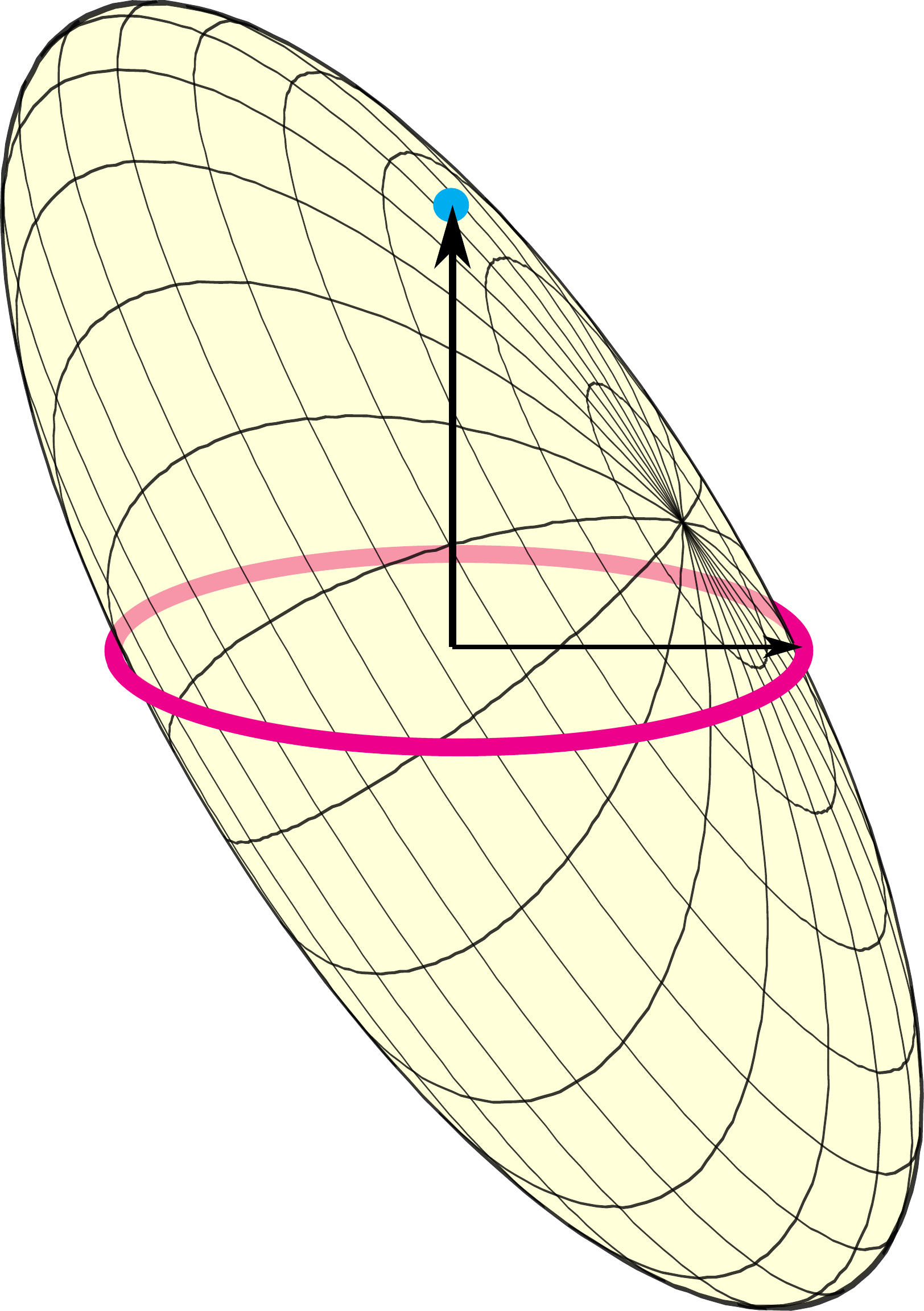}
	}
	\caption{Projections of the dark-field projection model from Eq.~\ref{eqn:3Ddfmodel} according to the trajectories in Fig.~\ref{fig:bvm-traj}.}
	\label{fig:bvm-projections}  
\end{figure}

The resulting dark-field signals are shown in Fig.~\ref{fig:bvm-result} over the rotation angle $\omega$. 
In magenta, the dark-field signals for both fibers on the circular trajectory (a)
are shown. They oscillate with a regular sinosoidal. The fiber that lies within
the rotation plane (magenta, dotted line) reaches the minimum and maximum
theoretically possible dark-field values. The elevated fiber (magenta, solid line)
creates an overall stronger signal that, due to the elevation, never reaches
the minimum. This is also illustrated in the example scattering spheroid in
Fig.~\ref{fig:iso_aniso}. Here, the magenta circumference indicates the measured
scatter intersection for trajectory (a) for the elevated fiber vector $\vec{f}
= \{1, 1, 1\}$.

For the second 2-D trajectory (b), the sensitivity direction is aligned with
the rotation axis. As shown in the light blue lines in Fig.~\ref{fig:bvm-result},
this results in a constant signal for both fiber directions. In this case, the
elevated fiber $\vec{f} = \{1, 1, 1\}$ (solid, light blue) creates a weaker signal.
The scattering strength of the elevated fiber is shown as a light blue dot on
the spheroid in Fig.~\ref{fig:iso_aniso}.

The most complex trajectory is the helical 3-D trajectory (trajectory (c) in Fog.~\ref{fig:bvm-traj}).
The dark-field signals for both fibers are shown in black in Fig.~\ref{fig:bvm-result}.
Due to the constant change in angle between the fiber and the sensitivity direction, the signal change is not symmetric over the $360^\circ$.  
While this observation holds for both fibers, it is more pronounced for fiber vector $\vec{f} = \{1, 1, 1\}$ (solid black line).

The experiments demonstrate how the dark-field signal depends on the X-ray
direction and the sensitivity direction.  Furthermore, the experiment also
shows that the dark-field signal behaves differently for 2-D and 3-D
trajectories. These differences in the predicted signals demonstrate the need
of a 3-D projection model for performing a true 3-D reconstruction.

\subsection{Dark-field Signal of a Single Fiber}
\label{sec:cxi}

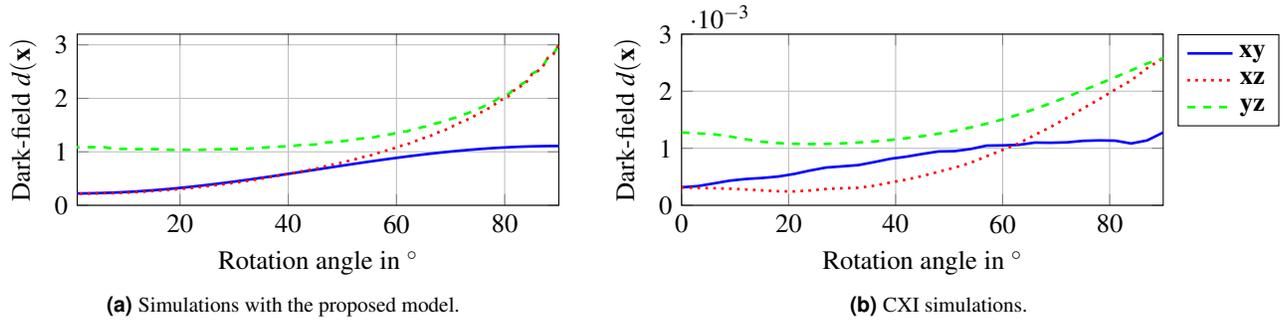
\begin{figure}[tb]
	\centering	
	\def\svgwidth{0.25\textwidth}  
	\subfloat[Simulations with the proposed model.\label{fig:sim-planes}]{	
		\input{planes_cxi_1.0_2.0_1.5_0.3.tikz}	}
	\quad			
	\subfloat[CXI simulations.\label{fig:sim-cxi}]{
		\input{cxi_simulations.tikz}}
	\caption{Dark-field signals (Eq.~\ref{eq:fiber_per_voxel}) for the rotation of a fiber in the three planes spanned by the axes of the coordinate system. 
		(a) generated signal with the proposed model. (b) wavefront simulations.}
	\label{fig:simulation_plane}
\end{figure}

To verify the proposed dark-field signal for a complete fiber (Eq.~\ref{eq:fiber_per_voxel}), we compare it to numerical simulations.
We simulate the dark-field with a simulation framework for coherent X-ray imaging (CXI) from Ritter~\etal~\cite{ritter2014simulation}.
The setup parameters for the simulation are chosen as follows
The $G_1$ is placed at \SI{0.01}{\meter}, with a period of \SI{4.37e-6}{\meter}, a height of \SI{5e-6}{\meter} and a duty-cycle of $0.5$.
The $G_2$ is placed at \SI{0.17}{\meter}, with a period of \SI{2.4e-6}{\meter}, a height of \SI{300e-6}{\meter} and a duty-cycle of $0.5$.
Both gratings are simulated as gold. The detector is positioned immediately behind $G_2$.
The size of the focal spot is set to \SI{10}{\micro\meter}, and the pixel width is set to \SI{50}{\micro\meter}.
The design energy of the system is \SI{25}{\kilo\electronvolt}.
The simulated object is a teflon fiber (PTFE) with a radius of \SI{1.79}{\micro\meter} and a length of \SI{15}{\micro\meter}.

We set the fiber parameters in the model to the eigenvalues $\sigma_1 = 1.5$ and $\sigma_2 = 0.3$, which correspond to the teflon fiber parameters.

The negative logarithm of the simulated signal is shown in Fig.~\ref{fig:simulation_plane}. Figure~\ref{fig:sim-planes} shows the result with our model, 
while Fig.~\ref{fig:sim-cxi} shows the result from the CXI simulations.
The dark-field signals are shown in the three planes spanned by the coordinate system, namely $\vec{xy}$-, $\vec{xz}$-, and $\vec{yz}$-plane.
Overall, the proposed model and the wavefront simulations agree very well.
The different magnitude between the two signals simulations, are due to the range chosen parameter spaces for the experiments.

The signal in the $\vec{xy}$-plane (blue, solid line) changes only slightly in both simulations.
In this plane, the change between the larger and the smaller scatter eigenvalues is observed.
While our model leads to a distinct sinusoidal change (Fig.~\ref{fig:sim-planes}), 
it is more noisy for the CXI simulation (Fig.~\ref{fig:sim-cxi}) due to numerical instabilities. 

The dark-field signal in the $\vec{xz}$- and $\vec{yz}$-plane (red and green
lines, respectively) increases with the rotation angle $\omega$, i.e., with
increasing inclination of the fiber into the beam direction.
Equation~\ref{eqn:3Ddfmodel} predicts for such an inclination no increase, but
instead a constant signal. However, the reason for the increasing signal lies
in the increased intersection length of the ray through the fiber, as denoted in Eq.~\ref{eq:fiber_per_voxel}.
It is also interesting to note that this increase is even stronger than the
difference of the scatter eigenvalues.
The green signal that shows the $\vec{yz}$-plane is affected by both effects,
the scattering eigenvalues and intersection length through the fiber.
Overall, the wavefront simulations and the predicted values of the proposed
model agree very well.

\subsection{Dark-field Measurements}
\label{sec:real}
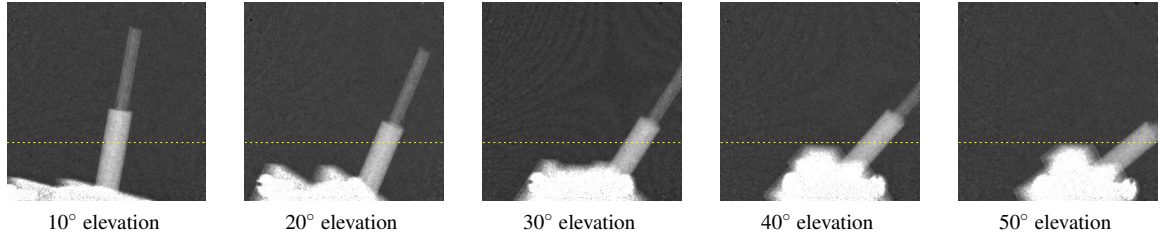
\begin{figure}
	\captionsetup[subfigure]{labelformat=empty}
	\centering	
 	\subfloat[$10^\circ$ elevation \label{10}]{
 					\begin{picture}(75,75)%
 					\put(0,0){\includegraphics[width=0.15\textwidth]{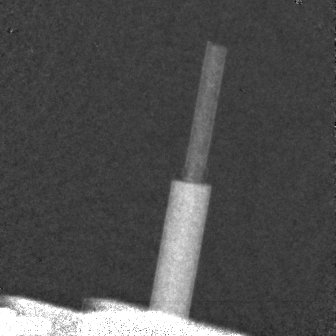}}%
 					\multiput(0,22)(2,0){38}{\color[rgb]{1,1,0}\line(1,0){0.7}}
 					\end{picture}
 		}
	\quad
	\subfloat[$20^\circ$ elevation \label{20}]{
					\begin{picture}(75,75)%
					\put(0,0){\includegraphics[width=0.15\textwidth]{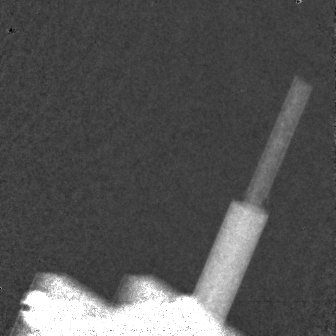}}%
					\multiput(0,22)(2,0){38}{\color[rgb]{1,1,0}\line(1,0){0.7}}
					\end{picture}
		}
	\quad
	\subfloat[$30^\circ$ elevation \label{30}]{
					\begin{picture}(75,75)%
					\put(0,0){\includegraphics[width=0.15\textwidth]{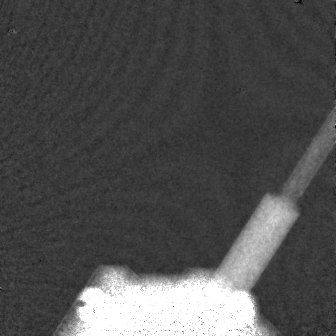}}%
					\multiput(0,22)(2,0){38}{\color[rgb]{1,1,0}\line(1,0){0.7}}
					\end{picture}
		}
	\quad
	\subfloat[$40^\circ$ elevation \label{40}]{
					\begin{picture}(75,75)%
					\put(0,0){\includegraphics[width=0.15\textwidth]{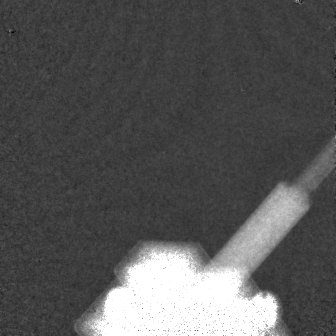}}%
					\multiput(0,22)(2,0){38}{\color[rgb]{1,1,0}\line(1,0){0.7}}
					\end{picture}
		}
	\quad
	\subfloat[$50^\circ$ elevation \label{50}]{
			\begin{picture}(75,75)%
			\put(0,0){\includegraphics[width=0.15\textwidth]{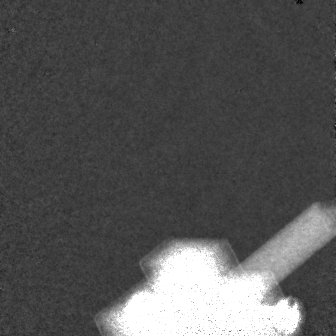}}%
		\multiput(0,22)(2,0){38}{\color[rgb]{1,1,0}\line(1,0){0.7}}
			\end{picture}
		}
	\caption{Carbon-rods at different elevation angles. First projection image with a windowing between $-0.31$ and $+1.0$. 
		For the measured dark-field signal we used the central pixels of the rod along the yellow, dashed line. }
	\label{fig:cfk-rods}
\end{figure}

We show that the proposed complete projection model (Eq.~\ref{eq:3D}) also
agrees with experimentally obtained dark-field measurements.
%
The real data consists of a carbon fiber reinforced polymer rod with a diameter of \SI{4}{\milli\meter}.
The fiber was measured at five different tilting (elevation) angles, namely
$10,~20,~30,~40,$ and $50^\circ$. A projection image at each angle is shown in
Fig.~\ref{fig:cfk-rods}. For each tilting angle, $100$ projection images are
taken over a rotation of $180^\circ$ at \SI{40}{kVp} and \SI{40}{\milli\ampere}. 
The measurements were performed with a Siemens MEGALIX CAT Plus 125/40/90-125GW 
medical X-ray tube using a tungsten anode. The used X-ray flat panel detector was a 
PerkinElmer Dexela 1512 with \SI{74.8}{\micro\meter} pixel pitch,
running in $2 \times 2$ binning mode for processing a faster read-out resulting in a \SI{150}{\micro\meter} pixel pitch. 
For each projection 30 phase-steps with \SI{0.1}{\second} acquisition time were used. The gratings have a
period of \SI{24,39}{\micro\meter} for $\text{G}_0$,
\SI{2,18}{\micro\meter} for $\text{G}_1$, and \SI{2,4}{\micro\meter} for
$\text{G}_2$. The setup is \SI{1,854}{\meter} long, with a $\text{G}_0 -
\text{G}_1$ distance of \SI{1473}{\milli\meter}, a $\text{G}_1 - \text{G}_2$
distance of \SI{142}{\milli\meter}, and a $\text{G}_0 - \text{object}$ distance
of \SI{1118}{\milli\meter}. The dark-field is given as $-\log(V/V_{\text{ref}})$.
For the extracted dark-field signal we used the central pixels of the rod along the yellow, dashed line in Fig.~\ref{fig:cfk-rods}.

To simulate the dark-field of the carbon rod, we have to estimate the corresponding parameters.
We simulated the dark-field signal of the rod with a diameter of \SI{40}{pixel} and used
scatter parameters $\sigma_1 = 1.4$ and $\sigma_3 = 1$ .
However, since the real measurements bring along a lot of other unknown parameters, 
such as the number of fibers within the rod, the dark-field values are not quantitatively comparable.

\begin{figure}[tb]
	\centering	
	\def\svgwidth{0.25\textwidth}  

	\input{legend_rotation.tikz}

	\subfloat[Simulations with the proposed model. The trajectory consists of 200 projections over \SI{180}{\degree}. \label{fig:sim-xy}]{
		\input{xz_0.7_20.0_1.4_1.0_s2_zoom.tikz}}
	\quad			
	\subfloat[Real measurements of carbon-rods. 100 projections over $180^\circ$.) \label{fig:sim-real}]{
		\input{xz_copy.tikz}}
	\caption{Dark-field signal of carbon-rods with a diameter of \SI{4}{\milli\meter}.}
	\label{fig:real}
\end{figure}
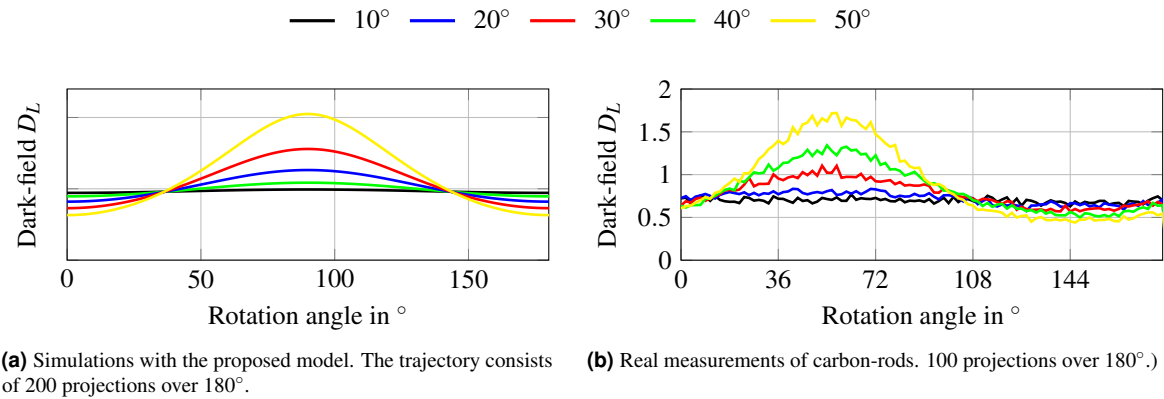

Figure~\ref{fig:real} shows the dark-field signal of the fiber over
\SI{180}{\degree} around the $y$-axis for different elevation angles, where an
elevation angle of $0^\circ$ represents a fiber that is aligned with the
rotation axis. The sensitivity direction is aligned with the rotation plane. 
Figure~\ref{fig:sim-xy} shows the dark-field simulations, while Fig.~\ref{fig:sim-real} shows the real dark-field measurements. 
Please note, that the rod were not aligned perfectly parallel to the detector plane on the beginning, which leads to shift in the rotation angle.
The simulations are in very good agreement with the real dataset.
In both cases, the signal fluctuation increases with increasing tilting angle of the fiber.
The influence of the path length of the projection ray through the object can be seen in the non-regular fluctuation along the rotation angles.
As a sidenote, the noise in the plot showing the real data corresponds to the
overall noise level of the setup. This can be observed in
Fig.~\ref{fig:cfk-rods} in the image background.

\section{Conclusions and Outlook}\label{sec:conclusions}
In this paper, we propose a X-ray dark-field imaging projection model.
It explicitly calculates structural quantities in 3-D using the direction of
the fiber, the ray direction and the sensitivity direction. To our knowledge,
this is the first true 3-D dark-field model.

We believe that this model is a powerful tool for further development of X-ray
dark-field imaging. In contrast to existing (2-D) projection models, where the
imaging trajectory is pre-defined, our model allows to image
arbitrary 3-D trajectories, like for example a helical trajectory.  The
proposed model is very general, but by addition of suitable constraints it can
be curbed to any of the existing 2-D models.
We evaluated the consistency of the model with itself, with a wave-front
simulation, and with experimental dark-field measurements.
In future work, we will investigate an algorithm for X-ray dark-field
reconstruction that can make full use of 3-D trajectories.

\section*{Acknowledgements}
The authors acknowledge funding from the German Research Foundation (DFG). \\ Project DFG GZ: AN 257/21-1~$\mid$~MA 4898/6-1.\\
L.F and V.L are supported by the International Max Planck Research School for Optics and Imaging.

\section*{Author Contributions Statement}
S.H. and A.M. conceived the model. S.H. developed the initial theoretical
framework. L.F. and C.R. expanded the theoretical framework. L.F. performed the
experiments and analyzed the data. J.B. carried out the wavefront simulations. V.L.
and G.A. performed the real data measurements. L.F., S.H., and C.R. wrote the
paper with input from all authors. 

\section*{Additional Information}
\textbf{Competing interests.} The authors declare no competing interests.

\bibliography{literature}
\end{document}

%% file: TLI_setup.pdf_tex
\begingroup%
  \makeatletter%
  \providecommand\color[2][]{%
    \errmessage{(Inkscape) Color is used for the text in Inkscape, but the package 'color.sty' is not loaded}%
    \renewcommand\color[2][]{}%
  }%
  \providecommand\transparent[1]{%
    \errmessage{(Inkscape) Transparency is used (non-zero) for the text in Inkscape, but the package 'transparent.sty' is not loaded}%
    \renewcommand\transparent[1]{}%
  }%
  \providecommand\rotatebox[2]{#2}%
  \ifx\svgwidth\undefined%
    \setlength{\unitlength}{1504.61500364bp}%
    \ifx\svgscale\undefined%
      \relax%
    \else%
      \setlength{\unitlength}{\unitlength * \real{\svgscale}}%
    \fi%
  \else%
    \setlength{\unitlength}{\svgwidth}%
  \fi%
  \global\let\svgwidth\undefined%
  \global\let\svgscale\undefined%
  \makeatother%
  \begin{picture}(1,0.39331411)%
    \put(0,0){\includegraphics[width=\unitlength,page=1]{TLI_setup.pdf}}%
    \put(0.33886437,0.28455997){\color[rgb]{0.8,0.8,0.8}\makebox(0,0)[lb]{\smash{$\vec{y}$}}}%
    \put(0.12741225,0.21217506){\color[rgb]{0,0,0}\makebox(0,0)[lb]{\smash{$G_0$}}}%
    \put(0.03047744,0.11800366){\color[rgb]{0,0,0}\makebox(0,0)[lb]{\smash{$S$}}}%
    \put(0.44357258,0.31852461){\color[rgb]{0,0,0}\makebox(0,0)[lb]{\smash{$G_1$}}}%
    \put(0.55664221,0.34696405){\color[rgb]{0,0,0}\makebox(0,0)[lb]{\smash{$G_2$}}}%
    \put(0.66734887,0.36295128){\color[rgb]{0,0,0}\makebox(0,0)[lb]{\smash{$D$}}}%
    \put(0,0){\includegraphics[width=\unitlength,page=2]{TLI_setup.pdf}}%
    \put(0.40545625,0.21723917){\color[rgb]{0.50196078,0.2,0}\makebox(0,0)[lb]{\smash{$\vec{f}$}}}%
    \put(0,0){\includegraphics[width=\unitlength,page=3]{TLI_setup.pdf}}%
    \put(0.4923787,0.1325495){\color[rgb]{0.8,0.8,0.8}\makebox(0,0)[lb]{\smash{$\vec{x}$}}}%
    \put(0.45359878,0.18273744){\color[rgb]{0.8,0.8,0.8}\makebox(0,0)[lb]{\smash{$\vec{z}$}}}%
    \put(0,0){\includegraphics[width=\unitlength,page=4]{TLI_setup.pdf}}%
    \put(0.85,0.17){\color[rgb]{0,0,0}\makebox(0,0)[lb]{\smash{$\vec{s}$}}}%
    \put(0,0){\includegraphics[width=\unitlength,page=5]{TLI_setup.pdf}}%
  \end{picture}%
\endgroup%

%% file: revol_projection_model2.pdf_tex
\begingroup%
  \makeatletter%
  \providecommand\color[2][]{%
    \errmessage{(Inkscape) Color is used for the text in Inkscape, but the package 'color.sty' is not loaded}%
    \renewcommand\color[2][]{}%
  }%
  \providecommand\transparent[1]{%
    \errmessage{(Inkscape) Transparency is used (non-zero) for the text in Inkscape, but the package 'transparent.sty' is not loaded}%
    \renewcommand\transparent[1]{}%
  }%
  \providecommand\rotatebox[2]{#2}%
  \ifx\svgwidth\undefined%
    \setlength{\unitlength}{575.81950817bp}%
    \ifx\svgscale\undefined%
      \relax%
    \else%
      \setlength{\unitlength}{\unitlength * \real{\svgscale}}%
    \fi%
  \else%
    \setlength{\unitlength}{\svgwidth}%
  \fi%
  \global\let\svgwidth\undefined%
  \global\let\svgscale\undefined%
  \makeatother%
  \begin{picture}(1,0.65126359)%
    \put(0,0){\includegraphics[width=\unitlength,page=1]{revol_projection_model2.pdf}}%
    \put(0.44448108,0.33452286){\color[rgb]{0.50196078,0.2,0}\makebox(0,0)[lb]{\smash{$\vec{f}$}}}%
    \put(0.49816887,0.11753452){\color[rgb]{0,0,0}\makebox(0,0)[lb]{\smash{$\omega$}}}%
    \put(0.24394446,0.58922543){\color[rgb]{0.8,0.8,0.8}\makebox(0,0)[lb]{\smash{y}}}%
    \put(0.68327848,0.00976722){\color[rgb]{0.8,0.8,0.8}\makebox(0,0)[lb]{\smash{x}}}%
    \put(0.57383086,0.21192059){\color[rgb]{0.8,0.8,0.8}\makebox(0,0)[lb]{\smash{z}}}%
    \put(0,0){\includegraphics[width=\unitlength,page=2]{revol_projection_model2.pdf}}%
    \put(0.92,0.03){\color[rgb]{0,0,0}\makebox(0,0)[lb]{\smash{$\vec{s}$}}}%
    \put(0,0){\includegraphics[width=\unitlength,page=3]{revol_projection_model2.pdf}}%
    \put(0.29,0.25){\color[rgb]{0,0,0}\makebox(0,0)[lb]{\smash{$\theta$}}}%
    \put(0,0){\includegraphics[width=\unitlength,page=4]{revol_projection_model2.pdf}}%
  \end{picture}%
\endgroup%

%% file: bayer_projection_model2.pdf_tex
\begingroup%
  \makeatletter%
  \providecommand\color[2][]{%
    \errmessage{(Inkscape) Color is used for the text in Inkscape, but the package 'color.sty' is not loaded}%
    \renewcommand\color[2][]{}%
  }%
  \providecommand\transparent[1]{%
    \errmessage{(Inkscape) Transparency is used (non-zero) for the text in Inkscape, but the package 'transparent.sty' is not loaded}%
    \renewcommand\transparent[1]{}%
  }%
  \providecommand\rotatebox[2]{#2}%
  \ifx\svgwidth\undefined%
    \setlength{\unitlength}{575.92502817bp}%
    \ifx\svgscale\undefined%
      \relax%
    \else%
      \setlength{\unitlength}{\unitlength * \real{\svgscale}}%
    \fi%
  \else%
    \setlength{\unitlength}{\svgwidth}%
  \fi%
  \global\let\svgwidth\undefined%
  \global\let\svgscale\undefined%
  \makeatother%
  \begin{picture}(1,0.64540663)%
    \put(0,0){\includegraphics[width=\unitlength,page=1]{bayer_projection_model2.pdf}}%
    \put(0.44386959,0.33446157){\color[rgb]{0.50196078,0.2,0}\makebox(0,0)[lb]{\smash{$\vec{f}$}}}%
    \put(0,0){\includegraphics[width=\unitlength,page=2]{bayer_projection_model2.pdf}}%
    \put(0.3874006,0.15134994){\color[rgb]{0,0,0}\makebox(0,0)[lb]{\smash{$\phi$}}}%
    \put(0.28238676,0.39291873){\color[rgb]{0,0,0}\makebox(0,0)[lb]{\smash{$\omega$}}}%
    \put(0,0){\includegraphics[width=\unitlength,page=3]{bayer_projection_model2.pdf}}%
    \put(0.24910725,0.58337984){\color[rgb]{0.8,0.8,0.8}\makebox(0,0)[lb]{\smash{y}}}%
    \put(0.68262323,0.00976539){\color[rgb]{0.8,0.8,0.8}\makebox(0,0)[lb]{\smash{x}}}%
    \put(0.57319567,0.21188177){\color[rgb]{0.8,0.8,0.8}\makebox(0,0)[lb]{\smash{z}}}%
    \put(0,0){\includegraphics[width=\unitlength,page=4]{bayer_projection_model2.pdf}}%
    \put(0.92,0.03){\color[rgb]{0,0,0}\makebox(0,0)[lb]{\smash{$\vec{s}$}}}%
  \end{picture}%
\endgroup%

%% file: schaff_projection_model2.pdf_tex
\begingroup%
  \makeatletter%
  \providecommand\color[2][]{%
    \errmessage{(Inkscape) Color is used for the text in Inkscape, but the package 'color.sty' is not loaded}%
    \renewcommand\color[2][]{}%
  }%
  \providecommand\transparent[1]{%
    \errmessage{(Inkscape) Transparency is used (non-zero) for the text in Inkscape, but the package 'transparent.sty' is not loaded}%
    \renewcommand\transparent[1]{}%
  }%
  \providecommand\rotatebox[2]{#2}%
  \ifx\svgwidth\undefined%
    \setlength{\unitlength}{575.92502817bp}%
    \ifx\svgscale\undefined%
      \relax%
    \else%
      \setlength{\unitlength}{\unitlength * \real{\svgscale}}%
    \fi%
  \else%
    \setlength{\unitlength}{\svgwidth}%
  \fi%
  \global\let\svgwidth\undefined%
  \global\let\svgscale\undefined%
  \makeatother%
  \begin{picture}(1,0.64540667)%
    \put(0,0){\includegraphics[width=\unitlength,page=1]{schaff_projection_model2.pdf}}%
    \put(0.27,0.27){\color[rgb]{0,0,0}\makebox(0,0)[lb]{\smash{$\Theta$}}}%
    \put(0,0){\includegraphics[width=\unitlength,page=2]{schaff_projection_model2.pdf}}%
    \put(0.44386956,0.33446154){\color[rgb]{0.50196078,0.2,0}\makebox(0,0)[lb]{\smash{$\vec{f}$}}}%
    \put(0.24910723,0.58337988){\color[rgb]{0.8,0.8,0.8}\makebox(0,0)[lb]{\smash{y}}}%
    \put(0.68262318,0.0097654){\color[rgb]{0.8,0.8,0.8}\makebox(0,0)[lb]{\smash{x}}}%
    \put(0.57319564,0.21188174){\color[rgb]{0.8,0.8,0.8}\makebox(0,0)[lb]{\smash{z}}}%
    \put(0,0){\includegraphics[width=\unitlength,page=3]{schaff_projection_model2.pdf}}%
    \put(0.92,0.03){\color[rgb]{0,0,0}\makebox(0,0)[lb]{\smash{$\vec{s}$}}}%
    \put(0,0){\includegraphics[width=\unitlength,page=4]{schaff_projection_model2.pdf}}%
    \put(0.5111016,0.10991381){\color[rgb]{0,0,0}\makebox(0,0)[lb]{\smash{$\omega$}}}%
    \put(0,0){\includegraphics[width=\unitlength,page=5]{schaff_projection_model2.pdf}}%
  \end{picture}%
\endgroup%

%% file: dark-field-signal2D.pdf_tex
\begingroup%
  \makeatletter%
  \providecommand\color[2][]{%
    \errmessage{(Inkscape) Color is used for the text in Inkscape, but the package 'color.sty' is not loaded}%
    \renewcommand\color[2][]{}%
  }%
  \providecommand\transparent[1]{%
    \errmessage{(Inkscape) Transparency is used (non-zero) for the text in Inkscape, but the package 'transparent.sty' is not loaded}%
    \renewcommand\transparent[1]{}%
  }%
  \providecommand\rotatebox[2]{#2}%
  \ifx\svgwidth\undefined%
    \setlength{\unitlength}{515.73991421bp}%
    \ifx\svgscale\undefined%
      \relax%
    \else%
      \setlength{\unitlength}{\unitlength * \real{\svgscale}}%
    \fi%
  \else%
    \setlength{\unitlength}{\svgwidth}%
  \fi%
  \global\let\svgwidth\undefined%
  \global\let\svgscale\undefined%
  \makeatother%
  \begin{picture}(1,0.6)%
    \put(0,0.05){\includegraphics[width=\unitlength,page=1]{dark-field-signal2D.pdf}}%
    \put(0.58,0){\color[rgb]{0,0,0}\makebox(0,0)[lb]{\smash{360}}}%
    \put(0.82218982, 0){\color[rgb]{0,0,0}\makebox(0,0)[lb]{\smash{degrees}}}%
    \put(0.0, 0){\color[rgb]{0,0,0}\makebox(0,0)[lb]{\smash{0}}}%
    \put(0,0.05){\includegraphics[width=\unitlength,page=2]{dark-field-signal2D.pdf}}%
    \put(0.34118953,0.17){\color[rgb]{0,0.78431373,1}\makebox(0,0)[lb]{\smash{$d_\text{iso}$}}}%
    \put(0,0.05){\includegraphics[width=\unitlength,page=3]{dark-field-signal2D.pdf}}%
    \put(0.50257918,0.38){\color[rgb]{0,0,1}\makebox(0,0)[lb]{\smash{$d_\text{aniso}$}}}%
    \put(0.0398392,0.58181571){\color[rgb]{0,0,0}\makebox(0,0)[lb]{\smash{dark-field}}}%
    \put(0.0398392,0.48181571){\color[rgb]{0,0,0}\makebox(0,0)[lb]{\smash{signal}}}%
    \put(0,0.05){\includegraphics[width=\unitlength,page=4]{dark-field-signal2D.pdf}}%
  \end{picture}%
\endgroup%

%% file: 2Dmodel.pdf_tex
\begingroup%
  \makeatletter%
  \providecommand\color[2][]{%
    \errmessage{(Inkscape) Color is used for the text in Inkscape, but the package 'color.sty' is not loaded}%
    \renewcommand\color[2][]{}%
  }%
  \providecommand\transparent[1]{%
    \errmessage{(Inkscape) Transparency is used (non-zero) for the text in Inkscape, but the package 'transparent.sty' is not loaded}%
    \renewcommand\transparent[1]{}%
  }%
  \providecommand\rotatebox[2]{#2}%
  \ifx\svgwidth\undefined%
    \setlength{\unitlength}{1072.6303068bp}%
    \ifx\svgscale\undefined%
      \relax%
    \else%
      \setlength{\unitlength}{\unitlength * \real{\svgscale}}%
    \fi%
  \else%
    \setlength{\unitlength}{\svgwidth}%
  \fi%
  \global\let\svgwidth\undefined%
  \global\let\svgscale\undefined%
  \makeatother%
  \begin{picture}(1,0.47679854)%
    \put(0,0){\includegraphics[width=\unitlength,page=1]{2Dmodel.pdf}}%
    \put(0.55368479,0.31524627){\color[rgb]{0,0.78431373,1}\makebox(0,0)[lb]{\smash{$d_\text{iso}$}}}%
    \put(0,0){\includegraphics[width=\unitlength,page=2]{2Dmodel.pdf}}%
    \put(0.79720007,0.27){\color[rgb]{0,0,1}\makebox(0,0)[lb]{\smash{$d_\text{aniso}$}}}%
    \put(0,0){\includegraphics[width=\unitlength,page=3]{2Dmodel.pdf}}%
  \end{picture}%
\endgroup%

%% file: fiber_and_scatter.pdf_tex
\begingroup%
  \makeatletter%
  \providecommand\color[2][]{%
    \errmessage{(Inkscape) Color is used for the text in Inkscape, but the package 'color.sty' is not loaded}%
    \renewcommand\color[2][]{}%
  }%
  \providecommand\transparent[1]{%
    \errmessage{(Inkscape) Transparency is used (non-zero) for the text in Inkscape, but the package 'transparent.sty' is not loaded}%
    \renewcommand\transparent[1]{}%
  }%
  \providecommand\rotatebox[2]{#2}%
  \ifx\svgwidth\undefined%
    \setlength{\unitlength}{673.68523007bp}%
    \ifx\svgscale\undefined%
      \relax%
    \else%
      \setlength{\unitlength}{\unitlength * \real{\svgscale}}%
    \fi%
  \else%
    \setlength{\unitlength}{\svgwidth}%
  \fi%
  \global\let\svgwidth\undefined%
  \global\let\svgscale\undefined%
  \makeatother%
  \begin{picture}(1,0.66062679)%
    \put(0,0){\includegraphics[width=\unitlength,page=1]{fiber_and_scatter.pdf}}%
    \put(0.31284348,0.62942791){\color[rgb]{0.50196078,0.2,0}\makebox(0,0)[lb]{\smash{$\vec{f}$}}}%
    \put(0,0){\includegraphics[width=\unitlength,page=2]{fiber_and_scatter.pdf}}%
    \put(0.78393099,0.52614304){\color[rgb]{0,0,0}\makebox(0,0)[lb]{\smash{$\vec{b}_1 \cdot \sigma_1$}}}%
    \put(0.35973712,0.4882516){\color[rgb]{0,0,0}\makebox(0,0)[lb]{\smash{$\vec{b}_3 \cdot \sigma_3$}}}%
    \put(0,0){\includegraphics[width=\unitlength,page=3]{fiber_and_scatter.pdf}}%
  \end{picture}%
\endgroup%

%% file: Spheroids_basis_vectors.pdf_tex
\begingroup%
  \makeatletter%
  \providecommand\color[2][]{%
    \errmessage{(Inkscape) Color is used for the text in Inkscape, but the package 'color.sty' is not loaded}%
    \renewcommand\color[2][]{}%
  }%
  \providecommand\transparent[1]{%
    \errmessage{(Inkscape) Transparency is used (non-zero) for the text in Inkscape, but the package 'transparent.sty' is not loaded}%
    \renewcommand\transparent[1]{}%
  }%
  \providecommand\rotatebox[2]{#2}%
  \ifx\svgwidth\undefined%
    \setlength{\unitlength}{1081.00471161bp}%
    \ifx\svgscale\undefined%
      \relax%
    \else%
      \setlength{\unitlength}{\unitlength * \real{\svgscale}}%
    \fi%
  \else%
    \setlength{\unitlength}{\svgwidth}%
  \fi%
  \global\let\svgwidth\undefined%
  \global\let\svgscale\undefined%
  \makeatother%
  \begin{picture}(1,0.32201247)%
    \put(0,0){\includegraphics[width=\unitlength,page=1]{Spheroids_basis_vectors.pdf}}%
    \put(0.71526869,0.1226557){\color[rgb]{0,0,0}\makebox(0,0)[lb]{\smash{$\vec{b}_1 \cdot \sigma_1$}}}%
    \put(0.30002471,0.0396132){\color[rgb]{0,0,0}\makebox(0,0)[lb]{\smash{$\vec{b}_2 \cdot \sigma_2$}}}%
    \put(0.34766033,0.30608465){\color[rgb]{0,0,0}\makebox(0,0)[lb]{\smash{$\vec{b}_3 \cdot \sigma_3$}}}%
  \end{picture}%
\endgroup%

%% file: projection_on_plane.pdf_tex
\begingroup%
  \makeatletter%
  \providecommand\color[2][]{%
    \errmessage{(Inkscape) Color is used for the text in Inkscape, but the package 'color.sty' is not loaded}%
    \renewcommand\color[2][]{}%
  }%
  \providecommand\transparent[1]{%
    \errmessage{(Inkscape) Transparency is used (non-zero) for the text in Inkscape, but the package 'transparent.sty' is not loaded}%
    \renewcommand\transparent[1]{}%
  }%
  \providecommand\rotatebox[2]{#2}%
  \ifx\svgwidth\undefined%
    \setlength{\unitlength}{2108.01714367bp}%
    \ifx\svgscale\undefined%
      \relax%
    \else%
      \setlength{\unitlength}{\unitlength * \real{\svgscale}}%
    \fi%
  \else%
    \setlength{\unitlength}{\svgwidth}%
  \fi%
  \global\let\svgwidth\undefined%
  \global\let\svgscale\undefined%
  \makeatother%
  \begin{picture}(1,0.23803738)%
    \put(0,0){\includegraphics[width=\unitlength,page=1]{projection_on_plane.pdf}}%
    \put(0.18596543,0.12592367){\color[rgb]{0.50196078,0.2,0}\makebox(0,0)[lb]{\smash{$\vec{f}$}}}%
    \put(0.02260185,0.04777224){\color[rgb]{0,0,0}\makebox(0,0)[lb]{\smash{$\vec{r}$}}}%
    \put(0.48812814,0.01007401){\color[rgb]{0,0,1}\makebox(0,0)[lb]{\smash{plane $\vec{E}$}}}%
    \put(0,0){\includegraphics[width=\unitlength,page=2]{projection_on_plane.pdf}}%
    \put(0.80542282,0.20740813){\color[rgb]{0,0,0}\transparent{0.58823532}\makebox(0,0)[lb]{\smash{$\vec{s}$}}}%
    \put(0.5484447,0.17596355){\color[rgb]{0.50196078,0.2,0}\makebox(0,0)[lb]{\smash{$\vec{f}'$}}}%
    \put(0.58981658,0.20851112){\color[rgb]{0,0,0}\transparent{0.58823532}\makebox(0,0)[lb]{\smash{$\vec{s}'$}}}%
  \end{picture}%
\endgroup%

%% file: projection_in_plane.pdf_tex
\begingroup%
  \makeatletter%
  \providecommand\color[2][]{%
    \errmessage{(Inkscape) Color is used for the text in Inkscape, but the package 'color.sty' is not loaded}%
    \renewcommand\color[2][]{}%
  }%
  \providecommand\transparent[1]{%
    \errmessage{(Inkscape) Transparency is used (non-zero) for the text in Inkscape, but the package 'transparent.sty' is not loaded}%
    \renewcommand\transparent[1]{}%
  }%
  \providecommand\rotatebox[2]{#2}%
  \ifx\svgwidth\undefined%
    \setlength{\unitlength}{1392.90409638bp}%
    \ifx\svgscale\undefined%
      \relax%
    \else%
      \setlength{\unitlength}{\unitlength * \real{\svgscale}}%
    \fi%
  \else%
    \setlength{\unitlength}{\svgwidth}%
  \fi%
  \global\let\svgwidth\undefined%
  \global\let\svgscale\undefined%
  \makeatother%
  \begin{picture}(1,0.59321076)%
    \put(0,0){\includegraphics[width=\unitlength,page=1]{projection_in_plane.pdf}}%
    \put(0.31919506,0.03309415){\color[rgb]{0,0,1}\makebox(0,0)[lb]{\smash{plane $\vec{E}$}}}%
    \put(0,0){\includegraphics[width=\unitlength,page=2]{projection_in_plane.pdf}}%
    \put(0.34786577,0.35777239){\color[rgb]{0.50196078,0.2,0}\makebox(0,0)[lb]{\smash{$\vec{f}'$}}}%
    \put(0.4351304,0.27676912){\color[rgb]{0,0,0}\transparent{0.58823532}\makebox(0,0)[lb]{\smash{$\vec{s}'$}}}%
  \end{picture}%
\endgroup%

%% file: exp1.pdf_tex
\begingroup%
  \makeatletter%
  \providecommand\color[2][]{%
    \errmessage{(Inkscape) Color is used for the text in Inkscape, but the package 'color.sty' is not loaded}%
    \renewcommand\color[2][]{}%
  }%
  \providecommand\transparent[1]{%
    \errmessage{(Inkscape) Transparency is used (non-zero) for the text in Inkscape, but the package 'transparent.sty' is not loaded}%
    \renewcommand\transparent[1]{}%
  }%
  \providecommand\rotatebox[2]{#2}%
  \ifx\svgwidth\undefined%
    \setlength{\unitlength}{528.1052115bp}%
    \ifx\svgscale\undefined%
      \relax%
    \else%
      \setlength{\unitlength}{\unitlength * \real{\svgscale}}%
    \fi%
  \else%
    \setlength{\unitlength}{\svgwidth}%
  \fi%
  \global\let\svgwidth\undefined%
  \global\let\svgscale\undefined%
  \makeatother%
  \begin{picture}(1,0.70384807)%
    \put(0,0){\includegraphics[width=\unitlength,page=1]{exp1.pdf}}%
    \put(0.29516163,0.364747){\color[rgb]{0.50196078,0.2,0}\makebox(0,0)[lb]{\smash{$\vec{f}$}}}%
    \put(0.11905657,0.42849744){\color[rgb]{0,0,0}\makebox(0,0)[lb]{\smash{$\omega$}}}%
    \put(0,0){\includegraphics[width=\unitlength,page=2]{exp1.pdf}}%
    \put(0.0827636,0.63620477){\color[rgb]{0.8,0.8,0.8}\makebox(0,0)[lb]{\smash{y}}}%
    \put(0.55553437,0.01064964){\color[rgb]{0.8,0.8,0.8}\makebox(0,0)[lb]{\smash{x}}}%
    \put(0.43619816,0.23106762){\color[rgb]{0.8,0.8,0.8}\makebox(0,0)[lb]{\smash{z}}}%
    \put(0,0){\includegraphics[width=\unitlength,page=3]{exp1.pdf}}%
    \put(0.74,0.10){\color[rgb]{0,0,0}\makebox(0,0)[lb]{\smash{$\vec{s}$}}}%
    \put(0,0){\includegraphics[width=\unitlength,page=4]{exp1.pdf}}%
  \end{picture}%
\endgroup%

%% file: exp2.pdf_tex
\begingroup%
  \makeatletter%
  \providecommand\color[2][]{%
    \errmessage{(Inkscape) Color is used for the text in Inkscape, but the package 'color.sty' is not loaded}%
    \renewcommand\color[2][]{}%
  }%
  \providecommand\transparent[1]{%
    \errmessage{(Inkscape) Transparency is used (non-zero) for the text in Inkscape, but the package 'transparent.sty' is not loaded}%
    \renewcommand\transparent[1]{}%
  }%
  \providecommand\rotatebox[2]{#2}%
  \ifx\svgwidth\undefined%
    \setlength{\unitlength}{511.05466513bp}%
    \ifx\svgscale\undefined%
      \relax%
    \else%
      \setlength{\unitlength}{\unitlength * \real{\svgscale}}%
    \fi%
  \else%
    \setlength{\unitlength}{\svgwidth}%
  \fi%
  \global\let\svgwidth\undefined%
  \global\let\svgscale\undefined%
  \makeatother%
  \begin{picture}(1,0.72733087)%
    \put(0,0){\includegraphics[width=\unitlength,page=1]{exp2.pdf}}%
    \put(0.30500924,0.37691621){\color[rgb]{0.50196078,0.2,0}\makebox(0,0)[lb]{\smash{$\vec{f}$}}}%
    \put(0.1230287,0.4427936){\color[rgb]{0,0,0}\makebox(0,0)[lb]{\smash{$\omega$}}}%
    \put(0,0){\includegraphics[width=\unitlength,page=2]{exp2.pdf}}%
    \put(0.08552488,0.65743075){\color[rgb]{0.8,0.8,0.8}\makebox(0,0)[lb]{\smash{y}}}%
    \put(0.57406891,0.01100495){\color[rgb]{0.8,0.8,0.8}\makebox(0,0)[lb]{\smash{x}}}%
    \put(0.45075124,0.23877683){\color[rgb]{0.8,0.8,0.8}\makebox(0,0)[lb]{\smash{z}}}%
    \put(0.96,0.13){\color[rgb]{0,0,0}\makebox(0,0)[lb]{\smash{$\vec{s}$}}}%
    \put(0,0){\includegraphics[width=\unitlength,page=3]{exp2.pdf}}%
  \end{picture}%
\endgroup%

%% file: exp3.pdf_tex
\begingroup%
  \makeatletter%
  \providecommand\color[2][]{%
    \errmessage{(Inkscape) Color is used for the text in Inkscape, but the package 'color.sty' is not loaded}%
    \renewcommand\color[2][]{}%
  }%
  \providecommand\transparent[1]{%
    \errmessage{(Inkscape) Transparency is used (non-zero) for the text in Inkscape, but the package 'transparent.sty' is not loaded}%
    \renewcommand\transparent[1]{}%
  }%
  \providecommand\rotatebox[2]{#2}%
  \ifx\svgwidth\undefined%
    \setlength{\unitlength}{443.87767667bp}%
    \ifx\svgscale\undefined%
      \relax%
    \else%
      \setlength{\unitlength}{\unitlength * \real{\svgscale}}%
    \fi%
  \else%
    \setlength{\unitlength}{\svgwidth}%
  \fi%
  \global\let\svgwidth\undefined%
  \global\let\svgscale\undefined%
  \makeatother%
  \begin{picture}(1,0.83740601)%
    \put(0,0){\includegraphics[width=\unitlength,page=1]{exp3.pdf}}%
    \put(0.35116971,0.43395917){\color[rgb]{0.50196078,0.2,0}\makebox(0,0)[lb]{\smash{$\vec{f}$}}}%
    \put(0.14164802,0.50980652){\color[rgb]{0,0,0}\makebox(0,0)[lb]{\smash{$\omega$}}}%
    \put(0.09846832,0.75692712){\color[rgb]{0.8,0.8,0.8}\makebox(0,0)[lb]{\smash{y}}}%
    \put(0.6609492,0.01267045){\color[rgb]{0.8,0.8,0.8}\makebox(0,0)[lb]{\smash{x}}}%
    \put(0.51896848,0.2749136){\color[rgb]{0.8,0.8,0.8}\makebox(0,0)[lb]{\smash{z}}}%
    \put(0.78,0.1){\color[rgb]{0,0,0}\makebox(0,0)[lb]{\smash{$\vec{s}$}}}%
    \put(0,0){\includegraphics[width=\unitlength,page=2]{exp3.pdf}}%
  \end{picture}%
\endgroup%

%% file: line_plot_trajectories.tikz
\begin{tikzpicture}
\begin{axis}[%
width=\textwidth,
height=\axisdefaultheight,
scale = 0.5,
xmin=1,
xmax=240,
xlabel={Rotation angle in $^\circ$},
xlabel near ticks,
xticklabels={0, 37.5, 75, 150, 225, 300, 337.5}, 
ymin=0,
ymax=1.2,
ylabel={Dark-field $d$},
y label style={at={(axis description cs:0.1,.5)}},
ylabel near ticks,
xmajorgrids,
ymajorgrids,
	legend pos=outer north east
]
\addplot [color=magenta, line width=1.0pt, dotted] table[row sep=crcr]{%
 0    0.50000006\\ 1    0.5003536\\ 2    0.50139135\\ 3    0.5031106\\ 4    0.50550646\\ 5    0.5085725\\ 6    0.5123003\\ 7    0.5166796\\ 8    0.5216984\\ 9    0.527343\\ 10    0.5335978\\ 11    0.54044586\\ 12    0.54786825\\ 13    0.5558446\\ 14    0.5643532\\ 15    0.57337064\\ 16    0.58287215\\ 17    0.5928318\\ 18    0.6032222\\ 19    0.6140149\\ 20    0.62518036\\ 21    0.63668793\\ 22    0.6485061\\ 23    0.66060245\\ 24    0.67294383\\ 25    0.6854964\\ 26    0.6982258\\ 27    0.71109706\\ 28    0.724075\\ 29    0.73712397\\ 30    0.75020826\\ 31    0.76329195\\ 32    0.77633923\\ 33    0.78931427\\ 34    0.8021816\\ 35    0.8149059\\ 36    0.8274523\\ 37    0.83978635\\ 38    0.85187435\\ 39    0.86368316\\ 40    0.8751803\\ 41    0.88633436\\ 42    0.89711475\\ 43    0.90749186\\ 44    0.9174374\\ 45    0.9269239\\ 46    0.9359255\\ 47    0.9444175\\ 48    0.9523766\\ 49    0.959781\\ 50    0.9666104\\ 51    0.9728461\\ 52    0.978471\\ 53    0.98346967\\ 54    0.9878284\\ 55    0.99153525\\ 56    0.9945801\\ 57    0.99695456\\ 58    0.99865216\\ 59    0.9996682\\ 60    0.99999994\\ 61    0.9996464\\ 62    0.99860865\\ 63    0.9968894\\ 64    0.99449354\\ 65    0.9914275\\ 66    0.9876997\\ 67    0.9833204\\ 68    0.9783016\\ 69    0.972657\\ 70    0.9664022\\ 71    0.95955414\\ 72    0.95213175\\ 73    0.9441554\\ 74    0.9356468\\ 75    0.92662936\\ 76    0.91712785\\ 77    0.9071682\\ 78    0.8967778\\ 79    0.8859851\\ 80    0.87481964\\ 81    0.86331207\\ 82    0.8514939\\ 83    0.83939755\\ 84    0.82705617\\ 85    0.8145036\\ 86    0.8017742\\ 87    0.78890294\\ 88    0.775925\\ 89    0.76287603\\ 90    0.74979174\\ 91    0.73670805\\ 92    0.72366077\\ 93    0.71068573\\ 94    0.6978184\\ 95    0.6850941\\ 96    0.6725477\\ 97    0.66021365\\ 98    0.64812565\\ 99    0.63631684\\ 100    0.6248197\\ 101    0.61366564\\ 102    0.60288525\\ 103    0.59250814\\ 104    0.5825626\\ 105    0.5730761\\ 106    0.5640745\\ 107    0.5555825\\ 108    0.5476234\\ 109    0.540219\\ 110    0.5333896\\ 111    0.5271539\\ 112    0.521529\\ 113    0.51653033\\ 114    0.5121716\\ 115    0.50846475\\ 116    0.5054199\\ 117    0.50304544\\ 118    0.50134784\\ 119    0.5003318\\ 120    0.50000006\\ 121    0.5003536\\ 122    0.50139135\\ 123    0.5031106\\ 124    0.50550646\\ 125    0.5085725\\ 126    0.5123003\\ 127    0.5166796\\ 128    0.5216984\\ 129    0.527343\\ 130    0.5335978\\ 131    0.54044586\\ 132    0.54786825\\ 133    0.5558446\\ 134    0.5643532\\ 135    0.57337064\\ 136    0.58287215\\ 137    0.5928318\\ 138    0.6032222\\ 139    0.6140149\\ 140    0.62518036\\ 141    0.63668793\\ 142    0.6485061\\ 143    0.66060245\\ 144    0.67294383\\ 145    0.6854964\\ 146    0.6982258\\ 147    0.71109706\\ 148    0.724075\\ 149    0.73712397\\ 150    0.75020826\\ 151    0.76329195\\ 152    0.77633923\\ 153    0.78931427\\ 154    0.8021816\\ 155    0.8149059\\ 156    0.8274523\\ 157    0.83978635\\ 158    0.85187435\\ 159    0.86368316\\ 160    0.8751803\\ 161    0.88633436\\ 162    0.89711475\\ 163    0.90749186\\ 164    0.9174374\\ 165    0.9269239\\ 166    0.9359255\\ 167    0.9444175\\ 168    0.9523766\\ 169    0.959781\\ 170    0.9666104\\ 171    0.9728461\\ 172    0.978471\\ 173    0.98346967\\ 174    0.9878284\\ 175    0.99153525\\ 176    0.9945801\\ 177    0.99695456\\ 178    0.99865216\\ 179    0.9996682\\ 180    0.99999994\\ 181    0.9996464\\ 182    0.99860865\\ 183    0.9968894\\ 184    0.99449354\\ 185    0.9914275\\ 186    0.9876997\\ 187    0.9833204\\ 188    0.9783016\\ 189    0.972657\\ 190    0.9664022\\ 191    0.95955414\\ 192    0.95213175\\ 193    0.9441554\\ 194    0.9356468\\ 195    0.92662936\\ 196    0.91712785\\ 197    0.9071682\\ 198    0.8967778\\ 199    0.8859851\\ 200    0.87481964\\ 201    0.86331207\\ 202    0.8514939\\ 203    0.83939755\\ 204    0.82705617\\ 205    0.8145036\\ 206    0.8017742\\ 207    0.78890294\\ 208    0.775925\\ 209    0.76287603\\ 210    0.74979174\\ 211    0.73670805\\ 212    0.72366077\\ 213    0.71068573\\ 214    0.6978184\\ 215    0.6850941\\ 216    0.6725477\\ 217    0.66021365\\ 218    0.64812565\\ 219    0.63631684\\ 220    0.6248197\\ 221    0.61366564\\ 222    0.60288525\\ 223    0.59250814\\ 224    0.5825626\\ 225    0.5730761\\ 226    0.5640745\\ 227    0.5555825\\ 228    0.5476234\\ 229    0.540219\\ 230    0.5333896\\ 231    0.5271539\\ 232    0.521529\\ 233    0.51653033\\ 234    0.5121716\\ 235    0.50846475\\ 236    0.5054199\\ 237    0.50304544\\ 238    0.50134784\\ 239    0.5003318\\ 
};
\addplot [color=magenta, line width=1.0pt] table[row sep=crcr]{%
  0    0.83347213\\ 1    0.84219456\\ 2    0.8508928\\ 3    0.8595428\\ 4    0.868121\\ 5    0.8766039\\ 6    0.88496816\\ 7    0.89319086\\ 8    0.9012495\\ 9    0.90912205\\ 10    0.91678685\\ 11    0.9242229\\ 12    0.9314098\\ 13    0.9383279\\ 14    0.9449582\\ 15    0.95128256\\ 16    0.9572836\\ 17    0.962945\\ 18    0.96825105\\ 19    0.9731873\\ 20    0.9777402\\ 21    0.98189735\\ 22    0.9856473\\ 23    0.98897976\\ 24    0.9918856\\ 25    0.9943568\\ 26    0.9963867\\ 27    0.9979697\\ 28    0.99910146\\ 29    0.9997788\\ 30    0.99999994\\ 31    0.99976426\\ 32    0.99907243\\ 33    0.9979263\\ 34    0.996329\\ 35    0.994285\\ 36    0.99179983\\ 37    0.9888803\\ 38    0.9855344\\ 39    0.98177135\\ 40    0.97760147\\ 41    0.9730361\\ 42    0.96808785\\ 43    0.9627703\\ 44    0.9570979\\ 45    0.9510863\\ 46    0.9447519\\ 47    0.9381122\\ 48    0.93118525\\ 49    0.9239901\\ 50    0.91654646\\ 51    0.90887475\\ 52    0.90099597\\ 53    0.89293176\\ 54    0.8847042\\ 55    0.8763358\\ 56    0.8678495\\ 57    0.85926867\\ 58    0.8506167\\ 59    0.8419174\\ 60    0.83319455\\ 61    0.82447207\\ 62    0.8157739\\ 63    0.8071239\\ 64    0.79854566\\ 65    0.7900628\\ 66    0.7816985\\ 67    0.7734758\\ 68    0.76541716\\ 69    0.75754464\\ 70    0.7498799\\ 71    0.74244386\\ 72    0.7352569\\ 73    0.72833884\\ 74    0.72170854\\ 75    0.7153841\\ 76    0.7093831\\ 77    0.70372176\\ 78    0.6984157\\ 79    0.6934794\\ 80    0.68892646\\ 81    0.68476933\\ 82    0.6810194\\ 83    0.677687\\ 84    0.67478114\\ 85    0.6723099\\ 86    0.67028\\ 87    0.66869706\\ 88    0.6675653\\ 89    0.66688794\\ 90    0.6666668\\ 91    0.6669025\\ 92    0.6675943\\ 93    0.66874045\\ 94    0.67033774\\ 95    0.67238176\\ 96    0.674867\\ 97    0.67778647\\ 98    0.6811324\\ 99    0.6848954\\ 100    0.68906534\\ 101    0.69363064\\ 102    0.6985789\\ 103    0.7038965\\ 104    0.70956886\\ 105    0.71558046\\ 106    0.7219148\\ 107    0.7285546\\ 108    0.73548156\\ 109    0.7426767\\ 110    0.75012034\\ 111    0.75779206\\ 112    0.76567084\\ 113    0.77373505\\ 114    0.78196263\\ 115    0.790331\\ 116    0.7988173\\ 117    0.80739814\\ 118    0.81605005\\ 119    0.82474935\\ 120    0.83347225\\ 121    0.8421947\\ 122    0.85089284\\ 123    0.8595429\\ 124    0.86812115\\ 125    0.87660396\\ 126    0.8849682\\ 127    0.893191\\ 128    0.90124965\\ 129    0.9091221\\ 130    0.9167869\\ 131    0.92422295\\ 132    0.93140984\\ 133    0.93832797\\ 134    0.94495827\\ 135    0.9512826\\ 136    0.9572837\\ 137    0.962945\\ 138    0.9682511\\ 139    0.9731873\\ 140    0.9777403\\ 141    0.9818974\\ 142    0.9856473\\ 143    0.98897976\\ 144    0.9918856\\ 145    0.9943569\\ 146    0.99638677\\ 147    0.99796975\\ 148    0.99910146\\ 149    0.9997788\\ 150    0.99999994\\ 151    0.99976426\\ 152    0.99907243\\ 153    0.9979263\\ 154    0.996329\\ 155    0.994285\\ 156    0.9917998\\ 157    0.9888803\\ 158    0.98553437\\ 159    0.98177135\\ 160    0.9776014\\ 161    0.97303605\\ 162    0.96808785\\ 163    0.9627702\\ 164    0.9570978\\ 165    0.9510862\\ 166    0.94475186\\ 167    0.9381121\\ 168    0.9311851\\ 169    0.92399\\ 170    0.91654634\\ 171    0.90887463\\ 172    0.90099585\\ 173    0.89293164\\ 174    0.88470405\\ 175    0.8763357\\ 176    0.8678494\\ 177    0.85926855\\ 178    0.85061663\\ 179    0.8419173\\ 180    0.83319443\\ 181    0.82447195\\ 182    0.8157738\\ 183    0.8071237\\ 184    0.79854554\\ 185    0.79006267\\ 186    0.7816984\\ 187    0.7734757\\ 188    0.76541704\\ 189    0.7575445\\ 190    0.7498797\\ 191    0.7424437\\ 192    0.7352568\\ 193    0.72833866\\ 194    0.72170836\\ 195    0.715384\\ 196    0.70938295\\ 197    0.7037216\\ 198    0.6984155\\ 199    0.69347924\\ 200    0.68892634\\ 201    0.6847692\\ 202    0.68101925\\ 203    0.6776868\\ 204    0.67478096\\ 205    0.67230976\\ 206    0.67027986\\ 207    0.6686969\\ 208    0.66756517\\ 209    0.66688776\\ 210    0.6666666\\ 211    0.6669023\\ 212    0.6675942\\ 213    0.66874033\\ 214    0.67033756\\ 215    0.6723816\\ 216    0.6748668\\ 217    0.67778635\\ 218    0.6811322\\ 219    0.6848952\\ 220    0.68906516\\ 221    0.6936305\\ 222    0.6985787\\ 223    0.70389634\\ 224    0.70956874\\ 225    0.71558034\\ 226    0.7219147\\ 227    0.7285544\\ 228    0.7354814\\ 229    0.74267656\\ 230    0.75012016\\ 231    0.7577919\\ 232    0.76567066\\ 233    0.7737349\\ 234    0.7819625\\ 235    0.7903309\\ 236    0.79881716\\ 237    0.80739796\\ 238    0.81604993\\ 239    0.82474923\\ 
};
\addplot [color=cyan, line width=1.0pt, dotted] table[row sep=crcr]{%
	0    0.99999994\\ 1    0.99999994\\ 2    0.99999994\\ 3    0.99999994\\ 4    0.99999994\\ 5    0.99999994\\ 6    0.99999994\\ 7    0.99999994\\ 8    0.99999994\\ 9    0.99999994\\ 10    0.99999994\\ 11    0.99999994\\ 12    0.99999994\\ 13    0.99999994\\ 14    0.99999994\\ 15    0.99999994\\ 16    0.99999994\\ 17    0.99999994\\ 18    0.99999994\\ 19    0.99999994\\ 20    0.99999994\\ 21    0.99999994\\ 22    0.99999994\\ 23    0.99999994\\ 24    0.99999994\\ 25    0.99999994\\ 26    0.99999994\\ 27    0.99999994\\ 28    0.99999994\\ 29    0.99999994\\ 30    0.99999994\\ 31    0.99999994\\ 32    0.99999994\\ 33    0.99999994\\ 34    0.99999994\\ 35    0.99999994\\ 36    0.99999994\\ 37    1.0\\ 38    1.0\\ 39    1.0\\ 40    1.0\\ 41    1.0\\ 42    1.0\\ 43    1.0\\ 44    1.0\\ 45    1.0\\ 46    1.0\\ 47    1.0\\ 48    1.0\\ 49    1.0\\ 50    1.0\\ 51    1.0\\ 52    1.0\\ 53    1.0\\ 54    1.0\\ 55    1.0\\ 56    1.0\\ 57    1.0\\ 58    1.0\\ 59    1.0\\ 60    1.0\\ 61    1.0\\ 62    1.0\\ 63    1.0\\ 64    1.0\\ 65    1.0\\ 66    1.0\\ 67    1.0\\ 68    1.0\\ 69    1.0\\ 70    1.0\\ 71    1.0\\ 72    1.0\\ 73    1.0\\ 74    1.0\\ 75    1.0\\ 76    1.0\\ 77    1.0\\ 78    1.0\\ 79    1.0\\ 80    1.0\\ 81    1.0\\ 82    1.0\\ 83    1.0\\ 84    0.99999994\\ 85    0.99999994\\ 86    0.99999994\\ 87    0.99999994\\ 88    0.99999994\\ 89    0.99999994\\ 90    0.99999994\\ 91    0.99999994\\ 92    0.99999994\\ 93    0.99999994\\ 94    0.99999994\\ 95    0.99999994\\ 96    0.99999994\\ 97    0.99999994\\ 98    0.99999994\\ 99    0.99999994\\ 100    0.99999994\\ 101    0.99999994\\ 102    0.99999994\\ 103    0.99999994\\ 104    0.99999994\\ 105    0.99999994\\ 106    0.99999994\\ 107    0.99999994\\ 108    0.99999994\\ 109    0.99999994\\ 110    0.99999994\\ 111    0.99999994\\ 112    0.99999994\\ 113    0.99999994\\ 114    0.99999994\\ 115    0.99999994\\ 116    0.99999994\\ 117    0.99999994\\ 118    0.99999994\\ 119    0.99999994\\ 120    0.99999994\\ 121    0.99999994\\ 122    0.99999994\\ 123    0.99999994\\ 124    0.99999994\\ 125    0.99999994\\ 126    0.99999994\\ 127    0.99999994\\ 128    0.99999994\\ 129    0.99999994\\ 130    0.99999994\\ 131    0.99999994\\ 132    0.99999994\\ 133    0.99999994\\ 134    0.99999994\\ 135    0.99999994\\ 136    0.99999994\\ 137    0.99999994\\ 138    0.99999994\\ 139    0.99999994\\ 140    0.99999994\\ 141    0.99999994\\ 142    0.99999994\\ 143    0.99999994\\ 144    0.99999994\\ 145    0.99999994\\ 146    0.99999994\\ 147    0.99999994\\ 148    0.99999994\\ 149    0.99999994\\ 150    0.99999994\\ 151    0.99999994\\ 152    0.99999994\\ 153    0.99999994\\ 154    0.99999994\\ 155    0.99999994\\ 156    0.99999994\\ 157    1.0\\ 158    1.0\\ 159    1.0\\ 160    1.0\\ 161    1.0\\ 162    1.0\\ 163    1.0\\ 164    1.0\\ 165    1.0\\ 166    1.0\\ 167    1.0\\ 168    1.0\\ 169    1.0\\ 170    1.0\\ 171    1.0\\ 172    1.0\\ 173    1.0\\ 174    1.0\\ 175    1.0\\ 176    1.0\\ 177    1.0\\ 178    1.0\\ 179    1.0\\ 180    1.0\\ 181    1.0\\ 182    1.0\\ 183    1.0\\ 184    1.0\\ 185    1.0\\ 186    1.0\\ 187    1.0\\ 188    1.0\\ 189    1.0\\ 190    1.0\\ 191    1.0\\ 192    1.0\\ 193    1.0\\ 194    1.0\\ 195    1.0\\ 196    1.0\\ 197    1.0\\ 198    1.0\\ 199    1.0\\ 200    1.0\\ 201    1.0\\ 202    1.0\\ 203    1.0\\ 204    0.99999994\\ 205    0.99999994\\ 206    0.99999994\\ 207    0.99999994\\ 208    0.99999994\\ 209    0.99999994\\ 210    0.99999994\\ 211    0.99999994\\ 212    0.99999994\\ 213    0.99999994\\ 214    0.99999994\\ 215    0.99999994\\ 216    0.99999994\\ 217    0.99999994\\ 218    0.99999994\\ 219    0.99999994\\ 220    0.99999994\\ 221    0.99999994\\ 222    0.99999994\\ 223    0.99999994\\ 224    0.99999994\\ 225    0.99999994\\ 226    0.99999994\\ 227    0.99999994\\ 228    0.99999994\\ 229    0.99999994\\ 230    0.99999994\\ 231    0.99999994\\ 232    0.99999994\\ 233    0.99999994\\ 234    0.99999994\\ 235    0.99999994\\ 236    0.99999994\\ 237    0.99999994\\ 238    0.99999994\\ 239    0.99999994\\
};
\addplot [color=cyan, line width=1.0pt] table[row sep=crcr]{%
	0    0.8331944\\ 1    0.8331908\\ 2    0.8331873\\ 3    0.8331839\\ 4    0.8331806\\ 5    0.83317745\\ 6    0.83317435\\ 7    0.83317137\\ 8    0.8331685\\ 9    0.83316576\\ 10    0.83316314\\ 11    0.83316064\\ 12    0.83315825\\ 13    0.83315593\\ 14    0.8331538\\ 15    0.83315176\\ 16    0.83314985\\ 17    0.83314806\\ 18    0.83314645\\ 19    0.8331449\\ 20    0.83314353\\ 21    0.8331422\\ 22    0.8331411\\ 23    0.8331401\\ 24    0.83313924\\ 25    0.83313847\\ 26    0.83313787\\ 27    0.8331374\\ 28    0.8331371\\ 29    0.83313686\\ 30    0.8331368\\ 31    0.83313686\\ 32    0.8331371\\ 33    0.8331374\\ 34    0.83313787\\ 35    0.83313847\\ 36    0.83313924\\ 37    0.8331401\\ 38    0.8331411\\ 39    0.8331422\\ 40    0.83314353\\ 41    0.8331449\\ 42    0.83314645\\ 43    0.83314806\\ 44    0.83314985\\ 45    0.83315176\\ 46    0.8331538\\ 47    0.83315593\\ 48    0.83315825\\ 49    0.83316064\\ 50    0.83316314\\ 51    0.83316576\\ 52    0.8331685\\ 53    0.83317137\\ 54    0.83317435\\ 55    0.83317745\\ 56    0.8331806\\ 57    0.8331839\\ 58    0.8331873\\ 59    0.8331908\\ 60    0.8331944\\ 61    0.8331981\\ 62    0.8332018\\ 63    0.8332057\\ 64    0.8332097\\ 65    0.83321375\\ 66    0.83321786\\ 67    0.83322203\\ 68    0.8332263\\ 69    0.8332307\\ 70    0.8332351\\ 71    0.8332396\\ 72    0.83324414\\ 73    0.83324873\\ 74    0.83325344\\ 75    0.83325815\\ 76    0.8332629\\ 77    0.83326775\\ 78    0.83327264\\ 79    0.8332775\\ 80    0.8332825\\ 81    0.8332875\\ 82    0.8332925\\ 83    0.83329755\\ 84    0.8333026\\ 85    0.8333077\\ 86    0.8333128\\ 87    0.83331794\\ 88    0.83332306\\ 89    0.83332825\\ 90    0.8333334\\ 91    0.8333385\\ 92    0.8333436\\ 93    0.83334875\\ 94    0.8333539\\ 95    0.833359\\ 96    0.8333641\\ 97    0.8333692\\ 98    0.8333742\\ 99    0.8333792\\ 100    0.8333842\\ 101    0.83338916\\ 102    0.83339405\\ 103    0.83339894\\ 104    0.83340377\\ 105    0.83340853\\ 106    0.8334133\\ 107    0.83341795\\ 108    0.83342254\\ 109    0.83342713\\ 110    0.8334316\\ 111    0.833436\\ 112    0.83344036\\ 113    0.83344465\\ 114    0.8334488\\ 115    0.83345294\\ 116    0.833457\\ 117    0.8334609\\ 118    0.8334648\\ 119    0.8334686\\ 120    0.83347225\\ 121    0.8334759\\ 122    0.83347934\\ 123    0.83348274\\ 124    0.833486\\ 125    0.83348924\\ 126    0.8334923\\ 127    0.83349526\\ 128    0.8334981\\ 129    0.83350086\\ 130    0.8335035\\ 131    0.833506\\ 132    0.8335084\\ 133    0.8335107\\ 134    0.83351284\\ 135    0.83351487\\ 136    0.8335168\\ 137    0.83351856\\ 138    0.8335202\\ 139    0.8335217\\ 140    0.8335231\\ 141    0.8335244\\ 142    0.83352554\\ 143    0.8335265\\ 144    0.8335274\\ 145    0.8335281\\ 146    0.83352876\\ 147    0.8335292\\ 148    0.83352953\\ 149    0.8335297\\ 150    0.83352983\\ 151    0.8335297\\ 152    0.83352953\\ 153    0.8335292\\ 154    0.83352876\\ 155    0.8335281\\ 156    0.8335274\\ 157    0.8335265\\ 158    0.83352554\\ 159    0.8335244\\ 160    0.8335231\\ 161    0.8335217\\ 162    0.8335202\\ 163    0.83351856\\ 164    0.8335168\\ 165    0.83351487\\ 166    0.83351284\\ 167    0.8335107\\ 168    0.8335084\\ 169    0.833506\\ 170    0.8335035\\ 171    0.83350086\\ 172    0.8334981\\ 173    0.83349526\\ 174    0.8334923\\ 175    0.83348924\\ 176    0.833486\\ 177    0.83348274\\ 178    0.83347934\\ 179    0.8334759\\ 180    0.83347225\\ 181    0.8334686\\ 182    0.8334648\\ 183    0.8334609\\ 184    0.833457\\ 185    0.83345294\\ 186    0.8334488\\ 187    0.83344465\\ 188    0.83344036\\ 189    0.833436\\ 190    0.8334316\\ 191    0.83342713\\ 192    0.83342254\\ 193    0.83341795\\ 194    0.8334133\\ 195    0.83340853\\ 196    0.83340377\\ 197    0.83339894\\ 198    0.83339405\\ 199    0.83338916\\ 200    0.8333842\\ 201    0.8333792\\ 202    0.8333742\\ 203    0.8333692\\ 204    0.8333641\\ 205    0.833359\\ 206    0.8333539\\ 207    0.83334875\\ 208    0.8333436\\ 209    0.8333385\\ 210    0.8333334\\ 211    0.83332825\\ 212    0.83332306\\ 213    0.83331794\\ 214    0.8333128\\ 215    0.8333077\\ 216    0.8333026\\ 217    0.83329755\\ 218    0.8332925\\ 219    0.8332875\\ 220    0.8332825\\ 221    0.8332775\\ 222    0.83327264\\ 223    0.83326775\\ 224    0.8332629\\ 225    0.83325815\\ 226    0.83325344\\ 227    0.83324873\\ 228    0.83324414\\ 229    0.8332396\\ 230    0.8332351\\ 231    0.8332307\\ 232    0.8332263\\ 233    0.83322203\\ 234    0.83321786\\ 235    0.83321375\\ 236    0.8332097\\ 237    0.8332057\\ 238    0.8332018\\ 239    0.8331981\\ 
};
\addplot [color=black, line width=1.0pt, dotted] table[row sep=crcr]{%
	0    0.0\\ 1    0.5378621\\ 2    0.5361326\\ 3    0.5350608\\ 4    0.53465056\\ 5    0.5349035\\ 6    0.53581965\\ 7    0.5373969\\ 8    0.53963155\\ 9    0.54251766\\ 10    0.54604775\\ 11    0.5502123\\ 12    0.55500007\\ 13    0.5603979\\ 14    0.56639105\\ 15    0.5729629\\ 16    0.5800951\\ 17    0.5877679\\ 18    0.5959597\\ 19    0.6046476\\ 20    0.6138071\\ 21    0.62341225\\ 22    0.633436\\ 23    0.6438499\\ 24    0.6546243\\ 25    0.66572845\\ 26    0.6771307\\ 27    0.6887984\\ 28    0.7006982\\ 29    0.71279573\\ 30    0.7250563\\ 31    0.7374445\\ 32    0.7499246\\ 33    0.7624603\\ 34    0.77501535\\ 35    0.78755325\\ 36    0.80003744\\ 37    0.8124315\\ 38    0.82469916\\ 39    0.8368044\\ 40    0.84871167\\ 41    0.8603859\\ 42    0.8717927\\ 43    0.8828981\\ 44    0.8936693\\ 45    0.90407413\\ 46    0.9140816\\ 47    0.92366165\\ 48    0.9327855\\ 49    0.9414257\\ 50    0.9495559\\ 51    0.9571514\\ 52    0.964189\\ 53    0.970647\\ 54    0.9765053\\ 55    0.9817456\\ 56    0.98635143\\ 57    0.990308\\ 58    0.9936024\\ 59    0.99622375\\ 60    0.998163\\ 61    0.9994131\\ 62    0.99996907\\ 63    0.9998279\\ 64    0.9989887\\ 65    0.99745256\\ 66    0.99522257\\ 67    0.99230397\\ 68    0.988704\\ 69    0.98443204\\ 70    0.9794992\\ 71    0.97391886\\ 72    0.96770614\\ 73    0.9608782\\ 74    0.95345384\\ 75    0.945454\\ 76    0.93690103\\ 77    0.92781925\\ 78    0.91823435\\ 79    0.90817374\\ 80    0.8976662\\ 81    0.88674194\\ 82    0.8754325\\ 83    0.8637705\\ 84    0.85178983\\ 85    0.83952534\\ 86    0.82701266\\ 87    0.8142885\\ 88    0.80139\\ 89    0.788355\\ 90    0.7752218\\ 91    0.7620292\\ 92    0.74881595\\ 93    0.73562115\\ 94    0.722484\\ 95    0.7094433\\ 96    0.69653773\\ 97    0.6838058\\ 98    0.67128545\\ 99    0.659014\\ 100    0.647028\\ 101    0.63536346\\ 102    0.62405527\\ 103    0.6131373\\ 104    0.6026424\\ 105    0.5926022\\ 106    0.58304685\\ 107    0.5740052\\ 108    0.56550455\\ 109    0.5575707\\ 110    0.55022764\\ 111    0.5434977\\ 112    0.53740126\\ 113    0.531957\\ 114    0.5271815\\ 115    0.5230894\\ 116    0.5196933\\ 117    0.5170038\\ 118    0.51502913\\ 119    0.5137756\\ 120    0.51324725\\ 121    0.5134459\\ 122    0.5143714\\ 123    0.5160211\\ 124    0.51839024\\ 125    0.5214721\\ 126    0.5252576\\ 127    0.5297355\\ 128    0.5348927\\ 129    0.5407137\\ 130    0.5471813\\ 131    0.5542762\\ 132    0.56197727\\ 133    0.5702614\\ 134    0.5791039\\ 135    0.58847815\\ 136    0.59835625\\ 137    0.60870844\\ 138    0.6195038\\ 139    0.6307099\\ 140    0.6422931\\ 141    0.6542187\\ 142    0.66645086\\ 143    0.67895293\\ 144    0.6916873\\ 145    0.7046159\\ 146    0.71769977\\ 147    0.7308998\\ 148    0.7441763\\ 149    0.7574895\\ 150    0.7707995\\ 151    0.7840664\\ 152    0.7972504\\ 153    0.8103121\\ 154    0.82321227\\ 155    0.8359124\\ 156    0.8483745\\ 157    0.86056125\\ 158    0.8724362\\ 159    0.88396406\\ 160    0.89511025\\ 161    0.9058416\\ 162    0.916126\\ 163    0.925933\\ 164    0.93523324\\ 165    0.9439992\\ 166    0.95220464\\ 167    0.95982534\\ 168    0.9668386\\ 169    0.9732237\\ 170    0.97896165\\ 171    0.98403555\\ 172    0.98843044\\ 173    0.9921333\\ 174    0.9951333\\ 175    0.9974215\\ 176    0.9989914\\ 177    0.9998383\\ 178    0.9999599\\ 179    0.9993558\\ 180    0.998028\\ 181    0.99598044\\ 182    0.9932193\\ 183    0.9897528\\ 184    0.9855913\\ 185    0.980747\\ 186    0.9752343\\ 187    0.96906954\\ 188    0.9622709\\ 189    0.95485836\\ 190    0.94685376\\ 191    0.9382807\\ 192    0.92916423\\ 193    0.9195312\\ 194    0.9094098\\ 195    0.8988297\\ 196    0.8878218\\ 197    0.8764183\\ 198    0.8646524\\ 199    0.85255855\\ 200    0.8401719\\ 201    0.8275284\\ 202    0.814665\\ 203    0.801619\\ 204    0.78842825\\ 205    0.775131\\ 206    0.7617657\\ 207    0.7483712\\ 208    0.734986\\ 209    0.721649\\ 210    0.70839846\\ 211    0.6952727\\ 212    0.68230945\\ 213    0.669546\\ 214    0.65701914\\ 215    0.64476466\\ 216    0.6328178\\ 217    0.6212128\\ 218    0.6099827\\ 219    0.5991598\\ 220    0.58877474\\ 221    0.5788573\\ 222    0.5694356\\ 223    0.5605365\\ 224    0.5521851\\ 225    0.5444052\\ 226    0.5372187\\ 227    0.5306458\\ 228    0.5247051\\ 229    0.5194132\\ 230    0.5147849\\ 231    0.510833\\ 232    0.50756854\\ 233    0.50500035\\ 234    0.5031355\\ 235    0.5019788\\ 236    0.50153327\\ 237    0.50179976\\ 238    0.50277704\\ 239    0.50446206\\ 
};
\addplot [color=black, line width=1.0pt] table[row sep=crcr]{%
	0    0.0\\ 1    0.9009647\\ 2    0.9068681\\ 3    0.912766\\ 4    0.91863686\\ 5    0.92445916\\ 6    0.93021137\\ 7    0.9358721\\ 8    0.9414198\\ 9    0.9468334\\ 10    0.95209193\\ 11    0.9571748\\ 12    0.9620617\\ 13    0.96673286\\ 14    0.9711689\\ 15    0.9753511\\ 16    0.9792614\\ 17    0.9828822\\ 18    0.9861969\\ 19    0.98918945\\ 20    0.99184483\\ 21    0.99414885\\ 22    0.99608815\\ 23    0.99765056\\ 24    0.9988247\\ 25    0.9996004\\ 26    0.9999686\\ 27    0.9999213\\ 28    0.99945176\\ 29    0.9985543\\ 30    0.9972245\\ 31    0.9954593\\ 32    0.99325675\\ 33    0.99061614\\ 34    0.9875383\\ 35    0.98402494\\ 36    0.9800794\\ 37    0.9757062\\ 38    0.970911\\ 39    0.965701\\ 40    0.9600844\\ 41    0.9540707\\ 42    0.9476708\\ 43    0.94089663\\ 44    0.9337613\\ 45    0.9262791\\ 46    0.91846544\\ 47    0.91033685\\ 48    0.9019107\\ 49    0.8932056\\ 50    0.88424087\\ 51    0.87503684\\ 52    0.86561465\\ 53    0.8559962\\ 54    0.8462042\\ 55    0.8362619\\ 56    0.8261933\\ 57    0.81602263\\ 58    0.805775\\ 59    0.79547554\\ 60    0.78515\\ 61    0.77482414\\ 62    0.76452404\\ 63    0.7542759\\ 64    0.744106\\ 65    0.7340403\\ 66    0.72410506\\ 67    0.71432596\\ 68    0.70472866\\ 69    0.69533837\\ 70    0.68617994\\ 71    0.67727774\\ 72    0.6686555\\ 73    0.6603364\\ 74    0.65234286\\ 75    0.64469653\\ 76    0.6374183\\ 77    0.6305282\\ 78    0.6240451\\ 79    0.6179869\\ 80    0.61237067\\ 81    0.60721207\\ 82    0.6025257\\ 83    0.598325\\ 84    0.59462196\\ 85    0.59142745\\ 86    0.58875084\\ 87    0.5866003\\ 88    0.5849824\\ 89    0.5839024\\ 90    0.5833641\\ 91    0.58336985\\ 92    0.5839204\\ 93    0.5850153\\ 94    0.5866522\\ 95    0.5888276\\ 96    0.5915364\\ 97    0.59477204\\ 98    0.59852654\\ 99    0.60279036\\ 100    0.60755277\\ 101    0.6128014\\ 102    0.6185227\\ 103    0.62470174\\ 104    0.6313224\\ 105    0.6383672\\ 106    0.6458176\\ 107    0.6536538\\ 108    0.66185504\\ 109    0.67039955\\ 110    0.6792644\\ 111    0.68842614\\ 112    0.6978602\\ 113    0.7075412\\ 114    0.71744347\\ 115    0.7275403\\ 116    0.7378048\\ 117    0.7482093\\ 118    0.75872606\\ 119    0.7693269\\ 120    0.7799835\\ 121    0.79066736\\ 122    0.80135\\ 123    0.81200284\\ 124    0.8225977\\ 125    0.8331063\\ 126    0.8435009\\ 127    0.853754\\ 128    0.8638386\\ 129    0.8737283\\ 130    0.8833971\\ 131    0.8928199\\ 132    0.90197235\\ 133    0.9108308\\ 134    0.9193726\\ 135    0.92757607\\ 136    0.93542045\\ 137    0.94288623\\ 138    0.9499549\\ 139    0.9566093\\ 140    0.96283334\\ 141    0.9686123\\ 142    0.9739329\\ 143    0.9787831\\ 144    0.9831522\\ 145    0.9870311\\ 146    0.99041206\\ 147    0.9932889\\ 148    0.9956568\\ 149    0.99751246\\ 150    0.9988543\\ 151    0.9996819\\ 152    0.9999966\\ 153    0.99980116\\ 154    0.9990998\\ 155    0.99789834\\ 156    0.9962038\\ 157    0.9940248\\ 158    0.99137145\\ 159    0.9882549\\ 160    0.98468786\\ 161    0.98068434\\ 162    0.97625947\\ 163    0.9714296\\ 164    0.96621215\\ 165    0.96062577\\ 166    0.95469004\\ 167    0.9484255\\ 168    0.94185346\\ 169    0.93499637\\ 170    0.9278771\\ 171    0.9205195\\ 172    0.91294765\\ 173    0.9051866\\ 174    0.8972616\\ 175    0.8891984\\ 176    0.881023\\ 177    0.8727616\\ 178    0.8644406\\ 179    0.8560865\\ 180    0.84772575\\ 181    0.8393847\\ 182    0.83108956\\ 183    0.8228663\\ 184    0.8147407\\ 185    0.80673784\\ 186    0.7988826\\ 187    0.79119927\\ 188    0.7837114\\ 189    0.776442\\ 190    0.76941335\\ 191    0.7626468\\ 192    0.75616294\\ 193    0.74998134\\ 194    0.7441208\\ 195    0.7385987\\ 196    0.73343176\\ 197    0.7286353\\ 198    0.72422355\\ 199    0.72020954\\ 200    0.71660507\\ 201    0.71342057\\ 202    0.7106653\\ 203    0.70834714\\ 204    0.7064726\\ 205    0.7050468\\ 206    0.70407355\\ 207    0.70355535\\ 208    0.7034931\\ 209    0.7038865\\ 210    0.7047339\\ 211    0.7060321\\ 212    0.70777667\\ 213    0.7099618\\ 214    0.7125804\\ 215    0.715624\\ 216    0.71908295\\ 217    0.7229464\\ 218    0.7272022\\ 219    0.731837\\ 220    0.73683643\\ 221    0.74218506\\ 222    0.7478664\\ 223    0.75386286\\ 224    0.76015615\\ 225    0.7667269\\ 226    0.77355504\\ 227    0.7806198\\ 228    0.7878996\\ 229    0.7953723\\ 230    0.80301523\\ 231    0.81080514\\ 232    0.81871843\\ 233    0.82673115\\ 234    0.834819\\ 235    0.8429576\\ 236    0.8511223\\ 237    0.85928845\\ 238    0.86743134\\ 239    0.8755264\\ 
};
\addlegendentry{(a) $\vec{f} = \{1, 0, 0\}$}
\addlegendentry{(a) $\vec{f} = \{1, 1, 1\}$}
\addlegendentry{(b) $\vec{f} = \{1, 0, 0\}$}
\addlegendentry{(b) $\vec{f} = \{1, 1, 1\}$}
\addlegendentry{(c) $\vec{f} = \{1, 0, 0\}$}
\addlegendentry{(c) $\vec{f} = \{1, 1, 1\}$}
0\end{axis}
\end{tikzpicture}%

%% file: planes_cxi_1.0_2.0_1.5_0.3.tikz
\begin{tikzpicture}
 	\begin{axis}[%
		width=\textwidth,
		height=\axisdefaultheight,
		scale = 0.4,
		xmin=1,
		xmax=90,
		xlabel={Rotation angle in $^\circ$},
		xlabel near ticks,
		ymin=0,
		ymax=3.2,
		ylabel={Dark-field $d(\vec{x})$},
ylabel near ticks,
	xmajorgrids,
	ymajorgrids,
legend pos=north west
	]
	\addplot [color=blue, line width=1.0pt] table[row sep=crcr]{%
0    0.22175195895695685\\ 1    0.2220221234234333\\ 2    0.2228323084159374\\ 3    0.22418152262649535\\ 4    0.22606811387515066\\ 5    0.22848976910996435\\ 6    0.2314435804942131\\ 7    0.23492591323199272\\ 8    0.23893253774261475\\ 9    0.24345856357340812\\ 10    0.2484985054869175\\ 11    0.2540461733155727\\ 12    0.26009487022328376\\ 13    0.26663717241477963\\ 14    0.27366512739720344\\ 15    0.2811701878929138\\ 16    0.28914316778135296\\ 17    0.2975744183315754\\ 18    0.30645360791158677\\ 19    0.31576996430807114\\ 20    0.32551209849386215\\ 21    0.3356681588314056\\ 22    0.34622576498556135\\ 23    0.3571720740108013\\ 24    0.36849373629307747\\ 25    0.3801769616370201\\ 26    0.39220747520780563\\ 27    0.40457067173461914\\ 28    0.4172514833362579\\ 29    0.43023444560832974\\ 30    0.4435037857395172\\ 31    0.4570432462790489\\ 32    0.47083639354362483\\ 33    0.4848663973267555\\ 34    0.4991162071312904\\ 35    0.5135684199950218\\ 36    0.528205412665081\\ 37    0.5430094297142029\\ 38    0.5579623632500649\\ 39    0.5730459732059479\\ 40    0.5882419754570007\\ 41    0.6035318215295792\\ 42    0.6188968307756424\\ 43    0.6343183225471496\\ 44    0.6497775280797958\\ 45    0.6652555904930114\\ 46    0.6807336529062271\\ 47    0.6961928143807411\\ 48    0.7116143502103806\\ 49    0.7269793594564438\\ 50    0.7422692055290222\\ 51    0.7574652077800751\\ 52    0.7725487736778259\\ 53    0.78750175127182\\ 54    0.8023057242628097\\ 55    0.8169427169328689\\ 56    0.8313949297966003\\ 57    0.8456447396011352\\ 58    0.859674787442398\\ 59    0.8734678906488418\\ 60    0.8870073952465057\\ 61    0.9002767353776931\\ 62    0.913259697649765\\ 63    0.9259405092514038\\ 64    0.9383036617200851\\ 65    0.9503341752908706\\ 66    0.9620174226638794\\ 67    0.9733390408880234\\ 68    0.9842853719423293\\ 69    0.9948430221546173\\ 70    1.0049990384340286\\ 71    1.0147411726198197\\ 72    1.0240575290163039\\ 73    1.0329367406253813\\ 74    1.0413679691465376\\ 75    1.049340993093109\\ 76    1.0568460315597534\\ 77    1.0638740085712433\\ 78    1.070416288733673\\ 79    1.076464941583252\\ 80    1.0820126534700394\\ 81    1.0870525513254166\\ 82    1.0915786432434083\\ 83    1.0955852897830962\\ 84    1.0990675564336776\\ 85    1.1020213898469924\\ 86    1.1044430891399384\\ 87    1.1063296583595275\\ 88    1.1076788064828873\\ 89    1.1084890355335235\\ 90    1.1087592\\  }; 	
	\addplot [color=green, line width=1.0pt, dashed] table[row sep=crcr]{%
 0    1.1087592\\ 1    1.08634776\\ 2    1.0862653500000001\\ 3    1.08650931\\ 4    1.0870833\\ 5    1.08799515\\ 6    1.08739176\\ 7    1.0685711549999999\\ 8    1.05879369\\ 9    1.0545231\\ 10    1.0554497999999999\\ 11    1.0551428999999999\\ 12    1.05005535\\ 13    1.0500192\\ 14    1.05041055\\ 15    1.05010875\\ 16    1.0498287\\ 17    1.04269194\\ 18    1.038313155\\ 19    1.0381770750000001\\ 20    1.0380543\\ 21    1.0382666999999999\\ 22    1.0383747\\ 23    1.03819245\\ 24    1.0381586999999999\\ 25    1.04298279\\ 26    1.04761809\\ 27    1.047579375\\ 28    1.04773515\\ 29    1.05237975\\ 30    1.05233775\\ 31    1.0525554\\ 32    1.0570935000000001\\ 33    1.061757065619123\\ 34    1.066761515221405\\ 35    1.0766284244372535\\ 36    1.0815919390427888\\ 37    1.0944549630205274\\ 38    1.0944984630170702\\ 39    1.099566512614298\\ 40    1.104835517195555\\ 41    1.1100173567837395\\ 42    1.126277085491532\\ 43    1.1262620104927301\\ 44    1.1316053850680767\\ 45    1.1372921996161294\\ 46    1.148223808747363\\ 47    1.174224431681019\\ 48    1.1806886061672925\\ 49    1.180516706180954\\ 50    1.1985827047451973\\ 51    1.211189828743273\\ 52    1.2172921282583058\\ 53    1.243370901185751\\ 54    1.2537071003643037\\ 55    1.2602397498451352\\ 56    1.2741470887398791\\ 57    1.3025442464830757\\ 58    1.317254205314033\\ 59    1.328634344409621\\ 60    1.35568459225986\\ 61    1.3757449906656025\\ 62    1.383856750020938\\ 63    1.4218441870019674\\ 64    1.439521685597086\\ 65    1.4667772384310092\\ 66    1.476034082695341\\ 67    1.5241903788682223\\ 68    1.5496785536851883\\ 69    1.5707909003294707\\ 70    1.6139082434761525\\ 71    1.6426056389148234\\ 72    1.6903282813295128\\ 73    1.7152428273694513\\ 74    1.7539860212113856\\ 75    1.8011145137205125\\ 76    1.8447628567827938\\ 77    1.889157899726391\\ 78    1.9436536410645244\\ 79    2.0123524301451443\\ 80    2.0468150746674536\\ 81    2.101645165952444\\ 82    2.212877198272586\\ 83    2.2549349415876865\\ 84    2.3222364808903935\\ 85    2.4579558593183757\\ 86    2.5100180010433197\\ 87    2.593880287713766\\ 88    2.76414811065042\\ 89    2.830000800183415\\ 90    2.999999523162842\\  };		
	\addplot [color=red, line width=1.0pt, dotted] table[row sep=crcr]{%
 0    0.22175195895695685\\ 1    0.21752174564497948\\ 2    0.21828634581093193\\ 3    0.21964490208090126\\ 4    0.2215980981804073\\ 5    0.22414778544479905\\ 6    0.22690791840938568\\ 7    0.22632457978817822\\ 8    0.22806786365950046\\ 9    0.23144088236300348\\ 10    0.23643025137884616\\ 11    0.24162945587807894\\ 12    0.24618157811490296\\ 13    0.2523577283466339\\ 14    0.2590989975328654\\ 15    0.26612191356740894\\ 16    0.27358970881011485\\ 17    0.27964850757422804\\ 18    0.28677926351545097\\ 19    0.2954553149145663\\ 20    0.3045318538561583\\ 21    0.31409540750336645\\ 22    0.3240066545152068\\ 23    0.3341908804521471\\ 24    0.344772518691951\\ 25    0.35735672596408724\\ 26    0.37030440495669364\\ 27    0.38196436442390086\\ 28    0.393997051368463\\ 29    0.4080597773611844\\ 30    0.420631238799423\\ 31    0.4335654323952198\\ 32    0.44857958355879785\\ 33    0.4639886724555731\\ 34    0.47988080136146544\\ 35    0.4983483239119935\\ 36    0.51492009830257\\ 37    0.5356531230864346\\ 38    0.5504317194491744\\ 39    0.5679361940612316\\ 40    0.5857974525175155\\ 41    0.6038500963021434\\ 42    0.6283015577166379\\ 43    0.6439569650020242\\ 44    0.6627891455327898\\ 45    0.6819961052021813\\ 46    0.7045806357529283\\ 47    0.7369079840114058\\ 48    0.7573878692056417\\ 49    0.7736386553659439\\ 50    0.8020086704126358\\ 51    0.8270470388597488\\ 52    0.8477772628212213\\ 53    0.8827117380468846\\ 54    0.9067933406147003\\ 55    0.9281598702425122\\ 56    0.95501589433038\\ 57    0.9930467580448866\\ 58    1.0209361789989781\\ 59    1.0462917945688606\\ 60    1.0841561676872968\\ 61    1.11667145083673\\ 62    1.1394685368840074\\ 63    1.187018455597186\\ 64    1.2178378965137004\\ 65    1.256822147751708\\ 66    1.2803184149219513\\ 67    1.3376652474839688\\ 68    1.3753464130795479\\ 69    1.4090542065712095\\ 70    1.4625292883757353\\ 71    1.5029825082433221\\ 72    1.5608673919453027\\ 73    1.5976261919287205\\ 74    1.6470675209629058\\ 75    1.7042927870544433\\ 76    1.758100459028673\\ 77    1.8124042118046284\\ 78    1.8761750277322768\\ 79    1.9534884215885997\\ 80    1.9972064160223961\\ 81    2.060283898649335\\ 82    2.178384521814692\\ 83    2.227960651670432\\ 84    2.30177691994344\\ 85    2.442876916001737\\ 86    2.500130657635498\\ 87    2.588106012864852\\ 88    2.7613905330127357\\ 89    2.8292783947485685\\ 90    2.999999523162842\\ }; 
\end{axis} 
 \end{tikzpicture}

%% file: cxi_simulations.tikz
	\begin{tikzpicture}
	\begin{axis}[%
	width=\textwidth,
	height=\axisdefaultheight,
		scale = 0.4,
	xmin=0,
	xmax=90,
	xlabel={Rotation angle in $^\circ$},
	xlabel near ticks,
	ymin=0,
	ymax=0.003,
	ylabel={Dark-field $d(\vec{x})$},
	ylabel near ticks,
	xmajorgrids,
	ymajorgrids,
	legend pos=outer north east
	]
	\addplot [color=blue, line width=1.0pt] table[row sep=crcr]{%
0.0   0.0003153467224942557 \\
3.0   0.00033587484825656195 \\
6.0   0.00038364485933049884 \\
9.0   0.00043202659293114495 \\
12.0   0.00046147239130126633 \\
14.999999999999998   0.0004784079563352758 \\
18.0   0.0005037795643310398 \\
20.999999999999996   0.0005486110639046216 \\
24.0   0.0006099367045663981 \\
27.0   0.0006623651310339245 \\
29.999999999999996   0.0006829419530835184 \\
33.0   0.0007043870698069695 \\
36.0   0.0007575904823048348 \\
39.0   0.0008155497725245812 \\
41.99999999999999   0.0008525447066065016 \\
45.0   0.0009006788986747069 \\
48.0   0.0009439509572168174 \\
51.0   0.0009490594067484142 \\
54.0   0.000986735257405159 \\
57.0   0.0010462719220311948 \\
59.99999999999999   0.001050386208876005 \\
63.0   0.0010593970638085254 \\
66.0   0.0010995247566729462 \\
68.99999999999999   0.0010965468403687328 \\
72.0   0.0011035369076712585 \\
74.99999999999999   0.0011297071847552445 \\
78.0   0.0011379263166240029 \\
81.0   0.0011331281354389079 \\
83.99999999999999   0.0010838110087532534 \\
87.0   0.0011372509500877377 \\
90.0   0.001277047650892386 \\
	};\addlegendentry{$\vec{x}\vec{y}$}
	\addplot [color=red, line width=1.0pt, dotted] table[row sep=crcr]{%
0.0   0.0003153467224942557 \\
3.0   0.0003058780201366913 \\
6.0   0.0003004533826875025 \\
9.0   0.00029094459523767616 \\
12.0   0.00027796737090529503 \\
14.999999999999998   0.00026153320926727836 \\
18.0   0.00024662489492600887 \\
20.999999999999996   0.00024330496845685054 \\
24.0   0.0002593628961593987 \\
27.0   0.00028635814968085176 \\
29.999999999999996   0.000299016393328656 \\
33.0   0.00030696140065689386 \\
36.0   0.0003428825648433733 \\
39.0   0.00039929727506950944 \\
41.99999999999999   0.0004514015145389996 \\
45.0   0.0005135797715806644 \\
48.0   0.0005883017191372867 \\
51.0   0.0006597334485620085 \\
54.0   0.000744612138017308 \\
57.0   0.0008599650234724721 \\
59.99999999999999   0.0009700932190134416 \\
63.0   0.001085682597024281 \\
66.0   0.0012336554046610661 \\
68.99999999999999   0.0013755393169657287 \\
72.0   0.0015339963694550808 \\
74.99999999999999   0.0016965702048791127 \\
78.0   0.0018569346023029132 \\
81.0   0.002028269326223794 \\
83.99999999999999   0.002186946495333005 \\
87.0   0.0024040543652308626 \\
90.0   0.0025801332856611988 \\
	};\addlegendentry{$\vec{x}\vec{z}$}
	\addplot [color=green, line width=1.0pt, dashed] table[row sep=crcr]{%
0.0   0.001277047650892386 \\
3.0   0.001261243039362817 \\
6.0   0.0012460951492624418 \\
9.0   0.0012069134560726516 \\
12.0   0.0011574107323703287 \\
14.999999999999998   0.0011149357457821353 \\
18.0   0.0010904858511026792 \\
20.999999999999996   0.0010784285829161658 \\
24.0   0.0010739314453992498 \\
27.0   0.0010768518167709046 \\
29.999999999999996   0.0010855732736507106 \\
33.0   0.001100176975674435 \\
36.0   0.0011198769937725127 \\
39.0   0.0011453038100115891 \\
41.99999999999999   0.0011760310245929508 \\
45.0   0.0012126469942554262 \\
48.0   0.0012557411607878432 \\
51.0   0.0013058063268794648 \\
54.0   0.0013637582286306152 \\
57.0   0.0014301594557566924 \\
59.99999999999999   0.001506086965884765 \\
63.0   0.001591920194790369 \\
66.0   0.0016871207601496456 \\
68.99999999999999   0.001790214989656771 \\
72.0   0.0018992561001301806 \\
74.99999999999999   0.0020122185998882654 \\
78.0   0.002127390528453725 \\
81.0   0.002244276227288163 \\
83.99999999999999   0.00236387993233537 \\
87.0   0.00248524058896031 \\
90.0   0.0025801332856611988 \\
	};\addlegendentry{$\vec{y}\vec{z}$}
	\end{axis}
	\end{tikzpicture}%

%% file: legend_rotation.tikz
    \newenvironment{customlegend}[1][]{%
        \begingroup
        \csname pgfplots@init@cleared@structures\endcsname
        \pgfplotsset{#1}%
    }{%
        \csname pgfplots@createlegend\endcsname
        \endgroup
    }%
    \def\addlegendimage{\csname pgfplots@addlegendimage\endcsname}

\begin{tikzpicture}
\begin{customlegend}[legend columns=5,legend style={align=left,draw=none,column sep=1ex},
    legend entries={$10^\circ$, $20^\circ$, $30^\circ$, $40^\circ$, $50^\circ$}]
	\addlegendimage{color=black,line width=1.2pt}
	\addlegendimage{color=blue, line width=1.2pt}
    \addlegendimage{color=red, line width=1.2pt}
	\addlegendimage{color=green, line width=1.2pt}
    \addlegendimage{color=yellow, line width=1.2pt}
\end{customlegend}
\end{tikzpicture}

%% file: xz_0.7_20.0_1.4_1.0_s2_zoom.tikz
\begin{tikzpicture}
 	\begin{axis}[%
width=\textwidth, 
	height=\axisdefaultheight,
	 scale = 0.4,
xmin=0,
	xmax=180,
	xlabel={Rotation angle in $^\circ$},
	xlabel near ticks,
ymin=0.5,
	ymax=1.7,
	ylabel={Dark-field $D_L$},
ylabel near ticks,
		yticklabels={,,},
	xmajorgrids,
	ymajorgrids,
legend pos=north west
	]
	\addplot [color=black, line width=1.0pt] table[row sep=crcr]{%
0    0.97155696\\ 1    0.97156405\\ 2    0.97158533\\ 3    0.9716205\\ 4    0.9716698\\ 5    0.971733\\ 6    0.97181\\ 7    0.9719009\\ 8    0.9720053\\ 9    0.97212344\\ 10    0.9722548\\ 11    0.97239953\\ 12    0.9725574\\ 13    0.9727281\\ 14    0.9729116\\ 15    0.9731076\\ 16    0.9733158\\ 17    0.9735361\\ 18    0.9737682\\ 19    0.9740118\\ 20    0.9742667\\ 21    0.97453237\\ 22    0.9748089\\ 23    0.9750955\\ 24    0.9753923\\ 25    0.9756985\\ 26    0.97601414\\ 27    0.9763386\\ 28    0.9766716\\ 29    0.97701275\\ 30    0.9773616\\ 31    0.9777177\\ 32    0.9780808\\ 33    0.97845036\\ 34    0.97882587\\ 35    0.979207\\ 36    0.9795932\\ 37    0.97998416\\ 38    0.98037934\\ 39    0.98077804\\ 40    0.98118013\\ 41    0.981585\\ 42    0.9819921\\ 43    0.98240095\\ 44    0.98281115\\ 45    0.9832221\\ 46    0.9836333\\ 47    0.9840443\\ 48    0.9844545\\ 49    0.9848637\\ 50    0.9852711\\ 51    0.9856762\\ 52    0.98607856\\ 53    0.98647773\\ 54    0.9868732\\ 55    0.9872645\\ 56    0.9876511\\ 57    0.9880325\\ 58    0.9884083\\ 59    0.98877794\\ 60    0.98914105\\ 61    0.9894971\\ 62    0.9898457\\ 63    0.9901864\\ 64    0.99051875\\ 65    0.9908423\\ 66    0.99115664\\ 67    0.99146146\\ 68    0.9917563\\ 69    0.9920408\\ 70    0.9923146\\ 71    0.9925773\\ 72    0.9928288\\ 73    0.9930685\\ 74    0.9932961\\ 75    0.99351144\\ 76    0.9937143\\ 77    0.9939042\\ 78    0.99408114\\ 79    0.99424464\\ 80    0.9943947\\ 81    0.9945311\\ 82    0.9946535\\ 83    0.99476194\\ 84    0.9948561\\ 85    0.9949361\\ 86    0.9950016\\ 87    0.9950525\\ 88    0.99508905\\ 89    0.9951109\\ 90    0.9951181\\ 91    0.9951107\\ 92    0.9950885\\ 93    0.99505174\\ 94    0.9950006\\ 95    0.9949348\\ 96    0.9948547\\ 97    0.9947602\\ 98    0.9946516\\ 99    0.9945289\\ 100    0.99439234\\ 101    0.994242\\ 102    0.9940782\\ 103    0.99390113\\ 104    0.99371094\\ 105    0.9935079\\ 106    0.99329233\\ 107    0.9930645\\ 108    0.9928246\\ 109    0.992573\\ 110    0.99231\\ 111    0.9920361\\ 112    0.9917513\\ 113    0.9914564\\ 114    0.99115133\\ 115    0.99083686\\ 116    0.9905131\\ 117    0.9901806\\ 118    0.9898398\\ 119    0.9894911\\ 120    0.9891349\\ 121    0.98877174\\ 122    0.98840195\\ 123    0.98802614\\ 124    0.98764455\\ 125    0.98725784\\ 126    0.9868666\\ 127    0.98647094\\ 128    0.98607165\\ 129    0.9856693\\ 130    0.98526406\\ 131    0.9848567\\ 132    0.98444754\\ 133    0.98403734\\ 134    0.98362625\\ 135    0.9832151\\ 136    0.98280406\\ 137    0.982394\\ 138    0.98198515\\ 139    0.98157805\\ 140    0.98117316\\ 141    0.9807711\\ 142    0.9803724\\ 143    0.97997737\\ 144    0.9795866\\ 145    0.97920036\\ 146    0.9788193\\ 147    0.978444\\ 148    0.9780745\\ 149    0.9777115\\ 150    0.9773555\\ 151    0.97700685\\ 152    0.97666574\\ 153    0.9763329\\ 154    0.97600853\\ 155    0.97569317\\ 156    0.975387\\ 157    0.97509044\\ 158    0.9748039\\ 159    0.97452766\\ 160    0.9742622\\ 161    0.9740075\\ 162    0.97376406\\ 163    0.97353226\\ 164    0.97331214\\ 165    0.97310406\\ 166    0.9729083\\ 167    0.9727251\\ 168    0.9725546\\ 169    0.97239697\\ 170    0.9722525\\ 171    0.97212124\\ 172    0.9720034\\ 173    0.9718992\\ 174    0.97180855\\ 175    0.9717317\\ 176    0.9716689\\ 177    0.97161984\\ 178    0.9715849\\ 179    0.9715639\\ 180    0.97155696\\ 181    0.9715642\\ 182    0.97158533\\ 183    0.97162056\\ 184    0.9716698\\ 185    0.971733\\ 186    0.97181\\ 187    0.9719009\\ 188    0.97200537\\ 189    0.97212344\\ 190    0.97225493\\ 191    0.9723996\\ 192    0.9725574\\ 193    0.97272813\\ 194    0.97291166\\ 195    0.9731076\\ 196    0.9733159\\ 197    0.97353613\\ 198    0.9737683\\ 199    0.97401184\\ 200    0.9742667\\ 201    0.9745324\\ 202    0.9748089\\ 203    0.97509557\\ 204    0.9753923\\ 205    0.9756986\\ 206    0.97601414\\ 207    0.9763386\\ 208    0.9766716\\ 209    0.97701275\\ 210    0.9773616\\ 211    0.9777178\\ 212    0.97808087\\ 213    0.9784504\\ 214    0.978826\\ 215    0.97920704\\ 216    0.9795933\\ 217    0.97998416\\ 218    0.98037934\\ 219    0.9807781\\ 220    0.98118013\\ 221    0.9815851\\ 222    0.9819922\\ 223    0.982401\\ 224    0.98281115\\ 225    0.9832221\\ 226    0.9836333\\ 227    0.9840444\\ 228    0.98445463\\ 229    0.9848637\\ 230    0.9852711\\ 231    0.9856762\\ 232    0.98607856\\ 233    0.98647773\\ 234    0.9868733\\ 235    0.9872645\\ 236    0.9876511\\ 237    0.98803264\\ 238    0.9884083\\ 239    0.98877794\\ 240    0.98914105\\ 241    0.9894972\\ 242    0.9898457\\ 243    0.9901864\\ 244    0.99051875\\ 245    0.9908423\\ 246    0.99115664\\ 247    0.99146146\\ 248    0.9917563\\ 249    0.9920408\\ 250    0.9923146\\ 251    0.99257743\\ 252    0.9928288\\ 253    0.9930685\\ 254    0.9932961\\ 255    0.99351144\\ 256    0.9937143\\ 257    0.9939043\\ 258    0.99408114\\ 259    0.99424464\\ 260    0.9943948\\ 261    0.9945311\\ 262    0.9946535\\ 263    0.99476194\\ 264    0.9948561\\ 265    0.9949361\\ 266    0.9950016\\ 267    0.9950525\\ 268    0.99508905\\ 269    0.9951109\\ 270    0.9951181\\ 271    0.9951107\\ 272    0.9950885\\ 273    0.99505174\\ 274    0.9950006\\ 275    0.9949348\\ 276    0.9948547\\ 277    0.9947602\\ 278    0.9946516\\ 279    0.9945289\\ 280    0.99439234\\ 281    0.99424195\\ 282    0.9940782\\ 283    0.99390113\\ 284    0.99371094\\ 285    0.9935079\\ 286    0.9932923\\ 287    0.9930644\\ 288    0.99282455\\ 289    0.992573\\ 290    0.99231\\ 291    0.9920361\\ 292    0.9917513\\ 293    0.9914564\\ 294    0.99115133\\ 295    0.9908368\\ 296    0.9905131\\ 297    0.9901806\\ 298    0.9898398\\ 299    0.9894911\\ 300    0.9891349\\ 301    0.98877174\\ 302    0.9884019\\ 303    0.988026\\ 304    0.98764443\\ 305    0.9872578\\ 306    0.9868665\\ 307    0.98647094\\ 308    0.98607165\\ 309    0.9856692\\ 310    0.98526406\\ 311    0.9848566\\ 312    0.98444754\\ 313    0.9840372\\ 314    0.98362625\\ 315    0.983215\\ 316    0.98280406\\ 317    0.9823939\\ 318    0.98198503\\ 319    0.98157805\\ 320    0.98117316\\ 321    0.9807711\\ 322    0.9803724\\ 323    0.9799773\\ 324    0.9795865\\ 325    0.97920036\\ 326    0.9788193\\ 327    0.97844386\\ 328    0.97807443\\ 329    0.9777115\\ 330    0.9773555\\ 331    0.97700673\\ 332    0.97666574\\ 333    0.97633284\\ 334    0.97600853\\ 335    0.97569317\\ 336    0.975387\\ 337    0.97509044\\ 338    0.9748039\\ 339    0.97452766\\ 340    0.9742621\\ 341    0.9740074\\ 342    0.97376406\\ 343    0.97353214\\ 344    0.97331214\\ 345    0.973104\\ 346    0.9729083\\ 347    0.972725\\ 348    0.9725545\\ 349    0.97239697\\ 350    0.9722524\\ 351    0.97212124\\ 352    0.9720034\\ 353    0.9718991\\ 354    0.97180855\\ 355    0.9717317\\ 356    0.9716688\\ 357    0.97161984\\ 358    0.9715848\\ 359    0.9715638\\ 
	};
 \addplot [color=green, line width=1.0pt] table[row sep=crcr]{%
0    0.94724613\\ 1    0.9472735\\ 2    0.94735456\\ 3    0.94748914\\ 4    0.9476772\\ 5    0.94791853\\ 6    0.9482131\\ 7    0.9485603\\ 8    0.94895995\\ 9    0.9494118\\ 10    0.94991535\\ 11    0.9504699\\ 12    0.95107526\\ 13    0.95173055\\ 14    0.9524354\\ 15    0.9531891\\ 16    0.9539907\\ 17    0.9548395\\ 18    0.9557349\\ 19    0.9566757\\ 20    0.9576612\\ 21    0.9586903\\ 22    0.9597621\\ 23    0.96087533\\ 24    0.9620291\\ 25    0.9632221\\ 26    0.9644531\\ 27    0.9657208\\ 28    0.96702397\\ 29    0.96836126\\ 30    0.9697313\\ 31    0.97113246\\ 32    0.97256345\\ 33    0.97402257\\ 34    0.9755084\\ 35    0.9770192\\ 36    0.97855335\\ 37    0.980109\\ 38    0.9816847\\ 39    0.9832784\\ 40    0.9848885\\ 41    0.98651296\\ 42    0.98815006\\ 43    0.98979783\\ 44    0.99145424\\ 45    0.9931174\\ 46    0.9947854\\ 47    0.99645615\\ 48    0.99812764\\ 49    0.99979764\\ 50    1.0014642\\ 51    1.0031254\\ 52    1.0047791\\ 53    1.0064229\\ 54    1.008055\\ 55    1.0096731\\ 56    1.0112752\\ 57    1.0128592\\ 58    1.0144229\\ 59    1.0159643\\ 60    1.0174812\\ 61    1.0189718\\ 62    1.0204338\\ 63    1.0218654\\ 64    1.0232644\\ 65    1.024629\\ 66    1.0259571\\ 67    1.0272471\\ 68    1.028497\\ 69    1.029705\\ 70    1.0308694\\ 71    1.0319885\\ 72    1.0330606\\ 73    1.0340843\\ 74    1.0350579\\ 75    1.03598\\ 76    1.0368493\\ 77    1.0376645\\ 78    1.0384244\\ 79    1.0391277\\ 80    1.0397735\\ 81    1.0403607\\ 82    1.0408884\\ 83    1.0413561\\ 84    1.0417627\\ 85    1.0421078\\ 86    1.0423907\\ 87    1.0426111\\ 88    1.0427686\\ 89    1.042863\\ 90    1.0428941\\ 91    1.042862\\ 92    1.0427666\\ 93    1.042608\\ 94    1.0423867\\ 95    1.0421027\\ 96    1.0417566\\ 97    1.0413492\\ 98    1.0408804\\ 99    1.0403517\\ 100    1.0397636\\ 101    1.0391169\\ 102    1.0384126\\ 103    1.0376519\\ 104    1.0368358\\ 105    1.0359656\\ 106    1.0350425\\ 107    1.0340681\\ 108    1.0330436\\ 109    1.0319707\\ 110    1.0308509\\ 111    1.0296857\\ 112    1.0284771\\ 113    1.0272264\\ 114    1.0259358\\ 115    1.0246071\\ 116    1.0232419\\ 117    1.0218422\\ 118    1.0204102\\ 119    1.0189476\\ 120    1.0174567\\ 121    1.015939\\ 122    1.0143973\\ 123    1.0128331\\ 124    1.0112488\\ 125    1.0096464\\ 126    1.0080279\\ 127    1.0063957\\ 128    1.0047516\\ 129    1.0030979\\ 130    1.0014365\\ 131    0.9997697\\ 132    0.9980996\\ 133    0.996428\\ 134    0.99475735\\ 135    0.9930894\\ 136    0.9914262\\ 137    0.9897699\\ 138    0.9881222\\ 139    0.9864853\\ 140    0.984861\\ 141    0.9832511\\ 142    0.9816576\\ 143    0.9800823\\ 144    0.97852683\\ 145    0.976993\\ 146    0.9754827\\ 147    0.97399724\\ 148    0.9725386\\ 149    0.971108\\ 150    0.96970737\\ 151    0.9683379\\ 152    0.96700114\\ 153    0.96569854\\ 154    0.9644314\\ 155    0.963201\\ 156    0.96200866\\ 157    0.96085566\\ 158    0.9597431\\ 159    0.958672\\ 160    0.95764357\\ 161    0.95665896\\ 162    0.9557189\\ 163    0.95482427\\ 164    0.95397615\\ 165    0.9531754\\ 166    0.9524227\\ 167    0.9517187\\ 168    0.9510642\\ 169    0.95045984\\ 170    0.94990605\\ 171    0.94940346\\ 172    0.94895256\\ 173    0.9485537\\ 174    0.9482075\\ 175    0.9479138\\ 176    0.94767344\\ 177    0.94748634\\ 178    0.9473527\\ 179    0.9472726\\ 180    0.9472462\\ 181    0.9472735\\ 182    0.9473546\\ 183    0.94748914\\ 184    0.94767725\\ 185    0.9479186\\ 186    0.94821316\\ 187    0.9485603\\ 188    0.94896007\\ 189    0.94941187\\ 190    0.9499154\\ 191    0.95047003\\ 192    0.9510753\\ 193    0.95173067\\ 194    0.9524355\\ 195    0.9531891\\ 196    0.9539907\\ 197    0.9548396\\ 198    0.95573497\\ 199    0.9566758\\ 200    0.9576612\\ 201    0.9586904\\ 202    0.95976216\\ 203    0.96087545\\ 204    0.96202916\\ 205    0.9632221\\ 206    0.96445316\\ 207    0.96572083\\ 208    0.967024\\ 209    0.9683614\\ 210    0.96973133\\ 211    0.9711325\\ 212    0.9725635\\ 213    0.9740226\\ 214    0.97550845\\ 215    0.9770192\\ 216    0.97855335\\ 217    0.98010904\\ 218    0.9816847\\ 219    0.9832784\\ 220    0.98488855\\ 221    0.98651296\\ 222    0.98815006\\ 223    0.98979783\\ 224    0.99145424\\ 225    0.9931175\\ 226    0.9947855\\ 227    0.99645615\\ 228    0.99812764\\ 229    0.9997977\\ 230    1.0014644\\ 231    1.0031255\\ 232    1.0047791\\ 233    1.006423\\ 234    1.008055\\ 235    1.0096731\\ 236    1.0112752\\ 237    1.0128592\\ 238    1.0144229\\ 239    1.0159643\\ 240    1.0174813\\ 241    1.0189719\\ 242    1.0204339\\ 243    1.0218655\\ 244    1.0232644\\ 245    1.024629\\ 246    1.0259571\\ 247    1.0272472\\ 248    1.028497\\ 249    1.029705\\ 250    1.0308694\\ 251    1.0319886\\ 252    1.0330607\\ 253    1.0340843\\ 254    1.0350579\\ 255    1.03598\\ 256    1.0368494\\ 257    1.0376645\\ 258    1.0384244\\ 259    1.0391277\\ 260    1.0397735\\ 261    1.0403607\\ 262    1.0408884\\ 263    1.0413561\\ 264    1.0417627\\ 265    1.0421078\\ 266    1.0423907\\ 267    1.0426111\\ 268    1.0427686\\ 269    1.042863\\ 270    1.0428941\\ 271    1.042862\\ 272    1.0427666\\ 273    1.042608\\ 274    1.0423867\\ 275    1.0421027\\ 276    1.0417566\\ 277    1.0413492\\ 278    1.0408804\\ 279    1.0403517\\ 280    1.0397636\\ 281    1.0391169\\ 282    1.0384126\\ 283    1.0376519\\ 284    1.0368358\\ 285    1.0359656\\ 286    1.0350425\\ 287    1.0340681\\ 288    1.0330436\\ 289    1.0319707\\ 290    1.0308509\\ 291    1.0296857\\ 292    1.0284771\\ 293    1.0272264\\ 294    1.0259358\\ 295    1.024607\\ 296    1.0232418\\ 297    1.0218422\\ 298    1.0204101\\ 299    1.0189476\\ 300    1.0174565\\ 301    1.015939\\ 302    1.0143973\\ 303    1.0128331\\ 304    1.0112488\\ 305    1.0096464\\ 306    1.0080279\\ 307    1.0063957\\ 308    1.0047516\\ 309    1.0030979\\ 310    1.0014365\\ 311    0.9997696\\ 312    0.9980995\\ 313    0.996428\\ 314    0.99475724\\ 315    0.9930893\\ 316    0.9914261\\ 317    0.98976976\\ 318    0.9881222\\ 319    0.98648524\\ 320    0.9848609\\ 321    0.9832511\\ 322    0.98165756\\ 323    0.9800822\\ 324    0.97852683\\ 325    0.97699296\\ 326    0.97548264\\ 327    0.97399724\\ 328    0.9725386\\ 329    0.971108\\ 330    0.9697073\\ 331    0.96833783\\ 332    0.9670011\\ 333    0.9656985\\ 334    0.9644313\\ 335    0.9632009\\ 336    0.9620086\\ 337    0.96085554\\ 338    0.95974296\\ 339    0.95867187\\ 340    0.95764357\\ 341    0.95665884\\ 342    0.95571876\\ 343    0.95482415\\ 344    0.9539761\\ 345    0.95317537\\ 346    0.95242256\\ 347    0.9517186\\ 348    0.9510641\\ 349    0.9504597\\ 350    0.949906\\ 351    0.94940335\\ 352    0.94895244\\ 353    0.9485537\\ 354    0.9482074\\ 355    0.9479138\\ 356    0.94767344\\ 357    0.9474862\\ 358    0.9473526\\ 359    0.9472725\\ 
	};
 \addplot [color=blue, line width=1.0pt] table[row sep=crcr]{%
0    0.90999997\\ 1    0.91005695\\ 2    0.91022587\\ 3    0.9105065\\ 4    0.9108988\\ 5    0.9114026\\ 6    0.91201735\\ 7    0.91274273\\ 8    0.9135784\\ 9    0.9145238\\ 10    0.91557825\\ 11    0.9167411\\ 12    0.9180116\\ 13    0.91938895\\ 14    0.9208722\\ 15    0.92246044\\ 16    0.92415243\\ 17    0.92594725\\ 18    0.9278435\\ 19    0.92984\\ 20    0.93193525\\ 21    0.9341278\\ 22    0.9364159\\ 23    0.9387981\\ 24    0.94127244\\ 25    0.9438372\\ 26    0.94649035\\ 27    0.9492299\\ 28    0.9520534\\ 29    0.954959\\ 30    0.957944\\ 31    0.9610058\\ 32    0.9641421\\ 33    0.96735\\ 34    0.97062665\\ 35    0.9739692\\ 36    0.97737443\\ 37    0.98083925\\ 38    0.9843604\\ 39    0.98793435\\ 40    0.9915574\\ 41    0.99522614\\ 42    0.9989366\\ 43    1.0026848\\ 44    1.006467\\ 45    1.0102787\\ 46    1.0141156\\ 47    1.0179734\\ 48    1.0218477\\ 49    1.0257337\\ 50    1.0296266\\ 51    1.0335217\\ 52    1.0374138\\ 53    1.0412982\\ 54    1.0451695\\ 55    1.0490224\\ 56    1.052852\\ 57    1.0566525\\ 58    1.0604186\\ 59    1.0641448\\ 60    1.0678258\\ 61    1.0714558\\ 62    1.0750291\\ 63    1.0785404\\ 64    1.081984\\ 65    1.0853543\\ 66    1.0886457\\ 67    1.0918529\\ 68    1.0949702\\ 69    1.0979925\\ 70    1.1009145\\ 71    1.1037308\\ 72    1.1064365\\ 73    1.1090268\\ 74    1.1114968\\ 75    1.1138419\\ 76    1.1160578\\ 77    1.1181402\\ 78    1.1200854\\ 79    1.1218891\\ 80    1.1235484\\ 81    1.1250595\\ 82    1.1264197\\ 83    1.1276263\\ 84    1.1286769\\ 85    1.1295692\\ 86    1.1303016\\ 87    1.1308725\\ 88    1.1312807\\ 89    1.1315256\\ 90    1.1316065\\ 91    1.1315233\\ 92    1.1312761\\ 93    1.1308655\\ 94    1.1302922\\ 95    1.1295575\\ 96    1.128663\\ 97    1.1276101\\ 98    1.1264013\\ 99    1.1250389\\ 100    1.1235254\\ 101    1.1218641\\ 102    1.1200582\\ 103    1.118111\\ 104    1.1160265\\ 105    1.1138086\\ 106    1.1114616\\ 107    1.1089897\\ 108    1.1063976\\ 109    1.10369\\ 110    1.100872\\ 111    1.0979483\\ 112    1.0949246\\ 113    1.0918057\\ 114    1.088597\\ 115    1.0853043\\ 116    1.0819327\\ 117    1.0784878\\ 118    1.0749754\\ 119    1.0714009\\ 120    1.0677699\\ 121    1.064088\\ 122    1.0603608\\ 123    1.0565939\\ 124    1.0527927\\ 125    1.0489625\\ 126    1.0451089\\ 127    1.0412371\\ 128    1.0373523\\ 129    1.0334598\\ 130    1.0295645\\ 131    1.0256714\\ 132    1.0217853\\ 133    1.017911\\ 134    1.0140532\\ 135    1.0102164\\ 136    1.0064049\\ 137    1.002623\\ 138    0.99887514\\ 139    0.9951651\\ 140    0.99149686\\ 141    0.98787427\\ 142    0.98430103\\ 143    0.9807806\\ 144    0.97731656\\ 145    0.97391206\\ 146    0.97057045\\ 147    0.9672947\\ 148    0.9640878\\ 149    0.9609526\\ 150    0.9578918\\ 151    0.95490813\\ 152    0.95200384\\ 153    0.9491816\\ 154    0.9464434\\ 155    0.9437916\\ 156    0.94122833\\ 157    0.9387554\\ 158    0.9363748\\ 159    0.9340882\\ 160    0.9318972\\ 161    0.9298036\\ 162    0.92780894\\ 163    0.92591435\\ 164    0.9241214\\ 165    0.9224311\\ 166    0.92084473\\ 167    0.9193633\\ 168    0.91798776\\ 169    0.9167192\\ 170    0.9155583\\ 171    0.9145058\\ 172    0.9135624\\ 173    0.91272867\\ 174    0.9120053\\ 175    0.91139245\\ 176    0.9108908\\ 177    0.9105005\\ 178    0.9102219\\ 179    0.9100549\\ 180    0.91\\ 181    0.910057\\ 182    0.9102259\\ 183    0.9105066\\ 184    0.9108989\\ 185    0.91140264\\ 186    0.9120174\\ 187    0.9127428\\ 188    0.9135785\\ 189    0.9145239\\ 190    0.9155783\\ 191    0.91674113\\ 192    0.91801167\\ 193    0.919389\\ 194    0.9208723\\ 195    0.92246056\\ 196    0.92415255\\ 197    0.92594737\\ 198    0.9278436\\ 199    0.9298401\\ 200    0.93193525\\ 201    0.9341278\\ 202    0.93641603\\ 203    0.9387982\\ 204    0.94127256\\ 205    0.94383734\\ 206    0.9464904\\ 207    0.94923\\ 208    0.95205355\\ 209    0.95495903\\ 210    0.95794404\\ 211    0.96100587\\ 212    0.96414214\\ 213    0.9673501\\ 214    0.9706268\\ 215    0.9739693\\ 216    0.97737455\\ 217    0.9808393\\ 218    0.98436046\\ 219    0.98793435\\ 220    0.9915575\\ 221    0.99522626\\ 222    0.9989367\\ 223    1.002685\\ 224    1.006467\\ 225    1.0102787\\ 226    1.0141157\\ 227    1.0179734\\ 228    1.0218477\\ 229    1.0257337\\ 230    1.0296266\\ 231    1.0335217\\ 232    1.0374138\\ 233    1.0412982\\ 234    1.0451695\\ 235    1.0490226\\ 236    1.052852\\ 237    1.0566527\\ 238    1.0604187\\ 239    1.064145\\ 240    1.0678259\\ 241    1.0714558\\ 242    1.0750293\\ 243    1.0785404\\ 244    1.081984\\ 245    1.0853543\\ 246    1.0886457\\ 247    1.0918529\\ 248    1.0949702\\ 249    1.0979925\\ 250    1.1009145\\ 251    1.1037308\\ 252    1.1064365\\ 253    1.1090269\\ 254    1.1114968\\ 255    1.113842\\ 256    1.1160579\\ 257    1.1181402\\ 258    1.1200854\\ 259    1.1218892\\ 260    1.1235484\\ 261    1.1250596\\ 262    1.1264197\\ 263    1.1276263\\ 264    1.1286769\\ 265    1.1295692\\ 266    1.1303016\\ 267    1.1308725\\ 268    1.1312808\\ 269    1.1315256\\ 270    1.1316065\\ 271    1.1315233\\ 272    1.1312761\\ 273    1.1308655\\ 274    1.1302922\\ 275    1.1295575\\ 276    1.128663\\ 277    1.12761\\ 278    1.1264012\\ 279    1.1250387\\ 280    1.1235254\\ 281    1.1218641\\ 282    1.1200582\\ 283    1.118111\\ 284    1.1160265\\ 285    1.1138086\\ 286    1.1114615\\ 287    1.1089897\\ 288    1.1063975\\ 289    1.10369\\ 290    1.100872\\ 291    1.0979483\\ 292    1.0949246\\ 293    1.0918057\\ 294    1.0885969\\ 295    1.0853041\\ 296    1.0819325\\ 297    1.0784878\\ 298    1.0749754\\ 299    1.0714008\\ 300    1.0677699\\ 301    1.0640879\\ 302    1.0603608\\ 303    1.0565939\\ 304    1.0527927\\ 305    1.0489625\\ 306    1.0451088\\ 307    1.041237\\ 308    1.0373523\\ 309    1.0334598\\ 310    1.0295645\\ 311    1.0256714\\ 312    1.0217853\\ 313    1.017911\\ 314    1.0140532\\ 315    1.0102164\\ 316    1.0064048\\ 317    1.002623\\ 318    0.9988751\\ 319    0.99516505\\ 320    0.9914968\\ 321    0.9878742\\ 322    0.9843009\\ 323    0.98078054\\ 324    0.9773165\\ 325    0.973912\\ 326    0.9705703\\ 327    0.96729463\\ 328    0.9640877\\ 329    0.9609525\\ 330    0.9578918\\ 331    0.954908\\ 332    0.9520037\\ 333    0.9491815\\ 334    0.9464433\\ 335    0.94379157\\ 336    0.9412282\\ 337    0.93875533\\ 338    0.9363747\\ 339    0.9340881\\ 340    0.93189716\\ 341    0.9298036\\ 342    0.9278089\\ 343    0.9259143\\ 344    0.92412126\\ 345    0.922431\\ 346    0.9208446\\ 347    0.91936326\\ 348    0.9179877\\ 349    0.9167191\\ 350    0.9155582\\ 351    0.9145057\\ 352    0.91356236\\ 353    0.9127286\\ 354    0.9120052\\ 355    0.9113924\\ 356    0.9108907\\ 357    0.9105004\\ 358    0.91022176\\ 359    0.91005486\\ 
	};
 \addplot [color=red, line width=1.0pt] table[row sep=crcr]{%
0    0.8643107\\ 1    0.864402\\ 2    0.86467254\\ 3    0.86512226\\ 4    0.8657511\\ 5    0.8665588\\ 6    0.8675452\\ 7    0.86871016\\ 8    0.87005323\\ 9    0.87157416\\ 10    0.87327266\\ 11    0.87514806\\ 12    0.87720007\\ 13    0.8794281\\ 14    0.8818315\\ 15    0.8844096\\ 16    0.88716155\\ 17    0.89008653\\ 18    0.8931839\\ 19    0.89645237\\ 20    0.899891\\ 21    0.9034986\\ 22    0.90727395\\ 23    0.91121554\\ 24    0.915322\\ 25    0.9195916\\ 26    0.9240228\\ 27    0.9286134\\ 28    0.9333617\\ 29    0.9382654\\ 30    0.9433221\\ 31    0.9485293\\ 32    0.9538844\\ 33    0.9593845\\ 34    0.9650263\\ 35    0.9708068\\ 36    0.9767222\\ 37    0.98276883\\ 38    0.9889427\\ 39    0.99523956\\ 40    1.0016547\\ 41    1.0081836\\ 42    1.0148207\\ 43    1.0215609\\ 44    1.0283982\\ 45    1.0353266\\ 46    1.0423398\\ 47    1.0494306\\ 48    1.0565922\\ 49    1.063817\\ 50    1.0710971\\ 51    1.0784241\\ 52    1.0857894\\ 53    1.0931839\\ 54    1.1005979\\ 55    1.1080217\\ 56    1.1154454\\ 57    1.1228576\\ 58    1.1302477\\ 59    1.1376041\\ 60    1.1449152\\ 61    1.1521686\\ 62    1.1593522\\ 63    1.1664529\\ 64    1.1734577\\ 65    1.1803536\\ 66    1.187127\\ 67    1.1937641\\ 68    1.2002512\\ 69    1.2065743\\ 70    1.2127199\\ 71    1.218674\\ 72    1.2244227\\ 73    1.2299523\\ 74    1.2352496\\ 75    1.2403016\\ 76    1.2450953\\ 77    1.2496179\\ 78    1.2538583\\ 79    1.2578046\\ 80    1.2614462\\ 81    1.2647731\\ 82    1.2677759\\ 83    1.270446\\ 84    1.2727758\\ 85    1.2747585\\ 86    1.2763883\\ 87    1.2776606\\ 88    1.2785714\\ 89    1.2791182\\ 90    1.2792991\\ 91    1.2791137\\ 92    1.2785627\\ 93    1.2776474\\ 94    1.2763709\\ 95    1.2747366\\ 96    1.2727497\\ 97    1.2704157\\ 98    1.2677414\\ 99    1.2647346\\ 100    1.2614038\\ 101    1.2577581\\ 102    1.2538079\\ 103    1.2495638\\ 104    1.2450374\\ 105    1.2402402\\ 106    1.2351848\\ 107    1.2298841\\ 108    1.224351\\ 109    1.2185993\\ 110    1.2126423\\ 111    1.2064939\\ 112    1.2001679\\ 113    1.1936781\\ 114    1.1870385\\ 115    1.1802628\\ 116    1.1733648\\ 117    1.1663579\\ 118    1.1592553\\ 119    1.1520699\\ 120    1.1448148\\ 121    1.1375022\\ 122    1.1301445\\ 123    1.1227531\\ 124    1.1153399\\ 125    1.1079154\\ 126    1.1004907\\ 127    1.0930759\\ 128    1.085681\\ 129    1.0783155\\ 130    1.070988\\ 131    1.063708\\ 132    1.0564833\\ 133    1.0493218\\ 134    1.0422312\\ 135    1.0352185\\ 136    1.0282906\\ 137    1.0214541\\ 138    1.0147147\\ 139    1.0080785\\ 140    1.0015507\\ 141    0.99513656\\ 142    0.98884094\\ 143    0.98266846\\ 144    0.9766233\\ 145    0.9707094\\ 146    0.9649306\\ 147    0.9592906\\ 148    0.9537924\\ 149    0.94843924\\ 150    0.943234\\ 151    0.9381795\\ 152    0.933278\\ 153    0.928532\\ 154    0.9239437\\ 155    0.919515\\ 156    0.91524786\\ 157    0.911144\\ 158    0.907205\\ 159    0.9034325\\ 160    0.8998276\\ 161    0.8963918\\ 162    0.89312613\\ 163    0.8900318\\ 164    0.88710976\\ 165    0.88436073\\ 166    0.88178587\\ 167    0.87938553\\ 168    0.8771606\\ 169    0.8751118\\ 170    0.8732396\\ 171    0.8715443\\ 172    0.8700267\\ 173    0.86868685\\ 174    0.86752516\\ 175    0.86654204\\ 176    0.86573774\\ 177    0.8651123\\ 178    0.8646659\\ 179    0.8643988\\ 180    0.8643108\\ 181    0.8644021\\ 182    0.8646726\\ 183    0.8651223\\ 184    0.8657512\\ 185    0.86655885\\ 186    0.86754537\\ 187    0.8687102\\ 188    0.87005335\\ 189    0.8715743\\ 190    0.8732727\\ 191    0.87514824\\ 192    0.8772002\\ 193    0.8794282\\ 194    0.88183165\\ 195    0.88440967\\ 196    0.8871616\\ 197    0.8900867\\ 198    0.89318395\\ 199    0.8964525\\ 200    0.8998911\\ 201    0.90349865\\ 202    0.907274\\ 203    0.91121566\\ 204    0.91532207\\ 205    0.91959167\\ 206    0.92402285\\ 207    0.92861354\\ 208    0.9333618\\ 209    0.9382655\\ 210    0.94332224\\ 211    0.94852936\\ 212    0.95388454\\ 213    0.95938456\\ 214    0.9650264\\ 215    0.97080684\\ 216    0.97672224\\ 217    0.9827689\\ 218    0.9889428\\ 219    0.9952397\\ 220    1.0016549\\ 221    1.0081836\\ 222    1.0148208\\ 223    1.021561\\ 224    1.0283983\\ 225    1.0353267\\ 226    1.0423398\\ 227    1.0494306\\ 228    1.0565923\\ 229    1.0638171\\ 230    1.0710971\\ 231    1.0784242\\ 232    1.0857894\\ 233    1.0931839\\ 234    1.100598\\ 235    1.1080219\\ 236    1.1154455\\ 237    1.1228576\\ 238    1.1302477\\ 239    1.1376042\\ 240    1.1449152\\ 241    1.1521688\\ 242    1.1593522\\ 243    1.1664529\\ 244    1.1734579\\ 245    1.1803538\\ 246    1.187127\\ 247    1.1937641\\ 248    1.2002512\\ 249    1.2065744\\ 250    1.21272\\ 251    1.2186741\\ 252    1.2244227\\ 253    1.2299525\\ 254    1.2352498\\ 255    1.2403016\\ 256    1.2450953\\ 257    1.249618\\ 258    1.2538583\\ 259    1.2578046\\ 260    1.2614464\\ 261    1.2647731\\ 262    1.2677759\\ 263    1.270446\\ 264    1.2727758\\ 265    1.2747585\\ 266    1.2763883\\ 267    1.2776606\\ 268    1.2785714\\ 269    1.2791182\\ 270    1.2792991\\ 271    1.2791137\\ 272    1.2785627\\ 273    1.2776474\\ 274    1.2763709\\ 275    1.2747366\\ 276    1.2727497\\ 277    1.2704157\\ 278    1.2677413\\ 279    1.2647346\\ 280    1.2614037\\ 281    1.2577581\\ 282    1.2538079\\ 283    1.2495638\\ 284    1.2450374\\ 285    1.2402402\\ 286    1.2351847\\ 287    1.2298841\\ 288    1.224351\\ 289    1.2185992\\ 290    1.2126423\\ 291    1.2064939\\ 292    1.2001679\\ 293    1.193678\\ 294    1.1870385\\ 295    1.1802628\\ 296    1.1733648\\ 297    1.1663579\\ 298    1.1592551\\ 299    1.1520699\\ 300    1.1448147\\ 301    1.1375022\\ 302    1.1301445\\ 303    1.122753\\ 304    1.1153398\\ 305    1.1079153\\ 306    1.1004906\\ 307    1.0930759\\ 308    1.0856808\\ 309    1.0783154\\ 310    1.070988\\ 311    1.063708\\ 312    1.0564831\\ 313    1.0493217\\ 314    1.0422312\\ 315    1.0352184\\ 316    1.0282905\\ 317    1.021454\\ 318    1.0147146\\ 319    1.0080783\\ 320    1.0015506\\ 321    0.99513656\\ 322    0.9888408\\ 323    0.9826684\\ 324    0.9766232\\ 325    0.9707094\\ 326    0.96493053\\ 327    0.9592905\\ 328    0.95379233\\ 329    0.9484392\\ 330    0.9432339\\ 331    0.93817943\\ 332    0.93327796\\ 333    0.92853194\\ 334    0.92394364\\ 335    0.91951495\\ 336    0.9152478\\ 337    0.9111439\\ 338    0.9072049\\ 339    0.90343237\\ 340    0.89982754\\ 341    0.8963917\\ 342    0.8931261\\ 343    0.8900317\\ 344    0.88710964\\ 345    0.8843607\\ 346    0.8817857\\ 347    0.8793854\\ 348    0.8771605\\ 349    0.87511176\\ 350    0.8732395\\ 351    0.87154424\\ 352    0.8700265\\ 353    0.8686868\\ 354    0.8675251\\ 355    0.866542\\ 356    0.8657377\\ 357    0.8651122\\ 358    0.86466587\\ 359    0.86439866\\ 
	};
 \addplot [color=yellow, line width=1.0pt] table[row sep=crcr]{%
0    0.8156892\\ 1    0.81581455\\ 2    0.8161858\\ 3    0.81680316\\ 4    0.81766653\\ 5    0.8187762\\ 6    0.8201322\\ 7    0.8217349\\ 8    0.8235846\\ 9    0.82568145\\ 10    0.8280259\\ 11    0.8306182\\ 12    0.8334587\\ 13    0.83654815\\ 14    0.83988667\\ 15    0.8434748\\ 16    0.847313\\ 17    0.851402\\ 18    0.8557418\\ 19    0.8603334\\ 20    0.86517704\\ 21    0.87027323\\ 22    0.8756223\\ 23    0.88122493\\ 24    0.88708127\\ 25    0.8931918\\ 26    0.8995567\\ 27    0.9061762\\ 28    0.9130504\\ 29    0.92017937\\ 30    0.92756295\\ 31    0.9352009\\ 32    0.9430928\\ 33    0.95123816\\ 34    0.9596362\\ 35    0.9682859\\ 36    0.9771861\\ 37    0.9863352\\ 38    0.99573153\\ 39    1.005373\\ 40    1.0152568\\ 41    1.0253804\\ 42    1.0357401\\ 43    1.0463325\\ 44    1.0571529\\ 45    1.0681965\\ 46    1.0794578\\ 47    1.0909303\\ 48    1.1026074\\ 49    1.1144811\\ 50    1.1265429\\ 51    1.1387833\\ 52    1.1511922\\ 53    1.1637574\\ 54    1.1764671\\ 55    1.1893071\\ 56    1.2022629\\ 57    1.2153181\\ 58    1.2284553\\ 59    1.2416557\\ 60    1.2548991\\ 61    1.2681639\\ 62    1.2814269\\ 63    1.2946633\\ 64    1.3078475\\ 65    1.3209513\\ 66    1.3339461\\ 67    1.3468014\\ 68    1.3594853\\ 69    1.3719647\\ 70    1.3842053\\ 71    1.3961719\\ 72    1.4078287\\ 73    1.4191383\\ 74    1.430064\\ 75    1.4405677\\ 76    1.4506121\\ 77    1.4601601\\ 78    1.4691747\\ 79    1.4776205\\ 80    1.4854625\\ 81    1.4926682\\ 82    1.4992061\\ 83    1.5050474\\ 84    1.510166\\ 85    1.514538\\ 86    1.5181433\\ 87    1.5209647\\ 88    1.5229887\\ 89    1.5242056\\ 90    1.5246093\\ 91    1.524198\\ 92    1.5229738\\ 93    1.5209426\\ 94    1.5181137\\ 95    1.5145013\\ 96    1.5101221\\ 97    1.5049965\\ 98    1.4991482\\ 99    1.4926035\\ 100    1.4853911\\ 101    1.4775428\\ 102    1.4690907\\ 103    1.4600699\\ 104    1.4505161\\ 105    1.4404659\\ 106    1.4299567\\ 107    1.4190259\\ 108    1.4077113\\ 109    1.3960499\\ 110    1.3840786\\ 111    1.3718337\\ 112    1.3593503\\ 113    1.3466626\\ 114    1.3338039\\ 115    1.3208057\\ 116    1.3076988\\ 117    1.2945119\\ 118    1.2812729\\ 119    1.2680076\\ 120    1.2547407\\ 121    1.2414955\\ 122    1.2282935\\ 123    1.215155\\ 124    1.2020987\\ 125    1.189142\\ 126    1.1763014\\ 127    1.1635911\\ 128    1.1510257\\ 129    1.1386168\\ 130    1.1263766\\ 131    1.1143152\\ 132    1.1024418\\ 133    1.0907656\\ 134    1.079294\\ 135    1.0680338\\ 136    1.0569915\\ 137    1.0461725\\ 138    1.0355817\\ 139    1.0252236\\ 140    1.015102\\ 141    1.0052202\\ 142    0.9955811\\ 143    0.98618704\\ 144    0.9770404\\ 145    0.96814275\\ 146    0.95949584\\ 147    0.9511007\\ 148    0.94295835\\ 149    0.9350695\\ 150    0.9274347\\ 151    0.92005444\\ 152    0.91292894\\ 153    0.9060582\\ 154    0.8994423\\ 155    0.893081\\ 156    0.8869744\\ 157    0.8811218\\ 158    0.87552315\\ 159    0.870178\\ 160    0.86508596\\ 161    0.8602464\\ 162    0.8556591\\ 163    0.85132354\\ 164    0.8472389\\ 165    0.84340507\\ 166    0.83982134\\ 167    0.8364873\\ 168    0.83340245\\ 169    0.8305664\\ 170    0.82797873\\ 171    0.8256389\\ 172    0.82354677\\ 173    0.8217017\\ 174    0.82010376\\ 175    0.8187524\\ 176    0.8176476\\ 177    0.816789\\ 178    0.81617635\\ 179    0.8158099\\ 180    0.8156893\\ 181    0.81581473\\ 182    0.81618595\\ 183    0.8168032\\ 184    0.8176667\\ 185    0.81877625\\ 186    0.8201323\\ 187    0.82173496\\ 188    0.8235847\\ 189    0.8256815\\ 190    0.82802594\\ 191    0.83061826\\ 192    0.8334589\\ 193    0.8365482\\ 194    0.8398867\\ 195    0.84347486\\ 196    0.84731305\\ 197    0.85140204\\ 198    0.8557419\\ 199    0.8603335\\ 200    0.8651771\\ 201    0.8702733\\ 202    0.87562233\\ 203    0.881225\\ 204    0.8870813\\ 205    0.8931919\\ 206    0.89955676\\ 207    0.90617627\\ 208    0.9130505\\ 209    0.9201795\\ 210    0.9275631\\ 211    0.935201\\ 212    0.94309294\\ 213    0.9512383\\ 214    0.95963633\\ 215    0.968286\\ 216    0.9771861\\ 217    0.98633534\\ 218    0.9957316\\ 219    1.005373\\ 220    1.0152569\\ 221    1.0253804\\ 222    1.0357403\\ 223    1.0463326\\ 224    1.057153\\ 225    1.0681965\\ 226    1.0794578\\ 227    1.0909305\\ 228    1.1026075\\ 229    1.1144812\\ 230    1.126543\\ 231    1.1387835\\ 232    1.1511922\\ 233    1.1637576\\ 234    1.1764671\\ 235    1.1893072\\ 236    1.202263\\ 237    1.2153181\\ 238    1.2284553\\ 239    1.2416557\\ 240    1.2548993\\ 241    1.2681639\\ 242    1.2814269\\ 243    1.2946634\\ 244    1.3078475\\ 245    1.3209515\\ 246    1.3339461\\ 247    1.3468015\\ 248    1.3594853\\ 249    1.3719648\\ 250    1.3842053\\ 251    1.396172\\ 252    1.4078287\\ 253    1.4191383\\ 254    1.4300641\\ 255    1.4405677\\ 256    1.4506122\\ 257    1.4601601\\ 258    1.4691749\\ 259    1.4776206\\ 260    1.4854625\\ 261    1.4926682\\ 262    1.4992061\\ 263    1.5050474\\ 264    1.510166\\ 265    1.514538\\ 266    1.5181433\\ 267    1.5209647\\ 268    1.5229887\\ 269    1.5242056\\ 270    1.5246093\\ 271    1.524198\\ 272    1.5229738\\ 273    1.5209426\\ 274    1.5181137\\ 275    1.5145013\\ 276    1.5101221\\ 277    1.5049965\\ 278    1.4991482\\ 279    1.4926035\\ 280    1.4853911\\ 281    1.4775428\\ 282    1.4690907\\ 283    1.4600699\\ 284    1.450516\\ 285    1.4404659\\ 286    1.4299567\\ 287    1.4190259\\ 288    1.4077113\\ 289    1.3960497\\ 290    1.3840786\\ 291    1.3718337\\ 292    1.3593502\\ 293    1.3466625\\ 294    1.3338038\\ 295    1.3208057\\ 296    1.3076987\\ 297    1.2945118\\ 298    1.2812728\\ 299    1.2680076\\ 300    1.2547407\\ 301    1.2414955\\ 302    1.2282934\\ 303    1.2151549\\ 304    1.2020986\\ 305    1.189142\\ 306    1.1763012\\ 307    1.1635911\\ 308    1.1510255\\ 309    1.1386168\\ 310    1.1263765\\ 311    1.114315\\ 312    1.1024418\\ 313    1.0907655\\ 314    1.0792938\\ 315    1.0680337\\ 316    1.0569913\\ 317    1.0461724\\ 318    1.0355816\\ 319    1.0252235\\ 320    1.0151019\\ 321    1.00522\\ 322    0.995581\\ 323    0.9861869\\ 324    0.9770403\\ 325    0.9681427\\ 326    0.9594957\\ 327    0.9511006\\ 328    0.94295824\\ 329    0.93506944\\ 330    0.9274346\\ 331    0.9200544\\ 332    0.9129289\\ 333    0.9060581\\ 334    0.89944214\\ 335    0.89308095\\ 336    0.8869742\\ 337    0.88112175\\ 338    0.87552303\\ 339    0.8701779\\ 340    0.8650858\\ 341    0.86024636\\ 342    0.855659\\ 343    0.8513234\\ 344    0.8472388\\ 345    0.84340495\\ 346    0.8398213\\ 347    0.8364871\\ 348    0.83340234\\ 349    0.83056635\\ 350    0.8279786\\ 351    0.82563883\\ 352    0.82354665\\ 353    0.82170165\\ 354    0.8201037\\ 355    0.81875235\\ 356    0.81764746\\ 357    0.81678885\\ 358    0.8161763\\ 359    0.8158098\\ 
	};
 \end{axis} 
 \end{tikzpicture}

%% file: xz_copy.tikz
\begin{tikzpicture}
 	\begin{axis}[%
width=\textwidth, 
	height=\axisdefaultheight,
	 scale = 0.4,
xmin=0,
	xmax=99,
	xlabel={Rotation angle in $^\circ$},
	xlabel near ticks,
	xticklabels={0, 0,  36, 72, 108, 144, 180}, 
ymin=0,
	ymax=2,
	ylabel={Dark-field $D_L$},
ylabel near ticks,
	xmajorgrids,
	ymajorgrids,
legend pos=north east
	]
\addplot [color=black, line width=1.0pt] table[row sep=crcr]{%
0    0.72330314\\ 1    0.7305319\\ 2    0.6659607\\ 3    0.72794574\\ 4    0.7538664\\ 5    0.7395821\\ 6    0.6935322\\ 7    0.7509998\\ 8    0.70501405\\ 9    0.688788\\ 10    0.6821536\\ 11    0.6833153\\ 12    0.6673848\\ 13    0.73841417\\ 14    0.727765\\ 15    0.6962762\\ 16    0.704709\\ 17    0.693997\\ 18    0.664124\\ 19    0.74590147\\ 20    0.7181008\\ 21    0.7160953\\ 22    0.69266754\\ 23    0.67096126\\ 24    0.7288588\\ 25    0.691881\\ 26    0.70126504\\ 27    0.6640048\\ 28    0.684319\\ 29    0.7278742\\ 30    0.71490306\\ 31    0.7046164\\ 32    0.7335028\\ 33    0.6940148\\ 34    0.69369316\\ 35    0.76766\\ 36    0.7057871\\ 37    0.72423714\\ 38    0.7169922\\ 39    0.7147262\\ 40    0.7274877\\ 41    0.70268595\\ 42    0.7154975\\ 43    0.69027865\\ 44    0.68694484\\ 45    0.71779966\\ 46    0.67658347\\ 47    0.69539976\\ 48    0.69123197\\ 49    0.69830203\\ 50    0.681746\\ 51    0.7136055\\ 52    0.6616681\\ 53    0.6906756\\ 54    0.7308068\\ 55    0.6674759\\ 56    0.73865896\\ 57    0.6793306\\ 58    0.71523494\\ 59    0.71325505\\ 60    0.69292253\\ 61    0.6690807\\ 62    0.69845283\\ 63    0.75269365\\ 64    0.72931296\\ 65    0.6799059\\ 66    0.70978224\\ 67    0.7336942\\ 68    0.66050196\\ 69    0.718626\\ 70    0.636082\\ 71    0.7051227\\ 72    0.69499046\\ 73    0.6744322\\ 74    0.6372232\\ 75    0.67085135\\ 76    0.6996556\\ 77    0.67978805\\ 78    0.68507975\\ 79    0.6680822\\ 80    0.62976855\\ 81    0.6865103\\ 82    0.6529136\\ 83    0.6815526\\ 84    0.66778094\\ 85    0.67133623\\ 86    0.6931229\\ 87    0.6927394\\ 88    0.6509319\\ 89    0.66524166\\ 90    0.6696676\\ 91    0.6327717\\ 92    0.69236284\\ 93    0.6853103\\ 94    0.68118656\\ 95    0.65337545\\ 96    0.6964323\\ 97    0.724198\\ 98    0.74538094\\ 99    0.70103985\\ 100    0.70831186\\ 101    0.681899\\ 102    0.7508723\\ 103    0.75028133\\ 104    0.76328796\\ 105    0.6640267\\ 106    0.68084127\\ 107    0.7302426\\ 108    0.7092217\\ 109    0.6966586\\ 110    0.7428743\\ 111    0.69961995\\ 112    0.70457584\\ 113    0.7268293\\ 114    0.78794056\\ 115    0.7353402\\ 116    0.7307402\\ 117    0.7404803\\ 118    0.7129595\\ 119    0.7626951\\ 120    0.6948541\\ 121    0.69627875\\ 122    0.7561597\\ 123    0.7081702\\ 124    0.74908817\\ 125    0.73770267\\ 126    0.71361744\\ 127    0.6667645\\ 128    0.727188\\ 129    0.7576088\\ 130    0.76499355\\ 131    0.68628\\ 132    0.71722287\\ 133    0.74987465\\ 134    0.7540226\\ 135    0.72784704\\ 136    0.69726527\\ 137    0.71740174\\ 138    0.69992113\\ 139    0.70262367\\ 140    0.7209621\\ 141    0.7525192\\ 142    0.693105\\ 143    0.6995712\\ 144    0.6691307\\ 145    0.67548287\\ 146    0.701962\\ 147    0.70302474\\ 148    0.69636464\\ 149    0.7091764\\ 150    0.6868106\\ 151    0.6865683\\ 152    0.65224713\\ 153    0.7058845\\ 154    0.64802\\ 155    0.72227895\\ 156    0.66801935\\ 157    0.6388175\\ 158    0.6651824\\ 159    0.6699464\\ 160    0.67339367\\ 161    0.68043417\\ 162    0.66154057\\ 163    0.70099074\\ 164    0.7290698\\ 165    0.6918891\\ 166    0.67515963\\ 167    0.6816575\\ 168    0.70325094\\ 169    0.69077754\\ 170    0.67198336\\ 171    0.7107351\\ 172    0.70185244\\ 173    0.69519925\\ 174    0.6887031\\ 175    0.6823538\\ 176    0.68339366\\ 177    0.70507175\\ 178    0.7052244\\ 179    0.68266946\\ 180    0.714765\\ 181    0.673153\\ 182    0.66779613\\ 183    0.6854264\\ 184    0.73898965\\ 185    0.68554467\\ 186    0.66435915\\ 187    0.69942796\\ 188    0.6306424\\ 189    0.6465829\\ 190    0.6461945\\ 191    0.7036978\\ 192    0.6859344\\ 193    0.6823098\\ 194    0.7394487\\ 195    0.6844023\\ 196    0.7077563\\ 197    0.69523793\\ 198    0.6709694\\ 199    0.71890223\\ 
	};
\addplot [color=blue, line width=1.0pt] table[row sep=crcr]{%
0    0.7252714\\ 1    0.7476978\\ 2    0.70092595\\ 3    0.7423708\\ 4    0.75682145\\ 5    0.77363396\\ 6    0.75241774\\ 7    0.72057873\\ 8    0.76312125\\ 9    0.7820665\\ 10    0.796559\\ 11    0.7837557\\ 12    0.77270484\\ 13    0.7932514\\ 14    0.7601859\\ 15    0.7869595\\ 16    0.75844973\\ 17    0.78237295\\ 18    0.7755944\\ 19    0.8048207\\ 20    0.7700114\\ 21    0.79535383\\ 22    0.8030358\\ 23    0.834095\\ 24    0.83013743\\ 25    0.78575903\\ 26    0.7698038\\ 27    0.78611964\\ 28    0.7963768\\ 29    0.7843837\\ 30    0.80592656\\ 31    0.8201456\\ 32    0.7619479\\ 33    0.7701569\\ 34    0.80166304\\ 35    0.7766993\\ 36    0.8058027\\ 37    0.7705635\\ 38    0.80002713\\ 39    0.8332082\\ 40    0.82761836\\ 41    0.81351393\\ 42    0.806638\\ 43    0.7505792\\ 44    0.7351286\\ 45    0.75330627\\ 46    0.7428737\\ 47    0.7466384\\ 48    0.7753555\\ 49    0.75050765\\ 50    0.75374746\\ 51    0.78282124\\ 52    0.7419895\\ 53    0.7105662\\ 54    0.7580785\\ 55    0.7598398\\ 56    0.7487362\\ 57    0.7244977\\ 58    0.68685526\\ 59    0.7323056\\ 60    0.72429085\\ 61    0.6552241\\ 62    0.6904404\\ 63    0.71833575\\ 64    0.6405964\\ 65    0.6281866\\ 66    0.65224576\\ 67    0.6376215\\ 68    0.65330434\\ 69    0.6390984\\ 70    0.63119584\\ 71    0.5753724\\ 72    0.6386625\\ 73    0.63912827\\ 74    0.6520077\\ 75    0.6531849\\ 76    0.6881793\\ 77    0.6652121\\ 78    0.62627405\\ 79    0.62245405\\ 80    0.66618866\\ 81    0.66578394\\ 82    0.64282805\\ 83    0.6633131\\ 84    0.6285615\\ 85    0.64304435\\ 86    0.6171457\\ 87    0.6670722\\ 88    0.64836067\\ 89    0.64218354\\ 90    0.6318105\\ 91    0.59213907\\ 92    0.65029204\\ 93    0.6609462\\ 94    0.6348272\\ 95    0.69102645\\ 96    0.6448971\\ 97    0.6517497\\ 98    0.6713673\\ 99    0.7128595\\ 100    0.0\\ 101    0.0\\ 102    0.0\\ 103    0.0\\ 104    0.0\\ 105    0.0\\ 106    0.0\\ 107    0.0\\ 108    0.0\\ 109    0.0\\ 110    0.0\\ 111    0.0\\ 112    0.0\\ 113    0.0\\ 114    0.0\\ 115    0.0\\ 116    0.0\\ 117    0.0\\ 118    0.0\\ 119    0.0\\ 120    0.0\\ 121    0.0\\ 122    0.0\\ 123    0.0\\ 124    0.0\\ 125    0.0\\ 126    0.0\\ 127    0.0\\ 128    0.0\\ 129    0.0\\ 130    0.0\\ 131    0.0\\ 132    0.0\\ 133    0.0\\ 134    0.0\\ 135    0.0\\ 136    0.0\\ 137    0.0\\ 138    0.0\\ 139    0.0\\ 140    0.0\\ 141    0.0\\ 142    0.0\\ 143    0.0\\ 144    0.0\\ 145    0.0\\ 146    0.0\\ 147    0.0\\ 148    0.0\\ 149    0.0\\ 150    0.0\\ 151    0.0\\ 152    0.0\\ 153    0.0\\ 154    0.0\\ 155    0.0\\ 156    0.0\\ 157    0.0\\ 158    0.0\\ 159    0.0\\ 160    0.0\\ 161    0.0\\ 162    0.0\\ 163    0.0\\ 164    0.0\\ 165    0.0\\ 166    0.0\\ 167    0.0\\ 168    0.0\\ 169    0.0\\ 170    0.0\\ 171    0.0\\ 172    0.0\\ 173    0.0\\ 174    0.0\\ 175    0.0\\ 176    0.0\\ 177    0.0\\ 178    0.0\\ 179    0.0\\ 180    0.0\\ 181    0.0\\ 182    0.0\\ 183    0.0\\ 184    0.0\\ 185    0.0\\ 186    0.0\\ 187    0.0\\ 188    0.0\\ 189    0.0\\ 190    0.0\\ 191    0.0\\ 192    0.0\\ 193    0.0\\ 194    0.0\\ 195    0.0\\ 196    0.0\\ 197    0.0\\ 198    0.0\\ 199    0.0\\ 
	};
\addplot [color=red, line width=1.0pt] table[row sep=crcr]{%
0    0.6391205\\ 1    0.6889766\\ 2    0.6721418\\ 3    0.7159336\\ 4    0.7305463\\ 5    0.6937544\\ 6    0.7246565\\ 7    0.74706745\\ 8    0.7704523\\ 9    0.8168492\\ 10    0.8202761\\ 11    0.8130702\\ 12    0.8258472\\ 13    0.88361573\\ 14    0.9609176\\ 15    0.86316466\\ 16    0.89127696\\ 17    0.891133\\ 18    0.89350516\\ 19    1.0029764\\ 20    0.9544085\\ 21    0.9350386\\ 22    0.96596014\\ 23    0.98066396\\ 24    0.9588382\\ 25    1.0287973\\ 26    0.9937681\\ 27    1.0424995\\ 28    0.987352\\ 29    1.1044023\\ 30    1.0221236\\ 31    1.0209434\\ 32    1.102564\\ 33    0.9631023\\ 34    0.97398025\\ 35    0.97333574\\ 36    0.9702086\\ 37    1.0093348\\ 38    0.93757105\\ 39    0.9749074\\ 40    0.922587\\ 41    0.9168891\\ 42    0.9268385\\ 43    0.935143\\ 44    0.8880713\\ 45    0.91247857\\ 46    0.87233925\\ 47    0.87921625\\ 48    0.8710057\\ 49    0.87163013\\ 50    0.88537693\\ 51    0.86781514\\ 52    0.7909403\\ 53    0.7888036\\ 54    0.8369297\\ 55    0.8124194\\ 56    0.7675796\\ 57    0.74716216\\ 58    0.7382746\\ 59    0.73314244\\ 60    0.7217787\\ 61    0.68881756\\ 62    0.716902\\ 63    0.71136266\\ 64    0.70092666\\ 65    0.6782663\\ 66    0.6570181\\ 67    0.6890826\\ 68    0.68054074\\ 69    0.588802\\ 70    0.62442243\\ 71    0.5971787\\ 72    0.5874712\\ 73    0.5896142\\ 74    0.59337986\\ 75    0.5784652\\ 76    0.59936905\\ 77    0.5732124\\ 78    0.5704782\\ 79    0.6378875\\ 80    0.5686631\\ 81    0.5983836\\ 82    0.571758\\ 83    0.590529\\ 84    0.5991265\\ 85    0.6048789\\ 86    0.60784537\\ 87    0.5899165\\ 88    0.5980385\\ 89    0.5787344\\ 90    0.5986158\\ 91    0.63845676\\ 92    0.62765586\\ 93    0.64141566\\ 94    0.64872044\\ 95    0.66081184\\ 96    0.6896313\\ 97    0.63811594\\ 98    0.68014956\\ 99    0.6519896\\ 100    0.0\\ 101    0.0\\ 102    0.0\\ 103    0.0\\ 104    0.0\\ 105    0.0\\ 106    0.0\\ 107    0.0\\ 108    0.0\\ 109    0.0\\ 110    0.0\\ 111    0.0\\ 112    0.0\\ 113    0.0\\ 114    0.0\\ 115    0.0\\ 116    0.0\\ 117    0.0\\ 118    0.0\\ 119    0.0\\ 120    0.0\\ 121    0.0\\ 122    0.0\\ 123    0.0\\ 124    0.0\\ 125    0.0\\ 126    0.0\\ 127    0.0\\ 128    0.0\\ 129    0.0\\ 130    0.0\\ 131    0.0\\ 132    0.0\\ 133    0.0\\ 134    0.0\\ 135    0.0\\ 136    0.0\\ 137    0.0\\ 138    0.0\\ 139    0.0\\ 140    0.0\\ 141    0.0\\ 142    0.0\\ 143    0.0\\ 144    0.0\\ 145    0.0\\ 146    0.0\\ 147    0.0\\ 148    0.0\\ 149    0.0\\ 150    0.0\\ 151    0.0\\ 152    0.0\\ 153    0.0\\ 154    0.0\\ 155    0.0\\ 156    0.0\\ 157    0.0\\ 158    0.0\\ 159    0.0\\ 160    0.0\\ 161    0.0\\ 162    0.0\\ 163    0.0\\ 164    0.0\\ 165    0.0\\ 166    0.0\\ 167    0.0\\ 168    0.0\\ 169    0.0\\ 170    0.0\\ 171    0.0\\ 172    0.0\\ 173    0.0\\ 174    0.0\\ 175    0.0\\ 176    0.0\\ 177    0.0\\ 178    0.0\\ 179    0.0\\ 180    0.0\\ 181    0.0\\ 182    0.0\\ 183    0.0\\ 184    0.0\\ 185    0.0\\ 186    0.0\\ 187    0.0\\ 188    0.0\\ 189    0.0\\ 190    0.0\\ 191    0.0\\ 192    0.0\\ 193    0.0\\ 194    0.0\\ 195    0.0\\ 196    0.0\\ 197    0.0\\ 198    0.0\\ 199    0.0\\ 
	};
\addplot [color=green, line width=1.0pt] table[row sep=crcr]{%
0    0.63652676\\ 1    0.6155006\\ 2    0.6420986\\ 3    0.64427435\\ 4    0.6453096\\ 5    0.71717393\\ 6    0.72338593\\ 7    0.7530268\\ 8    0.78880507\\ 9    0.82079893\\ 10    0.8369301\\ 11    0.81704235\\ 12    0.9295885\\ 13    0.91210663\\ 14    0.96735406\\ 15    1.0268936\\ 16    1.0147493\\ 17    1.0225528\\ 18    1.1169301\\ 19    1.0457363\\ 20    1.0799127\\ 21    1.1361464\\ 22    1.1411252\\ 23    1.1884346\\ 24    1.1996088\\ 25    1.224912\\ 26    1.2899199\\ 27    1.2176933\\ 28    1.2824515\\ 29    1.2392269\\ 30    1.3421733\\ 31    1.2761033\\ 32    1.2264212\\ 33    1.3096504\\ 34    1.3260376\\ 35    1.2814265\\ 36    1.290696\\ 37    1.2356831\\ 38    1.2434224\\ 39    1.1630878\\ 40    1.2252662\\ 41    1.1133674\\ 42    1.1361926\\ 43    1.059137\\ 44    1.0646006\\ 45    1.043613\\ 46    0.98882014\\ 47    1.007787\\ 48    1.0325143\\ 49    0.9785752\\ 50    0.92747337\\ 51    0.88640136\\ 52    0.9004569\\ 53    0.8745978\\ 54    0.81377697\\ 55    0.8124259\\ 56    0.79824066\\ 57    0.77346396\\ 58    0.77411306\\ 59    0.7477254\\ 60    0.7008295\\ 61    0.68624693\\ 62    0.633495\\ 63    0.6692941\\ 64    0.6235748\\ 65    0.65006703\\ 66    0.65786564\\ 67    0.60275894\\ 68    0.64106643\\ 69    0.60194016\\ 70    0.60243356\\ 71    0.5732253\\ 72    0.585408\\ 73    0.5565243\\ 74    0.5877017\\ 75    0.5442505\\ 76    0.56525147\\ 77    0.52723783\\ 78    0.52424425\\ 79    0.5243146\\ 80    0.56358755\\ 81    0.5391629\\ 82    0.49722487\\ 83    0.5493146\\ 84    0.54148847\\ 85    0.51925474\\ 86    0.51157606\\ 87    0.52793366\\ 88    0.509339\\ 89    0.5448302\\ 90    0.5131939\\ 91    0.5713204\\ 92    0.5764506\\ 93    0.56675917\\ 94    0.60099494\\ 95    0.61772305\\ 96    0.5992689\\ 97    0.67020464\\ 98    0.6364901\\ 99    0.6425519\\ 100    0.0\\ 101    0.0\\ 102    0.0\\ 103    0.0\\ 104    0.0\\ 105    0.0\\ 106    0.0\\ 107    0.0\\ 108    0.0\\ 109    0.0\\ 110    0.0\\ 111    0.0\\ 112    0.0\\ 113    0.0\\ 114    0.0\\ 115    0.0\\ 116    0.0\\ 117    0.0\\ 118    0.0\\ 119    0.0\\ 120    0.0\\ 121    0.0\\ 122    0.0\\ 123    0.0\\ 124    0.0\\ 125    0.0\\ 126    0.0\\ 127    0.0\\ 128    0.0\\ 129    0.0\\ 130    0.0\\ 131    0.0\\ 132    0.0\\ 133    0.0\\ 134    0.0\\ 135    0.0\\ 136    0.0\\ 137    0.0\\ 138    0.0\\ 139    0.0\\ 140    0.0\\ 141    0.0\\ 142    0.0\\ 143    0.0\\ 144    0.0\\ 145    0.0\\ 146    0.0\\ 147    0.0\\ 148    0.0\\ 149    0.0\\ 150    0.0\\ 151    0.0\\ 152    0.0\\ 153    0.0\\ 154    0.0\\ 155    0.0\\ 156    0.0\\ 157    0.0\\ 158    0.0\\ 159    0.0\\ 160    0.0\\ 161    0.0\\ 162    0.0\\ 163    0.0\\ 164    0.0\\ 165    0.0\\ 166    0.0\\ 167    0.0\\ 168    0.0\\ 169    0.0\\ 170    0.0\\ 171    0.0\\ 172    0.0\\ 173    0.0\\ 174    0.0\\ 175    0.0\\ 176    0.0\\ 177    0.0\\ 178    0.0\\ 179    0.0\\ 180    0.0\\ 181    0.0\\ 182    0.0\\ 183    0.0\\ 184    0.0\\ 185    0.0\\ 186    0.0\\ 187    0.0\\ 188    0.0\\ 189    0.0\\ 190    0.0\\ 191    0.0\\ 192    0.0\\ 193    0.0\\ 194    0.0\\ 195    0.0\\ 196    0.0\\ 197    0.0\\ 198    0.0\\ 199    0.0\\ 
	};
\addplot [color=yellow, line width=1.0pt] table[row sep=crcr]{%
0    0.60410583\\ 1    0.63135785\\ 2    0.6380951\\ 3    0.6718698\\ 4    0.69982743\\ 5    0.7385068\\ 6    0.71578944\\ 7    0.7860866\\ 8    0.80446273\\ 9    0.861018\\ 10    0.9012621\\ 11    0.95404387\\ 12    0.9147667\\ 13    1.0152394\\ 14    1.0163045\\ 15    1.0878774\\ 16    1.1808654\\ 17    1.2266437\\ 18    1.296551\\ 19    1.3374051\\ 20    1.396473\\ 21    1.3561164\\ 22    1.3987482\\ 23    1.5529395\\ 24    1.4730207\\ 25    1.5861429\\ 26    1.6085976\\ 27    1.6872373\\ 28    1.5765132\\ 29    1.6582584\\ 30    1.6213582\\ 31    1.7167162\\ 32    1.7199222\\ 33    1.5704608\\ 34    1.6663691\\ 35    1.6397007\\ 36    1.6322137\\ 37    1.5752795\\ 38    1.5337358\\ 39    1.6121138\\ 40    1.4702308\\ 41    1.4234444\\ 42    1.3671877\\ 43    1.3249459\\ 44    1.3458968\\ 45    1.1531898\\ 46    1.1626123\\ 47    1.0964985\\ 48    1.0752298\\ 49    1.0810945\\ 50    0.95124906\\ 51    0.90532035\\ 52    0.8784473\\ 53    0.8904969\\ 54    0.86083865\\ 55    0.79717934\\ 56    0.77216107\\ 57    0.723768\\ 58    0.70630527\\ 59    0.6523685\\ 60    0.6401694\\ 61    0.58762527\\ 62    0.5914471\\ 63    0.5825321\\ 64    0.577143\\ 65    0.5694319\\ 66    0.5552221\\ 67    0.5077782\\ 68    0.5538574\\ 69    0.49913096\\ 70    0.51714337\\ 71    0.5243363\\ 72    0.44983947\\ 73    0.48953322\\ 74    0.50493795\\ 75    0.47081143\\ 76    0.4666257\\ 77    0.47752297\\ 78    0.46471196\\ 79    0.47889224\\ 80    0.4544196\\ 81    0.43991393\\ 82    0.4743402\\ 83    0.47319812\\ 84    0.49861544\\ 85    0.46257922\\ 86    0.47093642\\ 87    0.47007132\\ 88    0.48346308\\ 89    0.45770466\\ 90    0.45414203\\ 91    0.5122552\\ 92    0.47438487\\ 93    0.48967016\\ 94    0.52264845\\ 95    0.5423171\\ 96    0.5577392\\ 97    0.5581069\\ 98    0.52439696\\ 99    0.5458739\\ 100    0.0\\ 101    0.0\\ 102    0.0\\ 103    0.0\\ 104    0.0\\ 105    0.0\\ 106    0.0\\ 107    0.0\\ 108    0.0\\ 109    0.0\\ 110    0.0\\ 111    0.0\\ 112    0.0\\ 113    0.0\\ 114    0.0\\ 115    0.0\\ 116    0.0\\ 117    0.0\\ 118    0.0\\ 119    0.0\\ 120    0.0\\ 121    0.0\\ 122    0.0\\ 123    0.0\\ 124    0.0\\ 125    0.0\\ 126    0.0\\ 127    0.0\\ 128    0.0\\ 129    0.0\\ 130    0.0\\ 131    0.0\\ 132    0.0\\ 133    0.0\\ 134    0.0\\ 135    0.0\\ 136    0.0\\ 137    0.0\\ 138    0.0\\ 139    0.0\\ 140    0.0\\ 141    0.0\\ 142    0.0\\ 143    0.0\\ 144    0.0\\ 145    0.0\\ 146    0.0\\ 147    0.0\\ 148    0.0\\ 149    0.0\\ 150    0.0\\ 151    0.0\\ 152    0.0\\ 153    0.0\\ 154    0.0\\ 155    0.0\\ 156    0.0\\ 157    0.0\\ 158    0.0\\ 159    0.0\\ 160    0.0\\ 161    0.0\\ 162    0.0\\ 163    0.0\\ 164    0.0\\ 165    0.0\\ 166    0.0\\ 167    0.0\\ 168    0.0\\ 169    0.0\\ 170    0.0\\ 171    0.0\\ 172    0.0\\ 173    0.0\\ 174    0.0\\ 175    0.0\\ 176    0.0\\ 177    0.0\\ 178    0.0\\ 179    0.0\\ 180    0.0\\ 181    0.0\\ 182    0.0\\ 183    0.0\\ 184    0.0\\ 185    0.0\\ 186    0.0\\ 187    0.0\\ 188    0.0\\ 189    0.0\\ 190    0.0\\ 191    0.0\\ 192    0.0\\ 193    0.0\\ 194    0.0\\ 195    0.0\\ 196    0.0\\ 197    0.0\\ 198    0.0\\ 199    0.0\\ 
	};
\end{axis} 
 \end{tikzpicture}